\documentclass[journal]{IEEEtran}
\pdfoutput=1
\usepackage[T1]{fontenc}
\usepackage{ifpdf}
\usepackage{cite}
\usepackage[pdftex]{graphicx}
\usepackage[cmex10]{amsmath}
\usepackage{bm}
\usepackage{algorithmic}
\usepackage{array}
\usepackage[cmintegrals]{newtxmath}
\usepackage{algorithmic}
\usepackage{fixltx2e}
\usepackage{dblfloatfix}
\usepackage{multicol}
\usepackage{float}
\usepackage{color}

\usepackage{mathtools}

\usepackage{graphicx}
\usepackage{multicol}
\usepackage{caption}
\usepackage{subcaption}

\newtheorem{theorem}{Theorem}

\newtheorem{definition}{Definition}

\newtheorem{lemma}{Lemma}

\begin{document}

\title{Convergence Rates for Empirical Estimation of Binary Classification Bounds}

\author{Salimeh~Yasaei Sekeh, Member, IEEE,
        Morteza~Noshad, Kevin R. Moon,\\
        ~Alfred~O.~Hero, Fellow, IEEE~
\\
\\
$^{1}$Department of EECS, University of Michigan, Ann Arbor, MI, U.S.A\\
$^{2}$ Departments of Genetics and Applied Math,Yale University, New Haven, CT,  U.S.A
}

\author{Salimeh~Yasaei Sekeh,~\IEEEmembership{Member,~IEEE,}
         Morteza~Noshad, Kevin R. Moon,~\IEEEmembership{Member,~IEEE},\\
        and~Alfred~O.~Hero,~\IEEEmembership{Fellow,~IEEE}
\thanks{S. Yasaei Sekeh, M. Noshad and A. Hero are  with the Department
of Electrical Engineering and Computer Science, University of Michigan, Ann Arbor, MI, USA  e-mail: \{salimehy,noshad,hero\}$@$umich.edu}
\thanks{K. R. Moon is with the Departments of Genetics and Applied Math, Yale University, New Haven, CT,  USA  e-mail: kevin.moon$@$yale.edu}
}
\maketitle

\begin{abstract}
Bounding the best achievable error probability for binary classification problems is relevant to many applications including machine learning, signal processing, and information theory. Many bounds on the Bayes binary classification error rate depend on information divergences between the pair of class distributions. Recently, the Henze-Penrose (HP) divergence has been proposed for bounding classification error probability. We consider the problem of empirically estimating the HP-divergence from random samples. We derive a bound on the convergence rate for the Friedman-Rafsky (FR) estimator of the HP-divergence, which is related to a multivariate runs statistic for testing between two distributions. The FR estimator is derived from a multicolored Euclidean minimal spanning tree (MST) that spans the merged samples.
We obtain a concentration inequality for the Friedman-Rafsky estimator of the Henze-Penrose divergence.
We validate our results experimentally and illustrate their application to real datasets.
\end{abstract}


\IEEEpeerreviewmaketitle

\def\cmp{{\complement}}

\def\tA{{\tt A}}
\def\tB{{\tt B}}
\def\tC{{\tt C}}
\def\tD{{\tt D}}
\def\td{{\tt d}}
\def\tE{{\tt E}}
\def\tte{{\tt e}}
\def\tF{{\tt F}}
\def\tG{{\tt G}}
\def\tg{{\tt g}}
\def\ti{{\tt i}}
\def\tI{{\tt I}}
\def\tj{{\tt j}}
\def\tn{{\tt n}}
\def\tL{{\tt L}}
\def\tO{{\tt O}}
\def\tP{{\tt P}}
\def\tq{{\tt q}}
\def\ttr{{\tt r}}
\def\tP{{\tt P}}
\def\tR{{\tt R}}
\def\tS{{\tt S}}
\def\ttt{\tt t}
\def\tT{{\tt T}}
\def\ttg{{\tt g}}
\def\ttG{{\tt G}}
\def\bttg{\overline{\tg}}
\def\tu{{\tt u}}
\def\tv{{\tt v}}
\def\tV{{\tt V}}
\def\tw{{\tt w}}
\def\tx{{\tt x}}
\def\ty{{\tt y}}
\def\tz{{\tt z}}

\def\bgam{{\mbox{\boldmath$\gamma$}}}
\def\uGam{\underline\Gamma}

\def\boeta{{\mbox{\boldmath$\eta$}}}
\def\oboeta{\overline\boeta}
\def\ups{\upsilon}

\def\Om{\Omega}
\def\om{\omega}
\def\oom{\overline\omega}
\def\bttg{\mbox{\boldmath${\tt g}$}}
\def\btau{\mbox{\boldmath${\tau}$}}
\def\bom{{\mbox{\boldmath$\omega$}}}
\def\obom{\overline\bom}
\def\0bom{{\bom}^0}
\def\0obom{{\obom}^0}
\def\nbom{{\bom}_n}
\def\0nbom{{\bom}_{n,0}}
\def\n*bom{{\bom}^*_{(n)}}
\def\wt{\widetilde}
\def\wtbom{\widetilde\bom}
\def\whbom{\widehat\bom}
\def\oom{\overline\om}
\def\wtom{\widetilde\om}
\def\bOm{\mbox{\boldmath${\Om}$}}
\def\obOm{\overline\bOm}
\def\whbOm{\widehat\bOm}
\def\wtbOm{\widetilde\bOm}

\def\Gam{\Gamma}
\def\Lam{\Lambda}
\def\lam{\lambda}

\def\Ups{\Upsilon}
\def\utheta{\underline\theta}
\def\ovr{\overline r}

\def\oG{\overline G}
\def\oL{\overline L}

\def\bbC{\mathbb C}
\def\bbE{\mathbb E}
\def\bbP{\mathbb P}
\def\fB{\mathfrak B}
\def\fG{\mathfrak G}
\def\fW{\mathfrak W}
\def\bbQ{\mathbb Q}

\def\bi{\mathbf i}
\def\bj{\mathbf j}
\def\bn{\mathbf n}
\def\bt{\mathbf t}
\def\bu{\mathbf u}
\def\bw{\mathbf w}
\def\bX{\mathbf X}
\def\ubX{\underline\bX}
\def\bx{\mathbf x}
\def\ubx{\underline\bx}
\def\bY{\mathbf Y}
\def\by{\mathbf y}
\def\ubY{\underline\bY}
\def\uby{\underline\by}
\def\bZ{\mathbf Z}
\def\bz{\mathbf z}

\def\cl{\centerline}

\def\cA{\mathcal A}
\def\cB{\mathcal B}
\def\cC{\mathcal C}
\def\cD{\mathcal D}
\def\cE{\mathcal E}
\def\cF{\mathcal F}
\def\cH{\mathcal H}
\def\cK{\mathcal K}
\def\cL{\mathcal L}
\def\cN{\mathcal N}
\def\cS{\mathcal S}
\def\ocS{\overline\cS}
\def\cT{\mathcal T}
\def\cV{\mathcal V}
\def\cW{\mathcal W}
\def\ocH{\overline\cH}
\def\ocW{\overline\cW}

\def\bbB{\mathbb B}
\def\bbK{\mathbb K}
\def\bbL{\mathbb L}

\def\bbR{\mathbb R}
\def\bbS{\mathbb S}
\def\bbT{\mathbb T}
\def\bbZ{\mathbb Z}
\def\ba{\mathbf a}
\def\bg{\mathbf g}
\def\bX{\mathbf X}
\def\bx{\mathbf x}
\def\wtbx{\widetilde\bx}
\def\ui{{\underline i}}

\def\oA{{\overline A}}
\def\uA{{\underline A}}
\def\ua{{\underline a}}
\def\uua{{\underline{a_{}}}}
\def\oa{{\overline a}}
\def\uk{{\underline k}}
\def\ux{{\underline x}}
\def\wtux{\widetilde\ux}
\def\uX{{\underline X}}
\def\by{\mathbf y}
\def\uy{\underline y}
\def\bY{\mathbf Y}
\def\uY{\underline Y}

\def\uj{{\underline j}}
\def\unn{\underline n}
\def\unp{\underline p}
\def\ovp{\overline p}
\def\bx{\mathbf x}
\def\ox{\overline x}
\def\obx{\overline\bx}
\def\uz{\underline z}
\def\bz{\mathbf z}
\def\uv{\underline v}
\def\dist{\textrm{dist}}
\def\diy{\displaystyle}
\def\ov{\overline}
\def\u0{{\underline 0}}

\def\oomega{\overline\omega}
\def\oUpsilon{\overline\Upsilon}
\def\wtomega{\widetilde\omega}
\def\wtz{\widetilde z}
\def\wtheta{\widetilde\theta}
\def\wtalpha{\widetilde\alpha}
\def\wh{\widehat}
\def\oV{\overline {\mathcal V}}

\def\bI{\mathbf I}
\def\bN{\mathbf N}
\def\bbN{\mathbf N}
\def\bP{\mathbf P}
\def\bV{\mathbf V}
\def\oW{\overline W}
\def\ofW{\overline\fW}
\def\LT{{\mathbb{LT}}}
\def\mucr{{\mu_{cr}}}

\def\rA{{\rm A}}
\def\rB{{\rm B}}
\def\urB{\underline\rB}
\def\rc{{\rm c}}
\def\rC{{\rm C}}
\def\rd{{\rm d}}
\def\rD{{\rm D}}
\def\rd{{\rm d}}
\def\re{{\rm e}}
\def\rE{{\rm E}}
\def\rF{{\rm F}}
\def\rI{{\rm I}}

\def\rn{{\rm n}}

\def\rO{{\rm O}}
\def\rP{{\rm P}}
\def\rQ{{\rm Q}}
\def\rr{{\rm r}}
\def\rR{{\rm R}}

\def\rs{{\rm s}}
\def\rS{{\rm S}}
\def\rT{{\rm T}}
\def\rV{{\rm V}}

\def\rw{{\rm w}}

\def\rx{{\rm x}}
\def\ry{{\rm y}}
\def\rtr{\rm{tr}}

\def\oa{\overline a}
\def\ua{\underline a}

\def\uk{\underline k}
\def\un{\underline n}
\def\ux{\underline x}
\def\uy{\underline y}
\def\wtux{\widetilde\ux}
\def\uX{\underline X}

\def\oJ{\overline J}
\def\oP{\overline P}
\def\utC{{\underline\tC}}
\def\utD{{\underline\tD}}
\def\utE{{\underline\tE}}
\def\urB{{\underline\rB}}
\def\urC{{\underline\rC}}
\def\urD{{\underline\rD}}
\def\urE{{\underline\rE}}
\def\vng{{\varnothing}}
\def\ueta{\underline{\eta}}
\def\wt{\widetilde}
\def\fB{\mathfrak B}\def\fM{\mathfrak M}\def\fX{\mathfrak X}
 \def\cB{\mathcal B}\def\cM{\mathcal M}\def\cX{\mathcal X}
\def\mbe{\mathbf e}
\def\bu{\mathbf u}\def\bv{\mathbf v}\def\bx{\mathbf x} \def\by{\mathbf y} \def\bz{\mathbf z}
\def\om{\omega} \def\Om{\Omega}
\def\bbP{\mathbb P} \def\hw{h^{\rm w}} \def\hwi{{h^{\rm w}}}
\def\beq{\begin{eqnarray}} \def\eeq{\end{eqnarray}}
\def\beqq{\begin{eqnarray*}} \def\eeqq{\end{eqnarray*}}
\def\rd{{\rm d}} \def\Dwphi{{D^{\rm w}_\phi}}
\def\BX{\mathbf{X}}\def\Lam{\Lambda}\def\BY{\mathbf{Y}}
\def\BZ{\mathbf{Z}} \def\BN{\mathbf{N}}
\def\BV{\mathbf{V}}

\def\mwe{{D^{\rm w}_\phi}}
\def\DwPhi{{D^{\rm w}_\Phi}} \def\iw{i^{\rm w}_{\phi}}
\def\bE{\mathbb{E}}
\def\1{{\mathbf 1}} \def\fB{{\mathfrak B}}  \def\fM{{\mathfrak M}}
\def\diy{\displaystyle} \def\bbE{{\mathbb E}} \def\bu{\mathbf u}
\def\BC{{\mathbf C}} \def\lam{\lambda} \def\bbB{{\mathbb B}}
\def\bbR{{\mathbb R}}\def\bbS{{\mathbb S}}
 \def\bmu{{\mbox{\boldmath${\mu}$}}}
 \def\bPhi{{\mbox{\boldmath${\Phi}$}}}  \def\bPi{{\mbox{\boldmath{$\Pi$}}}}
 \def\bbZ{{\mathbb Z}} \def\fF{\mathfrak F}\def\mbt{\mathbf t}\def\B1{\mathbf 1}
\def\hwphi{h^{\rm w}_{\phi}}
\def\BW{\mathbf{W}} \def\bw{\mathbf{w}}

\def\beal{\begin{array}{l}}
\def\beac{\begin{array}{c}}
\def\beacl{\begin{array}{cl}}
\def\ena{\end{array}}
\def\WBJ{\mathbf{J}^{\rm w}_{\phi}}
\def\BS{\mathbf{S}}
\def\BK{\mathbf{K}}
\def\BB{\mathbf{B}}
\def\wtD{{\widetilde D}}

\def\mwe{{D^{\rm w}_\phi}}
\def\DwPhi{{D^{\rm w}_\Phi}} \def\iw{i^{\rm w}_{\phi}}
\def\bE{\mathbb{E}}
\def\1{{\mathbf 1}} \def\fB{{\mathfrak B}}  \def\fM{{\mathfrak M}}
\def\diy{\displaystyle} \def\bbE{{\mathbb E}} \def\bu{\mathbf u}
\def\BC{{\mathbf C}} \def\lam{\lambda}
\def\bbB{{\mathbb B}} \def\bbM{{\mathbb M}}
\def\bbR{{\mathbb R}}\def\bbS{{\mathbb S}}
\def\blam{{\mbox{\boldmath${\lambda}$}}}
\def\bmu{{\mbox{\boldmath${\mu}$}}} \def\bta{{\mbox{\boldmath${\eta}$}}}
\def\bzeta{{\mbox{\boldmath${\zeta}$}}}
 \def\bPhi{{\mbox{\boldmath${\Phi}$}}}  \def\bPi{{\mbox{\boldmath{$\Pi$}}}}
 \def\bbZ{{\mathbb Z}} \def\fF{\mathfrak F}\def\mbt{\mathbf t}\def\B1{\mathbf 1}
\def\hwphi{h^{\rm w}_{\phi}}
\def\BT{{\mathbf T}} \def\BW{\mathbf{W}} \def\bw{\mathbf{w}}
\def\bfe{{\mathbf e}}
\def\beps{{\mathbf \varepsilon}}

\def\beal{\begin{array}{l}}
\def\beac{\begin{array}{c}}
\def\beacl{\begin{array}{cl}}
\def\ena{\end{array}}
\def\WBJ{\mathbf{J}^{\rm w}_{\phi}}
\def\BS{\mathbf{S}}
\def\BK{\mathbf{K}}
\def\tL{\mathbf{L}}
\def\BB{\mathbf{B}}
\def\vphi{{\varphi}}
\def\rw{{\rm w}}
\def\bZ{\mathbf Z}
\def\wtf{{\widetilde f}} \def\wtg{{\widetilde g}} \def\wtG{{\widetilde G}}
\def\vphi{\varphi}
\def\rT{{\rm T}}
\def\tA{{\tt A}} \def\tB{{\tt B}} \def\tC{{\tt C}} \def\tI{{\tt I}} \def\tJ{{\tt J}} \def\tK{{\tt K}}
\def\tL{{\tt L}} \def\tP{{\tt P}} \def\tQ{{\tt Q}} \def\tS{{\tt S}}
\def\beac{\begin{array}{c}} \def\beal{\begin{array}{l}} \def\beacl{\begin{array}{cl}} \def\ena{\end{array}}
\section{Introduction}
\IEEEPARstart{D}{}ivergence measures between probability density functions are used in many signal processing applications including classification, segmentation, source separation, and clustering (see  \cite{ref6,ref7,ref8}). For more applications of divergence measures, we refer to \cite{Ba}. 

In classification problems, the Bayes error rate is the expected risk for the Bayes classifier, which assigns a given feature vector $\bx$ to the class with the highest posterior probability. The Bayes error rate is the lowest possible error rate of any classifier for a particular joint distribution. Mathematically, let $\bx_1,\bx_2,...,\bx_N \in \bbR^d$ be realizations of random vector $\BX$ and class labels $S\in\{0,1\}$, with prior probabilities $p=P(S=0)$ and $q=P(S=1)$, such that $p+q=1$. Given conditional probability densities $f_0(\bx)$ and $f_1(\bx)$, the Bayes error rate is given by
\begin{equation}\label{BER}
\epsilon=\diy\int_{\bbR^d}\min\big\{p f_0(\bx),q f_1(\bx)\big\}\rd \bx.
\end{equation}
The Bayes error rate provides a measure of classification difficulty. Thus when known, the Bayes error rate can be used to guide the user in the choice of classifier and tuning parameter selection. In practice, the Bayes error is rarely known and must be estimated from data. Estimation of the Bayes error rate is difficult due to the nonsmooth $\min$ function within the integral in (\ref{BER}). Thus, research has focused on deriving tight bounds on the Bayes error rate based on smooth relaxations of the $\min$ function. Many of these bounds can be expressed in terms of divergence measures such as the Bhattacharyya \cite{Bha} and Jensen-Shannon \cite{Lin1991}. 
Tighter bounds on the Bayes error rate can be obtained using an important divergence measure known as the Henze-Penrose (HP) divergence \cite{BH,BWHS}. 

Many techniques have been developed for estimating divergence measures. These methods can be broadly classified into two categories: (i) plug-in estimators in which we estimate the probability densities and then plug them in the divergence function, \cite{MH, Moon2014ISIT, Moon2016ISIT, MSGH} (ii) entropic graph approaches, in which the relationship between the divergence function and a graph functional in Euclidean space is derived, \cite{BWHS}, \cite{NMYH}. Examples of plug-in methods include k-nearest neighbor (K-NN) and Kernel density estimator (KDE) divergence estimators. Examples of entropic graph approaches include methods based on minimal spanning trees (MST), K-nearest neighbors graphs (K-NNG), minimal matching graphs (MMG), traveling salesman problem (TSP), and their power-weighted variants. 

Disadvantages of plug-in estimators are that these methods often require assumptions on the support set boundary and are more computationally complex than direct graph-based approaches. Thus for practical and computational reasons, the asymptotic behavior of entropic graph approaches has been of great interest. Asymptotic analysis has been used to justify graph based approaches. For instance in \cite{YOH}, the authors showed that a cross match statistic based on optimal weighted matching converges to the the HP-divergence. In \cite{MortezaICASSP2018}, a more complex approach based on the K-NNG was proposed that also converges to the HP-divergence.

The first contribution of our paper is that we obtain a bound on the convergence rates for the Friedman and Rafsky (FR) estimator of the HP-divergence, which is based on a multivariate extension of the non-parametric run length test of equality of distributions. This estimator is constructed using a multicolored MST on the labeled training set where MST edges connecting samples with dichotomous labels are colored differently from edges connecting identically labeled samples. 
While previous works have investigated the FR test statistic in the context of estimating the HP-divergence (see \cite{BWHS,WBWRS}), to the best of our knowledge its minimax MSE convergence rate has not been previously derived.
The bound on convergence rate is established by using the umbrella theorem of \cite{Yu}, for which we define a dual version of the multicolor MST. The proposed dual MST in this work is different than the standard dual MST introduced by Yukich in \cite{Yu}. 
We show that the bias rate of the FR estimator is bounded by a function of $N$, $\eta$ and $d$, as $O\big((N)^{-\eta^2\big/(d(\eta+1))}\big)$, where $N$ is the total sample size, $d$ is the dimension of the data samples $d\geq 2$, and $\eta$ is the H\"{o}lder smoothness parameter $0<\eta\leq 1$. 
We also obtain the variance rate bound as $O\big(\diy(N)^{-1}\big)$. 

The second contribution of our paper is a new concentration bound for the FR test statistic. 
The bound is obtained by establishing a growth bound and a smoothness condition for the multicolored MST.
Since the FR test statistic is not a Euclidean functional we cannot use the standard subadditivity and superadditivity approaches of \cite{Yu,Steele1986,AldousSteel1992}. Our concentration inequality is derived using a different Hamming distance approach and a dual graph to the multicolored MST. 

We experimentally validate our theoretic results. We compare the MSE theory and simulation in three experiments with various dimensions $d=2,4,8$. We observe that in all three experiments as sample size increases the MSE rate decreases and for higher dimension the rate is slower. In all sets of experiments our theory matches the experimental results. Furthermore, we illustrate the application of our results on estimation of the Bayes error rate on three real datasets.

\subsection{Related work}
Much research on minimal graphs has focused on the use of Euclidean functionals for signal processing and statistics applications such as image registration \cite{MHGM}, \cite{HHC}, pattern matching \cite{HMMG} and non-parametric divergence estimation \cite{HM}. 
A K-NNG-based estimator of R\'{e}nyi and $f$-divergence measures has been proposed in \cite{MortezaISIT2017}. 
Additional examples of direct estimators of divergence measures include statistic based on the nonparametric two sample problem, the Smirnov maximum deviation test \cite{Sm} and the Wald-Wolfowitz \cite{WW0} runs test, which have been studied in \cite{Gi}. 

Many entropic graph estimators such as MST, K-NNG, MMG and TSP have been considered for multivariate data from a single probability density $f$. In particular, the normalized weight function of graph constructions all converge almost surely to the R\'{e}nyi entropy of  $f$, \cite{St1997,Yu}. 
For $N$ uniformly distributed points, the MSE is $O(N^{-1/d})$ \cite{RY,RY1}. Later Hero et al. \cite{HCM}, \cite{HCB} reported bounds on $L_\gamma$-norm bias convergence rates of power-weighted Euclidean weight functionals of order $\gamma$ for densities $f$ belonging to the space of H\"{o}lder continuous functions $\Sigma_d(\eta,K)$ as $O\big(N^{-\alpha\eta/(\alpha\eta+1)\;1/d}\big)$, where $0<\eta\leq 1$, $d\geq 1$,  $\gamma\in(1,d)$, and $\alpha=(d-\gamma)/d$. We derive a bound on convergence rates when the density functions belong to the strong H\"{o}lder class, $\Sigma_d^S(\eta,K)$, for $0<\eta\leq 1$, $d\geq 2$ \cite{Lo}. Note that throughout the paper we assume the density functions are 
absolutely continuous and bounded with support on the unit cube $[0,1]^d$.

In \cite{RY}, Yukich introduced the general framework of continuous and quasi-additive Euclidean functionals. This has led to many convergence rate bounds of entropic graph divergence estimators.

The framework of \cite{RY} is as follows: Let $F$ be finite subset of points in $[0,1]^d$, $d\geq 2$, drawn from an underlying density. A real-valued function $L_\gamma$ defined on $F$ is called a Euclidean functional of order $\gamma$ if it is of the form $L_\gamma (F)=\min\limits_{E\in\mathcal{E}}\sum\limits_{e\in E}|e(F)|^\gamma$, where $\mathcal{E}$ is a set of graphs, $e$ is an edge in the graph $E$, $|e|$ is the Euclidean length of $e$, and $\gamma$ is called the edge exponent or power-weighting constant. The MST, TSP, and MMG are some examples for which $\gamma=1$.

Following this framework, we show that the FR test statistic satisfies the required continuity and quasi-additivity properties to obtain similar convergence rates to those predicted in \cite{RY}. What distinguishes our work from previous work is that the count of dichotomous edges in the multicolored MST is not Euclidean. Therefore, the results in \cite{St1997, Yu},\cite{HCM}, and \cite{HCB} are not directly applicable. 


Using the isoperimetric approach, Talagrand \cite{Ta} showed that when the Euclidean functional $L_\gamma$ is based on the MST or TSP, then the functional $L_\gamma$ for derived random vertices uniformly distributed in a hypercube $[0,1]^d$ is concentrated around its mean. 
Namely, with high probability the functional $L_\gamma$ and its mean do not differ by more than $C(N\log N)^{(d-\gamma)/2d}$. In this paper, we establish concentration bounds for the FR statistic:  with high probability $1-\delta$ the FR statistic differs from its mean by not more than $O\Big((N)^{(d-1)/d}\big(\log(C/\delta)\big)^{(d-1)/d}\Big)$, where $C$ is a function of $N$ and $d$.
\subsection{Organization}
This paper is organized as follows. In Section \ref{sec:HP-divergence}, we first introduce the HP-divergence and the FR multivariate test statistic. We then present the bias and variance rates of the FR-based estimator of HP-divergence followed by the concentration bounds and the minimax MSE convergence rate. Section \ref{experiments} provides simulations that validate the theory. All proofs and relevant lemmas are given in the Appendices and Supplementary Materials. 

Throughout the paper, we denote  expectation by $\bbE$ and variance by abbreviation ${\rm Var}$. Bold face type indicates random variables. 

\section{The Henze-Penrose divergence measure}\label{sec:HP-divergence}

Consider parameters $p\in(0,1)$ and $q=1-p$. We focus on estimating the HP-divergence measure between distributions $f_0$ and $f_1$ with domain $\bbR^d$ defined by
\begin{equation} \label{EQ:DP} D_p(f_0,f_1)=\diy\frac{1}{4pq}\left[\int \frac{\big(p f_0(\bx)-q f_1(\bx)\big)^2}{p f_0(\bx)+q f_1(\bx)}\;\rd \bx -(p-q)^2\right].\end{equation}
It can be verified that this measure is bounded between 0 and 1 and if $f_0(\bx)=f_1(\bx)$, then $D_p=0$. In contrast with some other divergences such as the Kullback-Liebler \cite{KL} and R\'{e}nyi divergences \cite{R}, the HP-divergence is symmetrical, i.e., $D_p(f_0,f_1)=D_q(f_1,f_0)$. By invoking (3) in \cite{BWHS}, one can rewrite $D_p$ in the alternative form:
\beqq D_p(f_0,f_1)=1-A_p(f_0,f_1)=\diy \frac{u_p(f_0,f_1)}{4 p q}-\diy \frac{(p-q)^2}{4 p q},\eeqq
where
\beqq\begin{array}{ccl} A_p(f_0,f_1):&=&\diy \int\frac{f_0(\bx) f_1(\bx)}{p f_0(\bx)+q f_1(\bx)}\;\rd \bx\\[15pt]
&=&\diy\bbE_{f_0}\Big[\big(p\;\frac{f_0(\BX)}{f_1(\BX)}+q\big)^{-1}\Big],\\[15pt]
\diy u_p(f_0,f_1)&=&1- 4 p q\; A_p(f_0,f_1).\ena\eeqq

Throughout the paper, we refer to $A_p(f_0,f_1)$ as the HP-integral. The HP-divergence measure belongs to the class of $\phi$-divergences \cite{AS}. For the special case $p=0.5$, the divergence (\ref{EQ:DP}) becomes the symmetric $\chi^2$-divergence and is similar to the Rukhin $f$-divergence. See \cite{Cha}, \cite {Ru}.

\subsection{The Multivariate Runs Test Statistic}

The MST is a graph of minimum weight among all graphs $\mathcal{E}$ that span $n$ vertices. 
The MST has many applications including pattern recognition \cite{To}, clustering \cite{Za}, nonparametric regression \cite{BLJN}, and testing of randomness \cite{HJ}.  
In this section we focus on the FR multivariate two sample test statistic constructed from the MST.

Assume that sample realizations from $f_0$ and $f_1$, denoted by $\mathfrak{X}_m\in \bbR^{m\times d}$ and $\mathfrak{Y}_n\in \bbR^{n\times d}$, respectively, are available. Construct an MST spanning the samples from both $f_0$ and $f_1$ and color the edges in the MST that connect dichotomous samples green and color the remaining edges black.
The FR test statistic $\mathfrak{R}_{m,n}:=\mathfrak{R}_{m,n}(\mathfrak{X}_m,\mathfrak{Y}_n)$ is the number of green edges in the MST.
Note that the test assumes a unique MST, therefore all inter point distances between data points must be distinct. We recall the following theorem from \cite{BH} and \cite{BWHS}:
\begin{theorem}\label{FR:approximation}
As $m\rightarrow \infty$ and $n\rightarrow \infty$ such that $\diy\frac{m}{n+m}\rightarrow p$ and $\diy\frac{n}{n+m}\rightarrow q$, 
\beq \label{FR.theorem} 1-\mathfrak{R}_{m,n}(\mathfrak{X}_m,\mathfrak{Y}_n)\;\diy\frac{m+n}{2mn} \rightarrow  D_p(f_0 , f_1), \; \; \;  a.s.\eeq
\end{theorem}
In the next section we obtain bounds on the MSE convergence rates of the FR approximation for HP-divergence between densities that belong to $\Sigma_d^S(\eta,K)$, the class of strong H\"{o}lder continuous functions with Lipschitz constant $K$ and smoothness parameter $0<\eta\leq 1$, \cite{Lo}:
\begin{definition}\label{def:strong.Holder}
(Strong H\"{o}lder class) Let $\mathcal{X}\subset \bbR^d$ be a compact space. The strong H\"{o}lder class $\Sigma_d^s(\eta,K)$, with $\eta$-H\"{o}lder parameter, of functions with the $L_d$-norm, consists of the functions $g$ that satisfy 
\beq\label{SHC}\begin{array}{l}\Big\{g: \big\|g(\bz)-p_{\bx}^{\lfloor \eta \rfloor}(\bz)\big\|_d\leq K\; g(\bx)\; \big\|\bx-\bz\big\|_d^\eta,\;\; \bx,\;\bz\in \mathcal{X}\Big\}, \end{array}\eeq
where $p_{\bx}^k(\bz)$ is the Taylor polynomial (multinomial) of $g$ of order $k$ expanded about the point $\bx$ and $\lfloor\eta\rfloor$ is defined as the greatest integer strictly less than $\eta$. Note that  for the standard H\"{o}lder class the term $g(\bx)$ in the RHS of  (\ref{SHC}) is omitted. 
\end{definition}
In what follows, we will use both notations $\mathfrak{R}_{m,n}$ and $\mathfrak{R}_{m,n}(\mathfrak{X}_m,\mathfrak{Y}_n)$ for the FR statistic over the combined samples. 
\subsection{Convergence Rates}\label{Convergence Rates}

In this subsection we obtain the mean convergence rate bounds for general non-uniform Lebesgue densities $f_0$ and $f_1$ belonging to the strong H\"{o}lder class $\Sigma^S_{d}(\eta, K)$. Since the expectation of $\mathfrak{R}_{m,n}$ can be closely
approximated by the sum of the expectation of the FR statistic constructed on a dense partition of $[0,1]^d$, then $\mathfrak{R}_{m,n}$ is a quasi-additive functional in mean. The family of bounds (\ref{bound:Bias1}) in Appendix B enables us to achieve the minimax convergence rate for the mean under the strong H\"{o}lder class assumption with smoothness parameter $0<\eta\leq 1$, $d\geq 2$: 
\begin{theorem} \label{thm:bias.opt}
(\rm Convergence Rate of the Mean) Let $d\geq 2$, and $\mathfrak{R}_{m,n}$ be the FR statistic for samples drawn from strong H\"{o}lder continuous and bounded density functions $f_0$ and $f_1$ in $\Sigma^{S}_{d}(\eta, K)$. Then for $d\geq 2$,
\begin{equation}\label{bound:Bias.opt}
 \begin{array}{l}\diy \left|\frac{\bbE\big[\mathfrak{R}_{m,n}\big]}{m+n}-2pq\int \frac{f_0(\bx) f_1(\bx)}{p f_0(\bx)+q f_1(\bx)}\;\rd\bx \right|\\[15pt]
\qquad\qquad\qquad \qquad \quad\leq O\left((m+n)^{-\eta^2\big/(d(\eta+1))}\right).
 \ena \end{equation}
\end{theorem}
This bound holds over the class of Lebesgue densities $f_0,f_1\in\Sigma_d^{s}(\eta,K)$, $0<\eta\leq 1$. Note that this assumption can be relax to $f_0 \in \Sigma_d^{s}(\eta,K_0)$ and $f_1\in\Sigma_d^s(\eta,K_1)$ that is Lebesgue densities $f_0$ and $f_1$ belong to the Strong H\"{o}lder class with the same H\"{o}lder parameter $\eta$ and different constants $K_0$ and $K_1$ respectively. 

The following variance bound uses the Efron-Stein inequality \cite{ES}. Note that in Theorem \ref{Tight.bound:Variance} we do not impose any strict assumptions. we only assume that the density functions are 
absolutely continuous and bounded with support on the unit cube $[0,1]^d$. Appendix C contains the proof.
\begin{theorem}\label{Tight.bound:Variance}
The variance of the HP-integral estimator based on the FR statistic, $\diy\mathfrak{R}_{m,n}\big/(m+n)$ is bounded by
\beq\label{var:bound} Var\Big(\diy\frac{\mathfrak{R}_{m,n}(\mathfrak{X}_m,\mathfrak{Y}_n)}{m+n}\Big)\leq \diy \frac{32\; c_d^2\; q}{(m+n)},\eeq
where the constant $c_d$ depends only on $d$.
\end{theorem}
By combining Theorem \ref{thm:bias.opt} and Theorem \ref{Tight.bound:Variance} we obtain the MSE rate of the form $O\left(m+n)^{-\eta^2/(d(\eta+1))}\right)+O\left((m+n)^{-1}\right)$.
Fig. \ref{fig2-0} indicates a heat map showing the MSE rate as a function of $d$ and $N=m=n$. The heat map shows that the MSE rate of the FR test statistic-based estimator given in (\ref{FR.theorem}) is small for large sample size $N$. 
\begin{figure}[h]
\centering
  \includegraphics[width=1\linewidth]{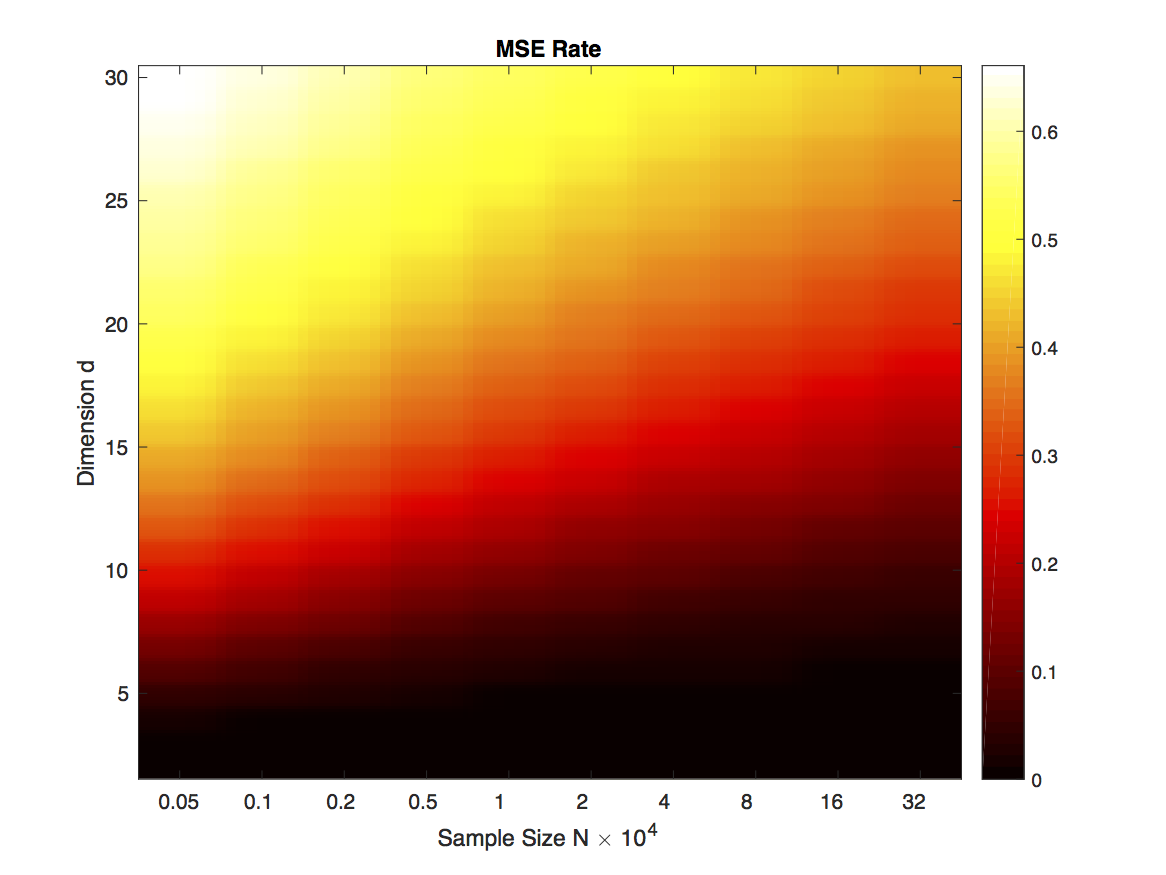}
  \caption{{\small Heat map of the theoretical MSE rate of the FR estimator of the HP-divergence based on Theorems \ref{thm:bias.opt} and \ref{Tight.bound:Variance} as a function of dimension and sample size when $N=m=n$. Note the color transition (MSE) as sample size increases for high dimension. For fixed sample size $N$ the MSE rate degrades in higher dimensions.}}
  \label{fig2-0}
\end{figure}

\subsection{Proof Sketch of Theorem \ref{thm:bias.opt}}

In this subsection, we first establish subadditivity and superadditivity properties of the FR statistic which will be employed to derive the MSE convergence rate bound. This will establish that the mean of the FR test statistic is a quasi-additive functional:
\begin{theorem}\label{lema:1.1}
Let $\mathfrak{R}_{m,n}(\mathfrak{X}_m,\mathfrak{Y}_n)$ be the number of edges that link nodes from differently labeled samples $\mathfrak{X}_m=\{\BX_1,\dots,\BX_m\}$ and $\mathfrak{Y}_n=\{\BY_1,\dots,\BY_n\}$ in $[0,1]^d$. Partition $[0,1]^d$ into $l^d$ equal volume subcubes $Q_i$ such that $m_i$ and $n_i$ are the number of samples from $\{\BX_1,\dots,\BX_m\}$ and $\{\BY_1,\dots,\BY_n\}$, respectively, that fall into the partition $Q_i$. Then there exists a constant $c_1$ such that
\begin{equation}\begin{array}{l}\label{Eq:3.2.1} \bbE\Big[\mathfrak{R}_{m,n}(\mathfrak{X}_m,\mathfrak{Y}_n)\Big]\\
\leq \diy\sum\limits_{i=1}^{l^d} \bbE\Big[\mathfrak{R}_{m_i,n_i}\big((\mathfrak{X}_m,\mathfrak{Y}_n)\cap Q_i\big)\Big]+2\;c_1\;l^{d-1}\;(m+n)^{1/d}. \ena \end{equation}
Here $\mathfrak{R}_{m_i,n_i}$ is the number of dichotomous edges in partition $Q_i$. 
Conversely, for the same conditions as above on partitions $Q_i$, there exists a constant $c_2$ such that
\begin{equation}\label{Eq:3.2.2}\begin{array}{l} \bbE\Big[\mathfrak{R}_{m,n}(\mathfrak{X}_m,\mathfrak{Y}_n)\Big]\\
 \geq \diy\sum\limits_{i=1}^{l^d} \bbE\Big[\mathfrak{R}_{m_i,n_i}\big((\mathfrak{X}_m,\mathfrak{Y}_n)\cap Q_i\big)\Big]-2\;c_2\;l^{d-1}\;(m+n)^{1/d}. \ena\end{equation}
\end{theorem}
The inequalities (\ref{Eq:3.2.1}) and (\ref{Eq:3.2.2}) are inspired by corresponding inequalities in \cite{HCM} and \cite{HCB}. The full proof is given in Appendix A. The key result in the proof is the inequality: 
$$\mathfrak{R}_{m,n}(\mathfrak{X}_m,\mathfrak{Y}_n)\leq \sum\limits_{i=1}^{l^d}\mathfrak{R}_{m_i,n_i}\big((\mathfrak{X}_m,\mathfrak{Y}_n)\cap Q_i\big)+2|D|,$$
where $|D|$ indicates the number of all edges of the MST which intersect two different partitions. 

Furthermore, we adapt the theory developed in \cite{Yu,HCM} to derive the MSE convergence rate of the FR statistic-based estimator by defining a dual MST and dual FR statistic, denoted by $\rm{MST}^*$ and $\mathfrak{R}^*_{m,n}$ respectively (see Fig. \ref{fig1}): 
\bigskip
\begin{definition}\label{def:dual} \def \bbF{\mathbb{F}}
{\rm (Dual MST, ${\rm MST}^*$ and dual FR statistic $\mathfrak{R}_{m,n}^*$)}  Let $\bbF_i$ be the set of corner points of the subsection $Q_i$ for $ 1\leq i \leq l^d$. Then we define ${\rm MST}^*(\mathfrak{X}_m\cup\mathfrak{Y}_n \cap Q_i)$ as the boundary MST graph of partition $Q_i$ \cite{Yu}, which contains $\mathfrak{X}_m$ and $\mathfrak{Y}_n$ points falling inside the section $Q_i$ and those corner points in $\bbF_i$ which minimize total MST length. Notice it is allowed to connect the MSTs in $Q_i$ and $Q_j$ through points strictly contained in $Q_i$ and $Q_j$ and corner points are taking into account under condition of minimizing total MST length. Another word, the dual MST can connect the points in $Q_i\cup Q_j$ by direct edges to pair to another point in $Q_i\cup Q_j$ or the corner the corner points (we assume that all corner points are connected) in order to minimize the total length. To clarify this, assume that there are two points in $Q_i\cup Q_j$, then the dual MST consists of the two edges connecting these points to the corner if they are closed to a corner point otherwise dual MST consists of an edge connecting one to another.  
Further, we define $\mathfrak{R}_{m,n}^*(\mathfrak{X}_m,\mathfrak{Y}_n \cap Q_i)$ as the number of edges in ${\rm MST}^*$ graph connecting nodes from different samples and number of edges connecting to the corner points. Note that the edges connected to the corner nodes (regardless of the type of points) are always counted in dual FR test statistic $\mathfrak{R}^*_{m,n}$. \end{definition}
\begin{figure}[h]
 \includegraphics[width=9cm]{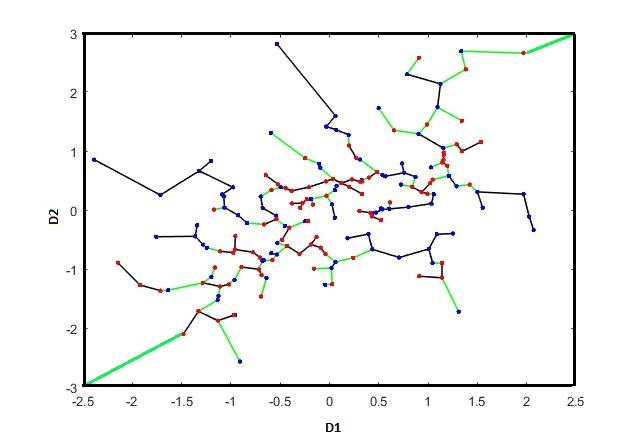}
\caption[]{{\small The dual MST spanning the merged set $\mathfrak{X}_m$ (blue points) and $\mathfrak{Y}_n$ (red points) drawn from two Gaussian distributions. The dual FR statistic ($\mathfrak{R}^*_{m,n}$) is the number of edges in the ${\rm MST}^*$ (contains nodes in $\mathfrak{X}_m\cup\mathfrak{Y}_n\cup\{2\;\hbox{corner points}\}$) that connect samples from different color nodes and corners (denoted in green). Black edges are the non-dichotomous edges in the ${\rm MST}^*$.}}
\label{fig1}
\end{figure}
In Appendix B, we show that the dual FR test statistic is a quasi-additive functional in mean and $\mathfrak{R}^*_{m,n}(\mathfrak{X}_m,\mathfrak{Y}_n)\geq \mathfrak{R}_{m,n} (\mathfrak{X}_m,\mathfrak{Y}_n)$. This property holds true since ${\rm MST}(\mathfrak{X}_m,\mathfrak{Y}_n)$ and ${\rm MST}^*(\mathfrak{X}_m,\mathfrak{Y}_n)$ graphs can only be different in the edges connected to the corner nodes, and in $\mathfrak{R}^*(\mathfrak{X}_m,\mathfrak{Y}_n)$ we take all of the edges between these nodes and corner nodes into account. 

To prove Theorem \ref{thm:bias.opt}, we partition $[0,1]^d$ into $l^d$ subcubes. Then by applying Theorem \ref{lema:1.1} and the dual MST we derive the bias rate in terms of partition parameter $l$ (see (\ref{bound:Bias1}) in Theorem \ref{thm:bias}). See Appendix B and Supplementary Materials for details. According to (\ref{bound:Bias1}), for $d\geq 2$, and $l=1,2,\dots$, the slowest rates as a function of $l$ are $l^d(m+n)^{\eta/d}$ and $l^{-\eta d}$. Therefore we obtain an $l$-independent bound by letting $l$ be a function of $m+n$ that minimizes the maximum of these rates i.e.
$$l(m+n)=arg\; \underset{l}\min\; \max\Big\{l^d(m+n)^{-\eta/d}, l^{-\eta d}\Big\}.$$
The full proof of the bound in (\ref{thm:bias.opt}) is given in Appendix B.  
\subsection{Concentration Bounds}\label{subsection:Con.Boun}

Another main contribution of our work in this part is to provide an exponential inequality convergence bound derived for the FR estimator of the HP-divergence. The error of this estimator can be decomposed into a bias term and a variance-like term via the triangle inequality:
\beqq\begin{array}{l} \left|\mathfrak{R}_{m,n}-\diy\int \frac{f_0(\bx)f_1(\bx)}{pf_0(\bx)+q f_1(\bx)}\rd\bx\right|\leq \mathop{\underbrace{\left|\mathfrak{R}_{m,n}-\bbE\big[\mathfrak{R}_{m,n}\big]\right|}}\limits_{\hbox{variance-like term}}\\
\\
\qquad\quad+\mathop{\underbrace{\left|\bbE\big[\mathfrak{R}_{m,n}\big]-\diy\int \frac{f_0(\bx)f_1(\bx)}{pf_0(\bx)+q f_1(\bx)}\rd\bx\right|}\limits_{\hbox{bias term}}}.
\ena\eeqq
The bias bound was given in Theorem \ref{thm:bias.opt}. Therefore we focus on  an exponential concentration bound for the variance-like term. One application of concentration bounds is to employ these bounds to compare confidence intervals on the HP-divergence measure in terms of the FR estimator. In \cite{SP} and \cite{SP2} the authors provided an exponential inequality convergence bound for an estimator of R\'{e}ny divergence for a smooth H\"{o}lder class of densities on the $d$-dimensional unite cube $[0,1]^d$. 
We show that if $\mathfrak{X}_m$ and $\mathfrak{Y}_n$ are the set of $m$ and $n$ points drawn from any two distributions $f_0$ and $f_1$ respectively, the FR criteria $\mathfrak{R}_{m,n}$ is tightly concentrated. Namely, we establish that with high probability, $\mathfrak{R}_{m,n}$ is within 
$$1-\;O\Big((m+n)^{-2/d} {\epsilon^*}^{2}\Big)$$
 of its expected value, where $\epsilon^*$ is the solution of the following convex optimization problem:
\begin{equation}\label{convexoptimization}\begin{aligned}
& \underset{\epsilon\geq 0}{\text{min}}
& & \diy C'_{m,n}(\epsilon) \; \exp\Big(\diy\frac{-t^{d/(d-1)}}{8(4\epsilon)^{d/(d-1)}(m+n)}\Big)\\
& \text{subject to}
& & \epsilon\geq O\big(7^{d+1}(m+n)^{1/d}\big),
\end{aligned}\end{equation}
where
\begin{equation}\label{eq:concen:ep} \begin{array}{l}
C'_{m,n}(\epsilon)\\
\qquad=8\bigg(1-\;O\Big((m+n)^{-2/d} {\epsilon}^{2}\Big)\bigg)^{-2}.
\ena\end{equation}
See Appendix D for more detail. 
 Indeed, we first show the concentration around the median. A median is by definition any real number $M_e$  that satisfies the inequalities $P(X\leq M_e)\geq \diy 1/2$  and $P(X\geq M_e)\geq \diy 1/2$. To derive the concentration results, the properties of growth bounds and smoothness for $\mathfrak{R}_{m,n}$, given in Appendix D, are exploited. 
\begin{theorem}\label{Concentration around the median}\def\ep{\epsilon}
{\rm (Concentration around the median)} Let $M_e$ be a median of $\mathfrak{R}_{m,n}$ which implies that $P\big(\mathfrak{R}_{m,n}\leq M_e\big)\geq \diy 1/2$. Recall $\epsilon^*$ from (\ref{convexoptimization}) then we have 
\beq\label{Concentration:Mediam} \begin{array}{l} \diy P\left(\big|\mathfrak{R}_{m,n}(\mathfrak{X}_m,\mathfrak{Y}_n)-M_e\big|\geq t\right)\\[9pt]
\qquad \qquad \leq \diy C'_{m,n}(\ep^*) \; \exp\Big(\diy\frac{-t^{d/(d-1)}}{8(4\epsilon^*)^{d/(d-1)}(m+n)}\Big).\ena\eeq
\end{theorem}
\def\ep{\epsilon}
\begin{theorem}\label{concentration}
{\rm (Concentration of $\mathfrak{R}_{m,n}$ around the mean)} Let $\mathfrak{R}_{m,n}$ be the FR statistic. Then 
\begin{equation}\label{Ineq:concentration} \begin{array}{l} P\left(\big|\mathfrak{R}_{m,n}-\bbE[\mathfrak{R}_{m,n}]\big| \geq t \right)\\[8pt]
\qquad \quad\leq C'_{m,n}(\ep^*) \exp\left(\diy\frac{-(t/(2\ep^*))^{d/(d-1)}}{(m+n)\;\tilde{C}}\right). 
\end{array}\end{equation}
Here $\tilde{C}=8 (4)^{d/(d-1)}$ and the explicit form for $C'_{m,n}(\ep^*)$ is given by (\ref{eq:concen:ep}) when $\ep=\ep^*$. 
\end{theorem}

See Appendix D for full proofs of Theorems \ref{Concentration around the median} and \ref{concentration}. Here we  sketch  the proofs. The proof of the concentration inequality for $\mathfrak{R}_{m,n}$, Theorem \ref{concentration}, requires 
involving the median $M_e$, where $P(\mathfrak{R}_{m,n}\leq M_e)\geq 1/2$, inside the probability term by using
\begin{equation*}
\big|\mathfrak{R}_{m,n}-\bbE[\mathfrak{R}_{m,n}]\big|\leq |\mathfrak{R}_{m,n}-M_e\big|+|\bbE[\mathfrak{R}_{m,n}]-M_e\big|.
\end{equation*}

To prove the expressions for the concentration around the median, Theorem \ref{Concentration around the median}, we first consider the $h^d$ uniform partitions of $[0,1]^d$, with edges parallel to the coordinate axes having edge lengths $h^{-1}$ and volumes $h^{-d}$. Then by applying the Markov inequality we  show that with at least probability $\diy1- \big(\delta^h_{m,n}/\ep\big)$, where $\delta^h_{m,n}=O\big(h^{d-1}(m+n)^{1/d}\big)$, the FR statistic $\mathfrak{R}_{m,n}$ is subadditive with $2\epsilon$ threshold. Afterward, owing to the induction method \cite{Yu}, the growth bound can be derived with at least probability $1-\big(\diy h\;\delta^h_{m,n}\big/\epsilon\big)$. The growth bound explains that with high probability there exists a constant depending on $\epsilon$ and $h$, $C_{\epsilon,h}$, such that $\mathfrak{R}_{m,n}\leq C_{\epsilon,h}\big(m\;n\big)^{1-1/d}$. 
Applying the law of total probability and semi-isoperimetric inequality (\ref{lem13:0.0}) in Lemma \ref{Eq:isoperimetry} gives us (\ref{eq:appendix.theorem5}). By considering the solution to convex optimization problem (\ref{convexoptimization}), i.e. $\ep^*$ and optimal $h=7$ the claimed results (\ref{Concentration:Mediam}) and (\ref{Ineq:concentration}) are derived. The only constraint here is that $\epsilon$ is lower bounded by a function of $\delta^h_{m,n}=O\big(h^{d-1}(m+n)^{1/d}\big)$. 

Next, we  provide a bound for the variance-like term with high probability at least $1-\delta$. According to the previous results we expect that this bound depends on $\epsilon^*$, $d$, $m$ and $n$. The proof is short and is given in Appendix D. 
\begin{theorem}\label{prob.variance}
{\rm (Variance-like bound for $\mathfrak{R}_{m,n}$)} Let $\mathfrak{R}_{m,n}$ be the FR statistic. 
With at least probability $1-\delta$ we have
\begin{equation}\label{main:var.bound} \begin{array}{l}\diy \left|\mathfrak{R}_{m,n}-\bbE[\mathfrak{R}_{m,n}]\right|\\[8pt]
\qquad\qquad \diy\leq O\left(\epsilon^*\;(m+n)^{(d-1)/d}\Big(\log\big(C'_{m,n}(\ep^*)\big/\delta\big)\Big)^{(d-1)/d}\right).\ena\end{equation}
Or equivalently 
\begin{equation} \begin{array}{l} \diy \left|\diy\frac{\mathfrak{R}_{m,n}}{m+n}-\frac{\bbE[\mathfrak{R}_{m,n}]}{m+n}\right|\\[10pt]
\qquad\qquad \leq \diy O\left(\epsilon^*\; (m+n)^{-1/d}\Big(\log\big(C'_{m,n}(\ep^*)\big/\delta\big)\Big)^{(d-1)/d}\right),\ena\end{equation}
where  $C'_{m,n}(\ep^*)$ depends on $m,\;n$, and $d$ is given in (\ref{eq:concen:ep}) when $\ep=\ep^*$. 
\end{theorem}

\section{Numerical Experiments}\label{experiments}
\subsection{Simulation Study}
In this section, we apply the FR statistic estimate of the HP-divergence to both simulated and real data sets. We present results of a simulation study that evaluates the proposed bound on the MSE. We numerically validate the theory stated in Subsection \ref{Convergence Rates} and \ref{subsection:Con.Boun} using multiple simulations. In the first set of simulations, We consider two multivariate Normal random vectors $\BX$, $\BY$ and perform three experiments $d=2,4,8$, to analyze the FR test statistic-based estimator performance as the sample sizes $m$, $n$ increase. For the three dimensions $d=2,4,8$ we generate samples from two normal distributions with identity covariance and shifted means: $\mu_1=[0,0]$, $\mu_2=[1,0]$ and $\mu_1=[0,0,0,0]$, $\mu_2=[1,0,0,0]$ and $\mu_1=[0,0,...,0]$, $\mu_2=[1,0,...,0]$ when $d=2$, $d=4$ and $d=8$ respectively. For all of the following experiments the sample sizes for each class are equal ($m=n$). 
\begin{figure}[h]
\centering
  \includegraphics[width=1\linewidth]{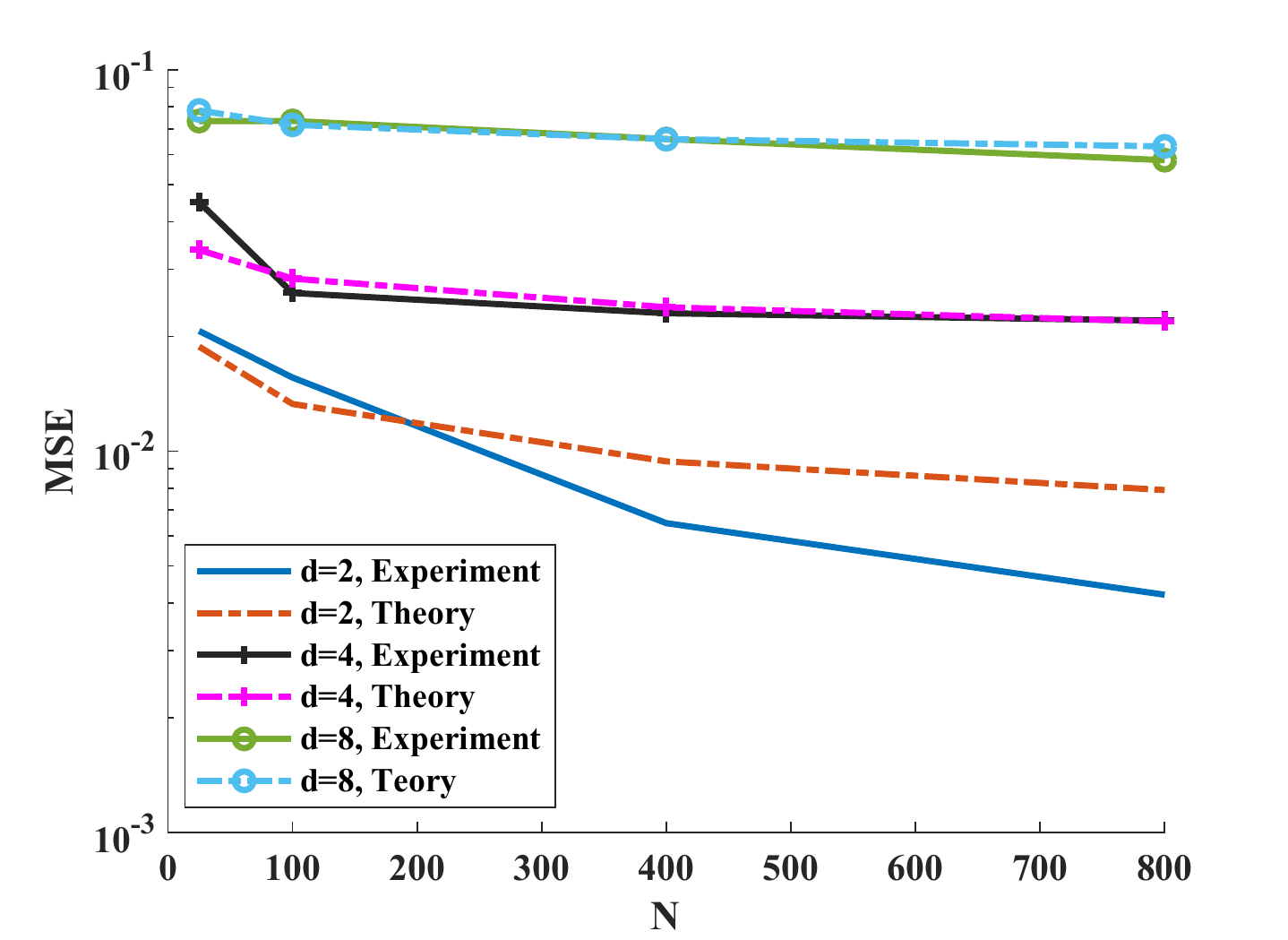}
  \caption{{\small Comparison of the bound on the MSE theory and experiments for $d=2,4,8$ standard Gaussian random vectors versus sample size from 100 trials.}}
  \label{fig2}
\end{figure}
We vary $N=m=n$ up to $800$. From Fig. \ref{fig2} we deduce that when the sample size increases the MSE decreases such that for higher dimensions the rate is slower. Furthermore we compare the experiments with the theory in Fig. \ref{fig2}. Our theory generally matches the experimental results. However, the MSE for the experiments tends to decrease to zero faster than the theoretical bound. Since the Gaussian distribution has a smooth density, this suggests that a tighter bound on the MSE may be possible by imposing stricter assumptions on the density smoothness as in \cite{MSGH}.
\begin{figure}[h]
\centering
  \includegraphics[width=1\linewidth]{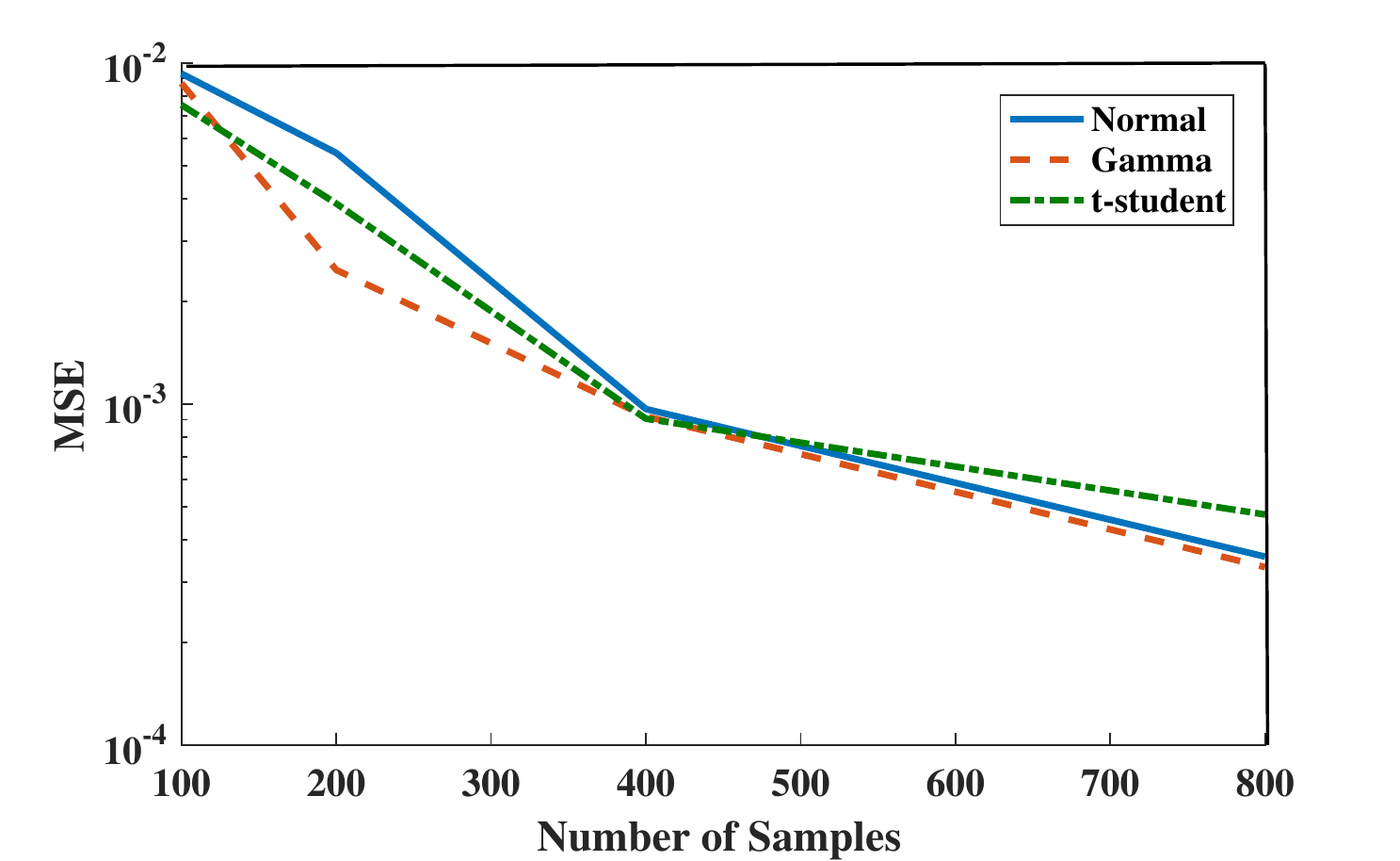}
  \caption{{\small Comparison of experimentally predicted MSE of the FR-statistic as a function of sample size $m=n$ in various distributions Standard Normal, Gamma ($\alpha_1=\alpha_2=1, \; \beta_1=\beta_2=1,\;\rho=0.5$) and Standard t-Student.}}
  \label{fig3}
\end{figure}

In our next simulation we compare three bivariate cases: First, we generate samples from a standard Normal distribution. Second, we consider a distinct smooth class of distributions i.e. binomial Gamma density with standard parameters and dependency coefficient $\rho=0.5$. Third, we generate samples from Standard t-student distributions. Our goal in this experiment is to compare the MSE of the HP-divergence estimator between two identical distributions, $f_0=f_1$, when $f_0$ is one of the Gamma, Normal, and t-student density function. In Fig. \ref{fig3}, we observe that the MSE decreases as $N$ increases for all three distributions.


\subsection{Real Datasets}
We now show the results of applying the FR test statistic to estimate the HP-divergence using three different real datasets, \cite{Lichman2013}: 
\begin{itemize}
\item Human Activity Recognition (HAR), Wearable Computing, Classification of Body Postures and Movements (PUC-Rio): This dataset contains 5 classes (sitting-down, standing-up, standing, walking, and sitting) collected on 8 hours of activities of 4 healthy subjects. 
\item Skin Segmentation dataset (SKIN): The skin dataset is collected by randomly sampling B,G,R values from face images of various age groups (young, middle, and old), race groups (white, black, and asian), and genders obtained from the FERET and PAL databases \cite{SKIN}.
\item Sensorless Drive Diagnosis (ENGIN) dataset: In this dataset features are extracted from electric current drive signals. The drive has intact and defective components. The dataset contains 11 different classes with different conditions. Each condition has been measured several times under 12 different operating conditions, e.g. different speeds, load moments and load forces. 
\end{itemize}

We focus on two classes from each of the HAR, SKIN, and ENGIN datasets.
\begin{figure}[h]
\centering
  \includegraphics[width=1.05\linewidth]{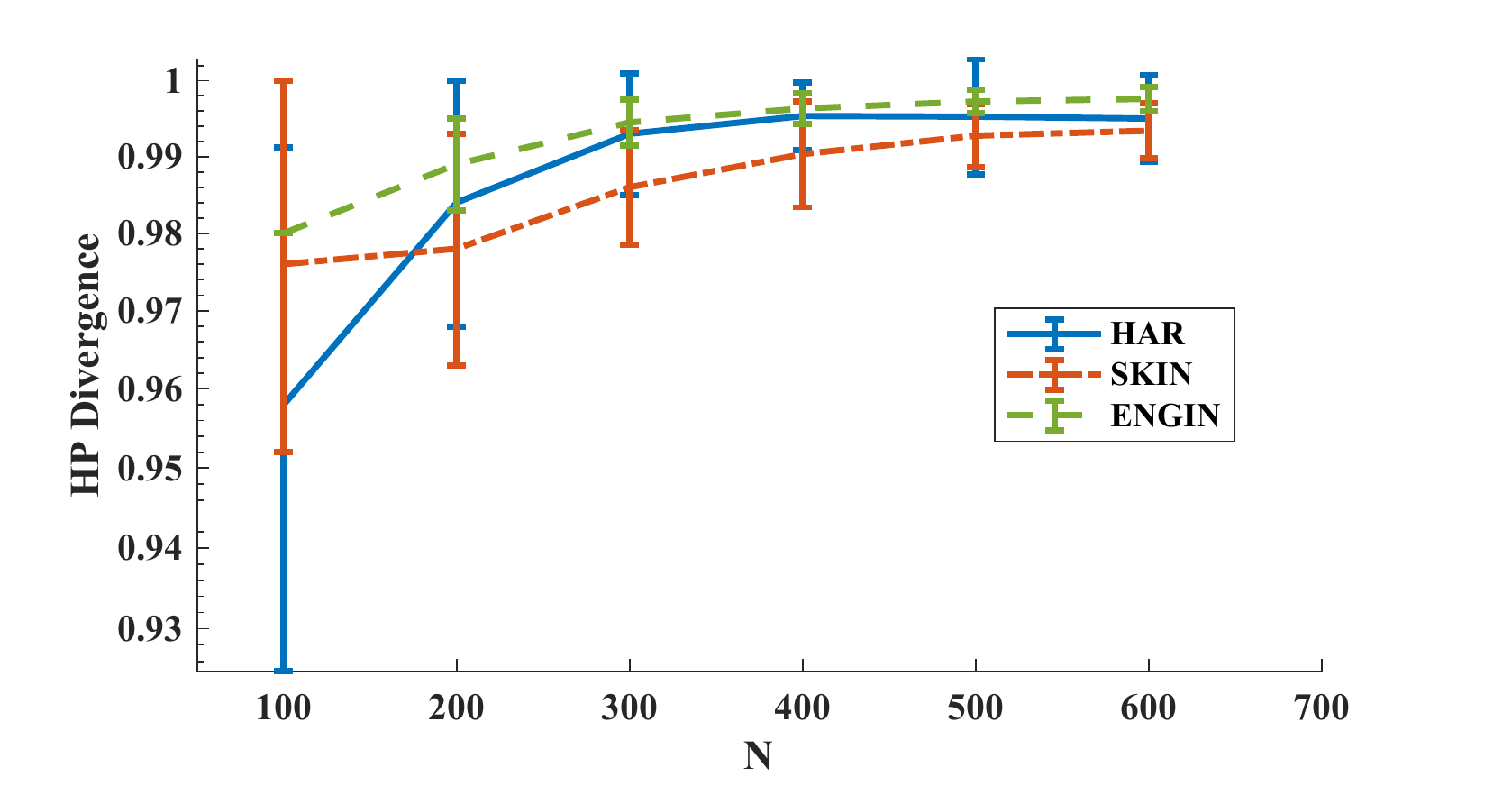}
  \caption{{\small HP-divergence vs. sample size for three real datasets HAR, SKIN, and ENGIN. }}
  \label{fig4}
\end{figure}
\begin{figure}[h]
\centering
  \includegraphics[width=1\linewidth]{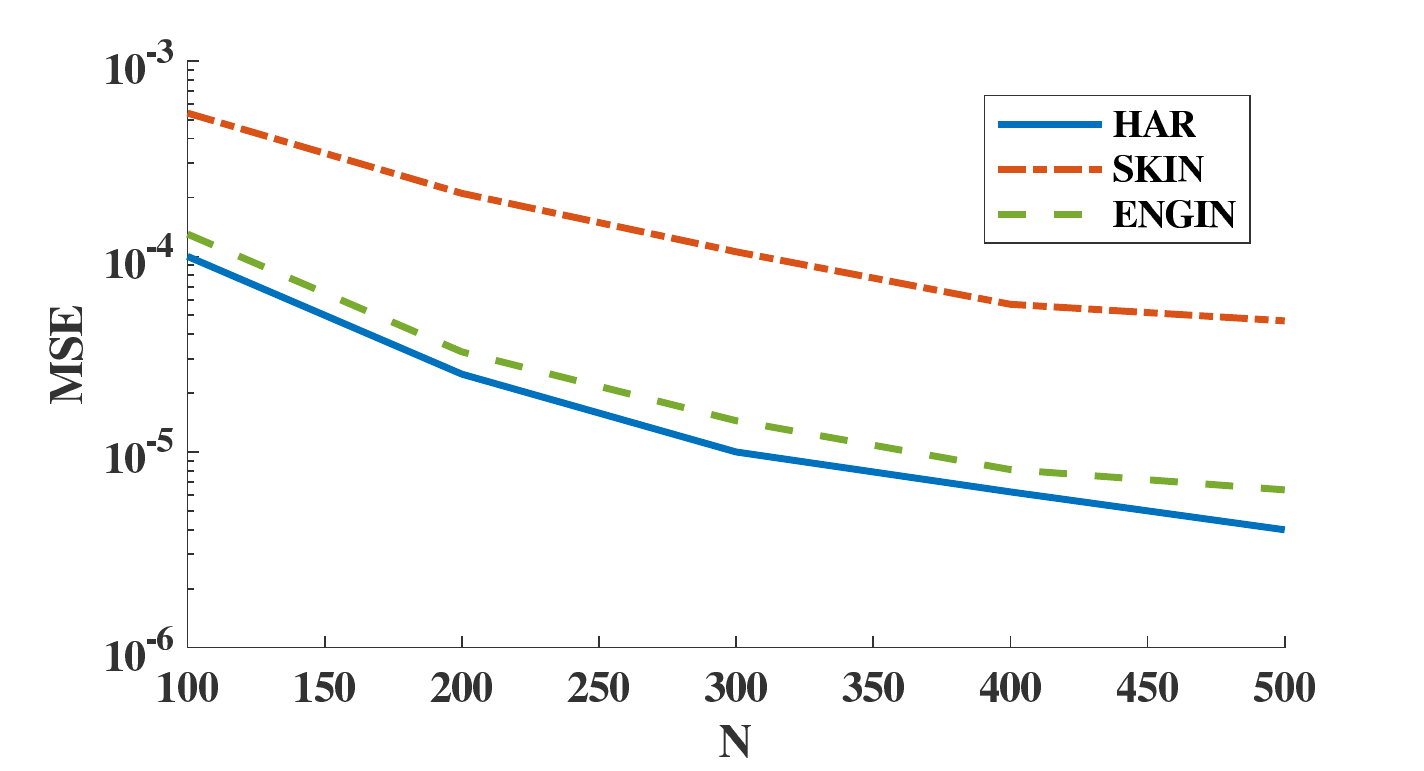}
  \caption{{\small The empirical MSE vs. sample size. The empirical MSE of the FR estimator for all three datasets HAR, SKIN, and ENGIN decreases for larger sample size $N$.}}
  \label{fig5}
\end{figure}
In the first experiment, we computed the HP-divergence and the MSE for the FR test statistic estimator as the sample size $N=m=n$ increases. We observe in Fig. \ref{fig4} that the estimated HP-divergence ranges in $[0,1]$, which is one of the HP-divergence properties, \cite{BWHS}. Interestingly, when $N$ increases the HP-divergence tends to 1 for all HAR, SKIN, and ENGIN datasets. Note that in this set of experiments we have repeated the experiments on independent parts of the datasets to obtain the error bars. Fig. \ref{fig5} shows that the MSE expectedly decreases as the sample size grows for all three datasets. Here we have used KDE plug-in estimator \cite{MSGH}, implemented on the all available samples, to determine the true HP-divergence. Furthermore, according to Fig. \ref{fig5} the FR test statistic-based estimator suggests that the Bayes error rate is larger for the SKIN dataset compared to the HAR and ENGIN datasets.

In our next experiment, we add the first 6 features (dimensions) in order to our datasets and evaluate the FR test statistic's performance as the HP-divergence estimator. Surprisingly, the estimated HP-divergence doesn't change for the HAR sample, however big changes are observed for the SKIN and ENGIN samples, (see Fig. \ref{fig6}).   
\begin{figure}[h]
\centering
  \includegraphics[width=1.05\linewidth]{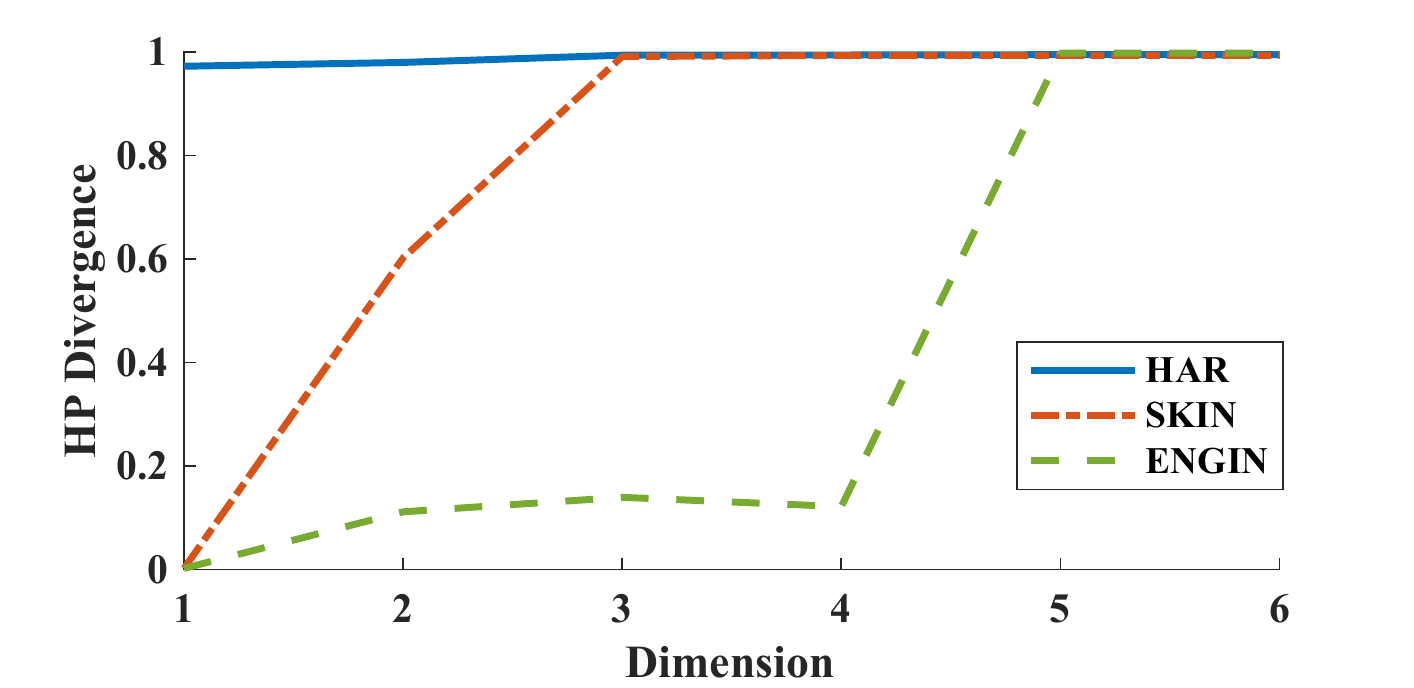}
  \caption{{\small HP-divergence vs. dimension for three datasets HAR, SKIN, and ENGIN.  }}
  \label{fig6}
\end{figure}

Finally, we apply the concentration bounds on the FR test statistic (i.e. Theorems \ref{concentration} and \ref{prob.variance}) and compute theoretical implicit variance-like bound for the FR criteria with $\delta=0.05$ error for the real datasets ENGIN, HAR, and SKIN. Since datasets ENGIN, HAR, and SKIN have the equal total sample size $N=m+n=1200$ and different dimensions $d=14,12,4$, respectively, here we first intend to compare the concentration bound (\ref{main:var.bound}) on the FR statistic in terms of dimension $d$ when $\delta=0.05$. For real datasets ENGIN, HAR, and SKIN we obtain
$$P\left(|\mathfrak{R}_{m,n}-\bbE[\mathfrak{R}_{m,n}]|\leq \xi\right)\geq 0.95,$$ 
where $\xi=\xi'.[0.257,0.005,0.6\times 10^{-11}]$, respectively and $\xi'$ is a constant not dependent on $d$.  
One observes that as the dimension decreases the interval becomes significantly tighter. However, this could not be generally correct and computing bound (\ref{main:var.bound}) precisely requires the knowledge of distributions and unknown constants. In Table 1 we compute the standard variance-like bound by applying the percentiles technique and observe that the bound threshold is not monotonic in terms of dimension $d$. Table 1 shows the FR test statistic, HP-divergence estimate (denoted by $\mathfrak{R}_{m,n}$, $\widehat{D}_p$, respectively), and standard variance-like interval for the FR statistic using the three real datasets HAR, SKIN, and ENGIN. 
\begin {table}[H]\label{Tabl1}
\begin{center}
\begin{tabular}{ |c||c|c|c|c|c|  }
 \hline
 \multicolumn{6}{|c|}{FR test statistic} \\
 \hline
 Dataset & $\mathbb{E}[\mathfrak{R}_{m,n}]$  & $\widehat{D}_p$ & $m$ & $n$ & Variance-like Interval \\
 \hline
HAR & 3  & 0.995  &600  &600 & (2.994,3.006)\\
SKIN & 4.2  & 0.993  &600   &600  &(4.196,4.204)\\
ENGIN & 1.8 &0.997 &600  &600  & (1.798,1.802) \\
 \hline
\end{tabular}
\caption*{{ Table 1: \small $\mathfrak{R}_{m,n}$, $\widehat{D}_p$, $m$, and $n$ are the FR test statistic, HP-divergence estimates using $\mathfrak{R}_{m,n}$, and sample sizes for two classes respectively. }}
\end{center}
\end{table}

\section{Conclusion}\label{conclusion}

We derived a bound on the MSE convergence rate for the Friedman-Rafsky estimator of the Henze-Penrose divergence assuming the densities are sufficiently smooth.  We employed a partitioning strategy to derive the bias rate which depends on the number of partitions, the sample size $m+n$, the H\"{o}lder smoothness parameter $\eta$, and the dimension $d$. However by using the optimal partition number, we derived the MSE convergence rate only in terms of $m+n$, $\eta$, and $d$. We validated our proposed MSE convergence rate using simulations and illustrated the approach for the meta-learning problem of estimating the HP-divergence for three real-world data sets. 
We also provided concentration bounds around the median and mean of the estimator. These bounds explicitly provide the  rate that the FR statistic approaches its median/mean with high probability, not only as a function of the number of samples, $m$, $n$, but also in terms of the dimension of the space $d$. 
By using these results we explored the asymptotic behavior of a variance-like rate in terms of $m$, $n$, and $d$. 

\appendices
\section{Proof of Theorem \ref{lema:1.1}}
In this section, we prove the subadditivity and superadditivity for the mean of FR test statistic. For this, first we need to illustrate the following lemma.  
\begin{lemma}\label{lema:1.0}
Let $\{Q_i\}_{i=1}^{l^d}$ be a uniform partition of $[0,1]^d$ into $l^d$ subcubes $Q_i$ with edges parallel to the coordinate axes having edge lengths $l^{-1}$ and volumes $l^{-d}$. Let $D_{ij}$ be the set of edges of MST graph between $Q_i$ and $Q_j$ with cardinality $|D_{ij}|$, then for $|D|$ defined as the sum of $|D_{ij}|$ for all $i,j=1,\dots,l^d$, $i\neq j$, we have $\bbE|D|=O(l^{d-1}\; n^{1/d})$, or more explicitly
\beq\mathbb{E}[|D|]\leq\diy C'l^{d-1}n^{1/d}+O(l^{d-1} n^{(1/d)-s}),\eeq
where $\eta>0$ is the H\"{o}lder smoothness parameter and 
$$s=\diy\frac{(1-1/d)\eta}{d\;((1-1/d)\eta+1)}.$$
\end{lemma}
\begin{IEEEproof}
Here and in what follows, denote $\Xi_{MST}(\mathfrak{X}_n)$ the length of the shortest spanning tree on $\mathfrak{X}_n=\{\BX_1,\dots,\BX_n\}$, namely 
\beqq \Xi_{MST}(\mathfrak{X}_n):=\diy\min\limits_{T}\sum\limits_{e\in T}|e|,\eeqq
where the minimum is over all spanning trees $T$ of the vertex set $\mathfrak{X}_n$. Using the subadditivity relation for $\Xi_{MST}$ in \cite{Yu}, with the uniform partition of $[0,1]^d$ into $l^d$ subcubes $Q_i$ with edges parallel to the coordinate axes having edge lengths $l^{-1}$ and volumes $l^{-d}$, we have 
\beq\Xi_{MST}(\mathfrak{X}_n)\leq \sum_{i=1}^{l^d} \Xi_{MST}(\mathfrak{X}_n\cap Q_i)+C\; l^{d-1},\eeq
where $C$ is constant. Denote $D$ the set of all edges of $ MST\Big(\bigcup\limits_{i=1}^M Q_i\Big)$ which intersect two different subcubes $Q_i$ and $Q_j$ with cardinality $|D|$. Let $|e_i|$ be the length of $i$-th edge in set $D$. We can write
\beqq \diy\sum_{i \in |D|} |e_i| \leq Cl^{d-1}\;\; \hbox{and}\;\;
 \mathbb{E}\diy\sum_{i \in |D|} |e_i| \leq Cl^{d-1},\eeqq
also we know that
\beq\label{ApA:A.1} \mathbb{E}\sum_{i \in |D|} |e_i| =\bbE_D \sum_{i \in |D|} \mathbb{E}\big[|e_i|\big|D\big].\eeq
Note that using the result from (\cite{HCB}, Proposition 3), for some constants $C_{i1}$ and $C_{i2}$, we have 
\beq \mathbb{E} |e_i| \leq C_{i1}n^{-1/d}+C_{i2}n^{-(1/d)-s},\;\;\; i\in |D|.\eeq
Now let $C_1=\max\limits_i\{C_{i1}\}$ and $C_2=\max\limits_i\{C_{i2}\}$,
hence we can bound the expectation (\ref{ApA:A.1}) as
\beqq \mathbb{E} |D|\; (C_1n^{-1/d}+C_2(n^{-(1/d)-s})) \leq Cl^{d-1},\eeqq
which implies
\beqq\begin{array}{l} \mathbb{E} |D| \leq (C_1n^{-1/d}+O(n^{-(1/d)-s}))\\
\\
\qquad\qquad\leq C'l^{d-1}n^{1/d}+O(l^{d-1} n^{(1/d)-s}).\ena\eeqq
\end{IEEEproof}

To aim toward the goal (\ref{Eq:3.2.1}), we partition $[0,1]^d$ into $M:=l^d$ subcubes $Q_i$ of side $1/l$. Recalling Lemma 2.1 in \cite{SSE} we therefore have the set inclusion:
\beq \label{Eq:3.1} MST\Big(\bigcup\limits_{i=1}^M Q_i\Big)\subset \bigcup\limits_{i=1}^M MST(Q_i) \cup  D, \eeq
where $D$ is defined as in Lemma \ref{lema:1.0}. Let $m_i$ and $n_i$ be the number of sample $\{\BX_1,\dots,\BX_m\}$ and $\{\BY_1,\dots,\BY_n\}$ respectively falling into the partition $Q_i$, such that $\sum\limits_i m_i=m$ and $\sum\limits_i n_i=n$. Introduce sets $A$ and $B$ as
\begin{equation*}A:=MST\Big(\bigcup\limits_{i=1}^M Q_i\Big),\;\;\; B:= \bigcup\limits_{i=1}^M MST(Q_i).\end{equation*}
Since set $B$ has fewer edges than set $A$, thus (\ref{Eq:3.1}) implies that the difference set of $B$ and $A$ contains at most $2|D|$ edges, where $|D|$ is the number of edges in $D$.  On the other word
\beqq\begin{array}{cl} \diy|A\Delta B|\leq |A-B|+|B-A|=|D|+|B-A|\\[8pt]
=|D|+(|B|-|B\cap A|\leq |D|+(|A|-|B\cap A|)=2|D|. \end{array}\eeqq
The number of edge linked nodes from different samples in set $A$
is bounded by the number of edge linked nodes from different samples in 
set $B$ 
plus $2|D|$:
\beq\label{subadd:R}\mathfrak{R}_{m,n}(\mathfrak{X}_m,\mathfrak{Y}_n)\leq \sum\limits_{i=1}^M \mathfrak{R}_{m_i,n_i}\big((\mathfrak{X}_m,\mathfrak{Y}_n)\cap Q_i\big)+2|D|. \eeq
Here $\mathfrak{R}_{m_i,n_i}$ stands with the number edge linked nodes from different samples in partition $Q_i$, $M$. Next, we address the reader to Lemma \ref{lema:1.0},  where it has been shown that there is a constant $c$ such that $\bbE|D|\leq c\;l^{d-1}\; (m+n)^{1/d}$. This concludes the claimed assertion (\ref{Eq:3.2.1}).  Now to accomplish the proof, the lower bound term in (\ref{Eq:3.2.2}) is obtained with similar methodology and the set inclusion:
\beq \bigcup\limits_{i=1}^M MST(Q_i) \subset MST\Big(\bigcup\limits_{i=1}^M Q_i\Big) \cup D. \eeq
This completes the proof. 
\section{Proof of Theorem \ref{thm:bias.opt}}

As many of continuous subadditive functionals on $[0,1]^d$, in the case of FR statistic there exist a dual superadditive functional $\mathfrak{R}^*_{m,n}$ based on dual MST, $\rm{MST}^*$, proposed in Definition \ref{def:dual}. Note that in MST* graph, the degrees of the corner points are bounded by $c_d$ where only depends on dimension $d$, and is the bound for degree of every node in MST graph. The following properties hold true for dual FR test statistic, $\mathfrak{R}^*_{m,n}$:
\begin{lemma}\label{lem:2.5}
Given samples $\mathfrak{X}_m=\{\BX_1,\dots,\BX_m\}$ and $\mathfrak{Y}_n=\{\BY_1,\dots,\BY_n\}$, the following inequalities hold true:
\begin{itemize}
\item[(i)] For constant $c_d$ which depends on $d$:
\beq\label{dual:eq:1.1}\label{RXYG}\begin{array}{cl}\diy \mathfrak{R}^*_{m,n}(\mathfrak{X}_m,\mathfrak{Y}_n)\leq \mathfrak{R}_{m,n}(\mathfrak{X}_m,\mathfrak{Y}_n)+c_d\;2^d,\\
\\
 \diy\mathfrak{R}_{m,n}(\mathfrak{X}_m,\mathfrak{Y}_n)\leq \mathfrak{R}^*_{m,n}(\mathfrak{X}_m,\mathfrak{Y}_n).\ena\eeq
\item[(ii)] ({\rm Subadditivity on $\bbE[\mathfrak{R}^*_{m,n}]$ and Superadditivity}) Partition $[0,1]^d$ into $l^d$ subcubes $Q_i$ such that $m_i$, $n_i$ be the number of sample $\mathfrak{X}_m=\{\BX_1,\dots,\BX_m\}$ and $\mathfrak{Y}_n=\{\BY_1,\dots,\BY_n\}$ respectively falling into the partition $Q_i$ with dual $\mathfrak{R}^*_{m_i,n_i}$. Then we have 
\begin{equation} \label{sub.sup}\begin{array}{l}\diy\bbE\Big[\mathfrak{R}^*_{m,n}(\mathfrak{X}_m, \mathfrak{Y}_n)\Big]\leq \diy\sum\limits_{i=1}^{l^d} \bbE\Big[\mathfrak{R}^*_{m_i,n_i}((\mathfrak{X}_m, \mathfrak{Y}_n)\cap Q_i)\Big]\\[15pt]
\qquad\qquad\qquad\qquad+ \diy c\;l^{d-1}\;(m+n)^{1/d},\\[7pt]
\diy \mathfrak{R}^*_{m,n}(\mathfrak{X}_m, \mathfrak{Y}_n)\geq \diy\sum\limits_{i=1}^{l^d} \mathfrak{R}^*_{m_i,n_i}((\mathfrak{X}_m, \mathfrak{Y}_n)\cap Q_i)-2^d c_d l^d.\end{array}\end{equation}
where $c$ is a constant.
\end{itemize}
\begin{IEEEproof} (i) Consider the nodes connected to the corner points. Since ${\rm MST}(\mathfrak{X}_m,\mathfrak{Y}_n)$ and ${\rm MST}^*(\mathfrak{X}_m,\mathfrak{Y}_n)$ can only be different in the edges connected to these nodes, and in $\mathfrak{R}^*(\mathfrak{X}_m,\mathfrak{Y}_n)$ we take all of the edges between these nodes and corner nodes into account, so we obviously have the second relation in \eqref{RXYG}. Also for the first  inequality in \eqref{RXYG} it is enough to say that the total number of edges connected to the corner nodes is upper bounded by $2^d \; c_d$.

(ii) Let $|D^*|$ be the set of edges of the $\rm{MST}^*$ graph which intersect two different partitions. Since MST and $\rm{MST}^*$ are only different in edges of points connected to the corners and edges crossing different partitions. Therefore $|D^*|\leq |D|$. By eliminating one edge in set $D$ in worse scenario we would face with two possibilities: either the corresponding node is connected to the corner which is counted anyways or any other point in MST graph which wouldn't change the FR test statistic. This implies the following subadditivity relation:
\beqq
\mathfrak{R}^*_{m,n}(\mathfrak{X}_m,\mathfrak{Y}_n)-|D| \leq \sum_{i=1}^{l^d} \mathfrak{R}^*_{m_i,n_i}\big((\mathfrak{X}_m,\mathfrak{Y}_n)\cap Q_i\big).
\eeqq
Further from Lemma \ref{lema:1.0}, we know that there is a constant $c$ such that $\bbE|D|\leq c\;l^{d-1}\;(m+n)^{1/d}$. Hence the first inequality in (\ref{sub.sup}) is obtained. Next consider $|D^*_c|$ which represents the total number of edges from both samples only connected to the all corners points in ${\rm MST}^*$ graph. Therefore one can easily claim: 
\beqq 
\mathfrak{R}^*_{m,n}(\mathfrak{X}_m,\mathfrak{Y}_n)\geq \sum_{i=1}^{l^d} \mathfrak{R}^*_{m_i,n_i}\big((\mathfrak{X}_m,\mathfrak{Y}_n)\cap Q_i\big)  -  |D^*_c|.\eeqq
Also we know that $|D^*_c|\leq 2^d l^d c_d$ where $c_d$ stands with the largest possible degree of any vertex. One can write 
\beqq 
\mathfrak{R}^*_{m,n}(\mathfrak{X}_m,\mathfrak{Y}_n)\geq \sum_{i=1}^{l^d} \mathfrak{R}^*_{m_i,n_i}\big((\mathfrak{X}_m,\mathfrak{Y}_n)\cap Q_i\big)  -  2^d c_d l^d.
\eeqq
\end{IEEEproof}
\end{lemma}

The following list of Lemmas \ref{lem:error1}, \ref{lem2:3} and \ref{lem:2.4} are inspired from \cite{HP} and are required to prove Theorem \ref{thm:bias}. See the Supplementary Materials for their proofs. 
\def\bz{\mathbf{z}}
\begin{lemma}\label{lem:error1}
 Let $g(\bx)$ be a density function with support $[0,1]^d$ and belong to the strong H\"{o}lder class $\Sigma_d^{\rm S}(\eta,L)$, $0<\eta\leq 1$, stated in Definition \ref{def:strong.Holder}. Also, assume that $P(\bx)$ is a $\eta$-H\"{o}lder smooth function, such that its absolute value is bounded from above by a constant. Define the quantized density function with parameter $l$ and constants $\phi_i$ as
\beq\label{hat:g} \widehat{g}(\bx)=\sum\limits_{i=1}^M \phi_i\mathbf{1}\{\bx\in Q_i\},\;\;\; \hbox{where}\; \phi_i=l^d\;\int\limits_{Q_i} g(\bx)\;\rd\bx. \eeq
Let $M=l^d$ and $Q_i=\{\bx,\bx_i:\|\bx-\bx_i\|<l^{-d}\}$. Then 
\beq \diy \int \Big\|\big(g(\bx)-\widehat{g}(\bx)\big) P(\bx)\Big\|\;\rd \bx\leq  O(l^{-d\eta}).\eeq
\end{lemma}

\begin{lemma}\label{lem2:3}
Denote $\Delta(\bx,\mathcal{S})$ the degree of vertex $\bx\in\mathcal{S}$ in the $MST$ over set $\mathcal{S}$ with the $n$ number of vertices. For given function $P(\bx,\bx)$, one obtains
\begin{equation}\label{eq2:(10)} \diy\int P(\bx,\bx)  g(\bx) \bbE[\Delta(\bx,\mathcal{S})]  \;\rd \bx=2\;\int P(\bx,\bx) g(\bx)\;\rd\bx+ \varsigma_\eta(l,n),\end{equation}
where for constant $\eta>0$, 
\beq\label{def.O(l)} \varsigma_{\eta}(l,n)=\diy\Big(O\big(l/n\big)-\diy 2\;l^{d}/n\Big)\diy\int g(\bx)P(\bx,\bx)\;\rd\bx+O(l^{-d\eta}).\eeq
\end{lemma}
\begin{lemma}\label{lem:2.4.0}
Assume that for given $k$, $g_k(\bx)$ is a bounded function belong to $\Sigma^s_d(\eta,L)$. Let $P:\bbR^d\times\bbR^d\mapsto[0,1]$ be a symmetric, smooth, jointly measurable function, such that, given $k$,  for almost every $\bx\in\bbR^d$, $P(\bx,.)$ is measurable with $\bx$ a Lebesgue point of the function $g_k(.)P(\bx,.)$. Assume that the first derivative $P$ is bounded.  For each $k$, let $\BZ_1^k,\BZ_2^k,\dots,\BZ_k^k$ be independent $d$-dimensional variable with common density function $g_k$. Set $\mathfrak{Z}_k=\{\BZ_1^k,\BZ_2^k\dots,\BZ_k^k\}$ and $\mathfrak{Z}_k^{\bx}=\{\bx,\BZ_2^k,\BZ_3^k\dots,\BZ_k^k\}$. Then
\begin{equation}\label{eq:13.1.0} \begin{array}{l}\diy \bbE\bigg[ \sum\limits_{j=2}^k P(\bx,\BZ_j^k)\mathbf{1}\big\{(\bx,\BZ_j^k)\in MST(\mathfrak{Z}_k^{\bx})\big\}\bigg]\\[10pt]
\qquad=\diy P(\bx,\bx)\;\bbE\big[\Delta(\bx,\mathfrak{Z}_k^{\bx})\big]+\Big\{O\big(k^{-\eta/d}\big)+O\big(k^{-1/d}\big)\Big\}.\ena\end{equation}
\end{lemma}
\begin{lemma}\label{lem:2.4}
Consider the notations and assumptions in Lemma \ref{lem:2.4.0}.  Then 
\begin{equation}\label{lema:eq:2.4}\begin{array}{l}\diy  \Big|\diy k^{-1} \mathop{\sum\sum}_{\ 1\leq i<j\leq k} P(\BZ_i^k,\BZ_j^k)\mathbf{1}\{(\BZ_i^k,\BZ_j^k)\in MST(\mathfrak{Z}_k)\}\\[10pt]
\qquad\qquad\qquad-\diy\int_{\bbR^d}P(\bx,\bx)g_k(\bx)\;\rd\bx\Big|\\[12pt]
\qquad\qquad \quad \leq\diy \varsigma_{\eta}(l,k)+O(k^{-\eta/d})+O(k^{-1/d}).\ena\end{equation}
Here $MST(\mathcal{S})$ denotes the MST graph over nice and finite set $\mathcal{S}\subset \bbR^d$ and $\eta$ is the smoothness H\"{o}lder parameter. Note that $\varsigma_{\eta}(l,k)$ is given as before in Lemma \ref{lem2:3} (\ref{def.O(l)}). 
\end{lemma}
\def\bbV{\mathbb{V}}
\begin{theorem}\label{thm:bias}
 Assume $\mathfrak{R}_{m,n}:=\mathfrak{R}(\mathfrak{X}_m,\mathfrak{Y}_n)$ denotes the FR test statistic and densities $f_0$ and $f_1$ belong to the strong H\"{o}lder class $\Sigma_d^{\rm S}(\eta,L)$, $0<\eta\leq 1$. Then the rate for the  bias of the $\mathfrak{R}_{m,n}$ estimator for $d\geq 2$ is of the form: 
 \begin{equation}\label{bound:Bias1} \begin{array}{l}\diy \left|\frac{\bbE\big[\mathfrak{R}_{m,n}\big]}{m+n}-2pq\int \frac{f_0(\bx) f_1(\bx)}{p f_0(\bx)+q f_1(\bx)}\;\rd\bx \right|\\
 \\
 \qquad\qquad\quad\qquad \leq O\big(l^d(m+n)^{-\eta/d}\big)+O(l^{-d\eta}).
\\
\ena \end{equation}
The proof and a more explicit form for the bound on the RHS are given in Supplementary Materials. 
\end{theorem}

 Now, we are at the position to prove the assertion in (\ref{bound:Bias.opt}). Without lose of generality assume that $(m+n)l^{-d}>1$. In the range $d\geq 2$ and $0<\eta\leq 1$, we select $l$ as a function of $m+n$ to be the sequence increasing in $m+n$ which minimizes the maximum of these rates:
 \beqq l(m+n)=arg\;\min\limits_{l}\max\Big\{l^d(m+n)^{-\eta/d},\; l^{-\eta d}\Big\}. \eeqq
 The solution $l=l(m+n)$ occurs when $l^d(m+n)^{-\eta/d}=l^{-\eta d}$, or equivalently $l=\lfloor(m+n)^{\eta/(d^2(\eta+1))}\rfloor$. Substitute this into $l$ in the bound (\ref{bound:Bias1}), the RHS expression in (\ref{bound:Bias.opt}) for $d\geq 2$ is established. 
\section{Proof of Theorems \ref{Tight.bound:Variance}}

To bound the variance we will apply one of the first concentration inequalities which was proved by Efron and Stein \cite{ES} and further was improved by Steele \cite{Steele1986}. 

\begin{lemma}\label{Ineq:Efron-Stein}
{\rm (The Efron-Stein Inequality)} Let $\mathfrak{X}_m=\{\BX_1,\dots,\BX_m\}$ be a random vector on the space $\mathcal{S}$. Let $\mathfrak{X}'=\{\BX'_1,\dots,\BX'_m\}$ be the copy of random vector $\mathfrak{X}_m$. Then if $f:\mathcal{S}\times\dots\times\mathcal{S}\rightarrow \bbR$, we have 
\begin{equation}\begin{array}{l} \bbV\big[f(\mathfrak{X}_m)\big]\\[5pt]
\;\; \leq \diy\frac{1}{2} \sum\limits_{i=1}^m \bbE\Big[ \big(f(\BX_1,\dots,\BX_m)-f(\BX_1,\dots,\BX'_i,\dots,\BX_m)\big)^2\Big].\ena\end{equation}
\end{lemma}

Consider two set of nodes $\BX_i$, $1\leq i \leq m$ and $\BY_j$ for $1\leq j \leq n$. Without loss of generality, assume that $m < n$.  Then  consider the  $n-m$ virtual random points $\BX_{m+1},...,\BX_n$ with the same distribution as $\BX_i$, and define $\BZ_i:=(\BX_i,\BY_i)$. Now for using the Efron-Stein inequality on set $\mathfrak{Z}_n=\{\BZ_1,...,\BZ_n\}$, we involve another independent copy of $\mathfrak{Z}_n$ as $\mathfrak{Z}'_n=\{\BZ'_1,...,\BZ'_n\}$, and define $\mathfrak{Z}^{(i)}_n:=(\BZ_1,...,\BZ_{i-1},\BZ'_i,\BZ_{i+1},...,\BZ_n)$, then $\mathfrak{Z}^{(1)}_n$ becomes $(\BZ'_1,\BZ_{2},...,\BZ_n)=\big\{ (\BX'_1,\BY_1'),(\BX_2,\BY_2),\dots,(\BX_m,\BY_n)\big\}=:(\mathfrak{X}^{(1)}_m,\mathfrak{Y}^{(1)}_n)$ where $(\BX'_1,\BY'_1)$ is independent copy of $(\BX_1,\BY_1)$. Next define the function $r_{m,n}(\mathfrak{Z}_n):=\diy\mathfrak{R}_{m,n}/(m+n)$, which means that we discard the random samples $\BX_{m+1},...,\BX_n$, and find the previously defined $\mathfrak{R}_{m,n}$ function on the nodes $\BX_i$, $1\leq i \leq m$ and $\BY_j$ for $1\leq j \leq n$, and multiply by some coefficient to normalize it. Then, according to the Efron-Stein inequality we have
\beqq Var(r_{m,n}(\mathfrak{Z}_n))\leq \frac{1}{2} \sum_{i=1}^n \mathbb{E}\left[(r_{m,n}(\mathfrak{Z}_n)-r_{m,n}(\mathfrak{Z}^{(i)}_n))^2\right].\eeqq
Now we can divide the RHS as
\begin{equation}\label{sumands:var}\begin{array}{l}
\diy \frac{1}{2} \sum_{i=1}^n \mathbb{E}\left[(r_{m,n}(\mathfrak{Z}_n)-r_{m,n}(\mathfrak{Z}^{(i)}_n))^2\right] \\
\qquad=\diy\frac{1}{2} \sum_{i=1}^m \mathbb{E}\left[(r_{m,n}(\mathfrak{Z}_n)-r_{m,n}(\mathfrak{Z}^{(i)}_n))^2\right] \\
\qquad  \quad+\diy\frac{1}{2} \sum_{i=m+1}^n \mathbb{E}\left[(r_{m,n}(\mathfrak{Z}_n)-r_{m,n}(\mathfrak{Z}^{(i)}_n))^2\right].
\ena\end{equation}
The first summand becomes 
\beqq \begin{array}{l}
=\diy\frac{1}{2} \sum_{i=1}^m \mathbb{E}\left[(r_{m,n}(\mathfrak{Z}_n)-r_{m,n}(\mathfrak{Z}^{(i)}_n))^2\right]\nonumber\\
=\diy\frac{m}{2\;(m+n)^2} \mathbb{E}\left[(\mathfrak{R}_{m,n}(\mathfrak{X}_m,\mathfrak{Y}_n)-\mathfrak{R}_{m,n}(\mathfrak{X}^{(1)}_m,\mathfrak{Y}^{(1)}_n))^2\right],
\ena\eeqq
which can also be upper bounded as follows:
\beq\label{eq1:Var1}\begin{array}{ccl}\diy \left|\mathfrak{R}_{m,n}(\mathfrak{X}_m,\mathfrak{Y}_n)-\mathfrak{R}_{m,n}(\mathfrak{X}^{(1)}_m,\mathfrak{Y}^{(1)}_n)\right|\\[10pt]
\qquad\qquad\;\;\leq \left|\mathfrak{R}_{m,n}(\mathfrak{X}_m,\mathfrak{Y}_n)-\mathfrak{R}_{m,n}(\mathfrak{X}^{(1)}_m,\mathfrak{Y}_n)\right| \\[10pt]
\qquad\qquad\qquad\qquad+\diy\left|\mathfrak{R}(\mathfrak{X}^{(1)}_m,\mathfrak{Y}_n)-\mathfrak{R}_{m,n}(\mathfrak{X}^{(1)}_m,\mathfrak{Y}^{(1)}_n)\right|.\ena\eeq

For deriving an upper bound on the second line in (\ref{eq1:Var1})
we should observe how much changing a point's position modifies the amount of $\mathfrak{R}_{m,n}(\mathfrak{X}_m,\mathfrak{Y}_n)$. We consider two steps of changing $\BX_1$'s position: we first remove it from the graph, and then add it to the new position. Removing it would change $\mathfrak{R}_{m,n}(\mathfrak{X}_m,\mathfrak{Y}_n)$ at most by $2\; c_d$, because $X_1$ has a degree of at most $c_d$, and $c_d$ edges will be removed from the MST graph, and $c_d$ edges will be added to it. Similarly, adding $\BX_1$ to the new position will affect $\mathfrak{R}_{m,n}(\mathfrak{X}_{m,n},\mathfrak{Y}_{m,n})$ at most by $2c_d$. So, we have $$\left|\mathfrak{R}_{m,n}(\mathfrak{X}_m,\mathfrak{Y}_n)-\mathfrak{R}_{m,n}(\mathfrak{X}^{(1)}_m,\mathfrak{Y}_n)\right|\leq 4\; c_d,$$
 and we can also similarly reason that  
 $$\left|\mathfrak{R}_{m,n}(\mathfrak{X}^{(1)}_m,\mathfrak{Y}_n)-\mathfrak{R}_{m,n}(\mathfrak{X}^{(1)}_m,\mathfrak{Y}^{(1)}_n)\right|\leq 4\; c_d.$$
Therefore totally we would have
\beqq \left|\mathfrak{R}_{m,n}(\mathfrak{X}_m,\mathfrak{Y}_n)-\mathfrak{R}_{m,n}(\mathfrak{X}^{(1)}_m,\mathfrak{Y}^{(1)}_n)\right|\leq 8\;c_d.\eeqq
Furthermore, the second summand in (\ref{sumands:var}) becomes
\beqq \begin{array}{l}
\qquad=\diy\frac{1}{2} \sum_{i=m+1}^n \mathbb{E}\left[(r_{m,n}(\mathfrak{Z}_n)-r_{m,n}(\mathfrak{Z}^{(i)}_n))^2\right]\nonumber\\[16pt]
=\diy K_{m,n} \mathbb{E}\left[(\mathfrak{R}_{m,n}(\mathfrak{X}_m,\mathfrak{Y}_n)-\mathfrak{R}_{m,n}(\mathfrak{X}^{(m+1)}_m,\mathfrak{Y}^{(m+1)}_n))^2\right],
\ena\eeqq
where $K_{m,n}=\frac{n-m}{2\;(m+n)^2}$. Since in $(\mathfrak{X}^{(m+1)}_m,\mathfrak{Y}^{(m+1)}_n)$, the point $\BX'_{m+1}$ is a copy of virtual random point $\BX_{m+1}$, therefore this point doesn't change the FR test statistic $\mathfrak{R}_{m,n}$. Also following the above arguments we have 
\beqq \left|\mathfrak{R}_{m,n}(\mathfrak{X}_m,\mathfrak{Y}_n)-\mathfrak{R}_{m,n}(\mathfrak{X}_m,\mathfrak{Y}^{(m+1)}_n)\right|\leq 4\;c_d.\eeqq
Hence we can bound the variance as below:
\beq Var(r_{m,n}(\mathfrak{Z}_n))\leq \diy \frac{8 c_d^2(n-m)}{(m+n)^2}+\diy \frac{32\; c_d^2\; m}{(m+n)^2} .\eeq
Combining all results with the fact that $\diy\frac{n}{m+n}\rightarrow q$ concludes the proof.

\section{Proof of Theorems \ref{Concentration around the median}, \ref{concentration} and \ref{prob.variance}}\label{AppendixD}

We will need the following prominent results for the proofs.
\def\ep{\epsilon}
\begin{lemma}\label{varianceD}
For $h=1,2,\dots$, let $\delta^h_{m,n}$ be the function $c\;h^{d-1}(m+n)^{1/d}$, where $c$ is a constant. Then for $\epsilon> 0$ 
, we have
\begin{equation} \label{variancD}P\Big(\mathfrak{R}_{m,n}(\mathfrak{X}_m,\mathfrak{Y}_n)\leq \sum\limits_{i=1}^{h^d}\mathfrak{R}_{m_i,n_i}(\mathfrak{X}_{m_i},\mathfrak{Y}_{n_i})+2\ep\Big)\geq \diy\frac{\ep-\delta^h_{m,n}}{\ep}.\end{equation}
Note that in the case $\ep\leq \delta^h_{m,n}$, the above claimed inequality becomes trivial. 
\end{lemma}

The subadditivity property for FR test statistic $\mathfrak{R}_{m,n}$ in Lemma \ref{varianceD},  as well as Euclidean functionals,  leads to several non-trivial consequences. The growth bound was first explored by Rhee (1993b), \cite{Rh1993b} and as is illustrated in \cite{Yu}, \cite{St1997}  has a wide range of applications. In this paper we investigate the probabilistic growth bound for $\mathfrak{R}_{m,n}$. This observation will lead us to our main goal in this appendix which is providing the proof of Theorem \ref{concentration}. For what follows we will use $\delta^h_{m,n}$ notation for the expression $O\big(h^{d-1}(m+n)^{1/d}\big)$. 
\def\bbB{\mathbb{B}}
\begin{lemma}\label{growth bound}
{\rm (Growth bounds for $\mathfrak{R}_{m,n}$)} Let $\mathfrak{R}_{m,n}$ be the FR test statistic. Then for given non-negative $\ep$,  such that $\ep\geq h^2\; \delta^h_{m,n}$, with at least probability $g(\ep):=1-\diy\frac{h\;\delta^h_{m,n}}{\ep}$,  $h=2,3,\dots$, we have
\beq\label{Growth bound} \mathfrak{R}_{m,n}(\mathfrak{X}_m,\mathfrak{Y}_n)\leq {c}''_{\ep,h}\;\big(\#\mathfrak{X}_m \;\#\mathfrak{Y}_n\big)^{1-1/d}.\eeq
Here  $ {c}''_{\ep,h}=O\left(\diy\frac{\ep}{h^{d-1}-1}\right)$ depending only on $\ep$ and $h$. 
\end{lemma}

The complexity of $\mathfrak{R}_{m,n}$'s behavior and the need to pursue the proof encouraged us to explore the smoothness condition for $\mathfrak{R}_{m,n}$. In fact, this is where both subadditivity and superadditivity for the FR statistic are used together and become more important. 
\begin{lemma}\label{smoothness}
{\rm (Smoothness for $\mathfrak{R}_{m,n}$)} Given observations of 
$$\mathfrak{X}_m:=(\mathfrak{X}_{m'},\mathfrak{X}_{m''})=\{\BX_1,\dots,\BX_{m'},\BX_{m'+1},\dots,\BX_m\},$$
 where $m'+m''=m$ and  $\mathfrak{Y}_n:=(\mathfrak{Y}_{n'},\mathfrak{Y}_{n''})=\{\BY_1,\dots,\BY_{n'},\BY_{n'+1},\dots,\BY_n\}$, where $n'+n''=n$, denote $\mathfrak{R}_{m,n}(\mathfrak{X}_m,\mathfrak{Y}_n)$ as before, the number of edges of ${\rm MST}(\mathfrak{X}_m,\mathfrak{Y}_n)$ which connect a point of $\mathfrak{X}_m$ to a point of $\mathfrak{Y}_n$. 
Then for given integer $h\geq 2$, for all $(\mathfrak{X}_n, \mathfrak{Y}_m)\in[0,1]^d$, $\ep\geq h^2\;\delta^h_{m,n}$ where $\delta^h_{m,n}=O\big(h^{d-1}(m+n)^{1/d}\big)$, we have
\begin{equation}\label{Smoothness:1}\begin{array}{l}P\bigg(\Big|\mathfrak{R}_{m,n}(\mathfrak{X}_m,\mathfrak{Y}_n)-\mathfrak{R}_{m',n'}(\mathfrak{X}_{m'},\mathfrak{Y}_{n'})\Big|
\leq \tilde{c}_{\ep,h} \;\big(\#\mathfrak{X}_{m''}\;\#\mathfrak{Y}_{n''}\Big)^{1-1/d}\bigg)\\
\qquad \qquad\qquad \geq \diy 1-\diy\frac{2h\;\delta^{h}_{m,n}}{\ep}, \ena\end{equation}
 where $\tilde{c}_{\ep,h}=O\left(\diy\frac{\ep}{h^{d-1}-1}\right)$. 
\end{lemma}

{\it Remark:} Using Lemma \ref{smoothness}, we can imply the continuty property, i.e.  for all observations $(\mathfrak{X}_m,\mathfrak{Y}_n)$ and $(\mathfrak{X}_{m'}, \mathfrak{Y}_{n'})$, with at least probability $2\;g(\ep)-1$, one obtains 
\beq \begin{array}{l} \Big|\mathfrak{R}_{m,n}(\mathfrak{X}_m,\mathfrak{Y}_n)-\mathfrak{R}_{m',n'}(\mathfrak{X}_{m'},\mathfrak{Y}_{n'})\Big|\\
\qquad \qquad \leq c^*_{\ep,h} \;\big(\#(\mathfrak{X}_m\Delta\;\mathfrak{X}_{m'})\;\#(\mathfrak{Y}_{n}\Delta\;\mathfrak{Y}_{n'})\big)^{1-1/d},\ena\eeq
for given $\ep>0$, $c^*_{\ep,h}=O\left(\diy\frac{\ep}{h^{d-1}-1}\right)$, $h\geq 2$. Here $\mathfrak{X}_m\Delta\;\mathfrak{X}_{m'}$ denotes symmetric difference of observations $\mathfrak{X}_m$ and $\mathfrak{X}_{m'}$.\\

The path to approach the assertions (\ref{Concentration:Mediam}) and (\ref{Ineq:concentration})  proceeds via semi-isoperimentic inequality for the $\mathfrak{R}_{m,n}$ involving the Hamming distance. 
\def\bbA{\mathbb{A}}\def\bbF{\mathbb{F}} \def\bbG{\mathbb{G}}
\begin{lemma}\label{Eq:isoperimetry}
{\rm (Semi-Isoperimetry)} Let $\mu$ be a measure on $[0,1]^d$; $\mu^n$ denotes the product measure on space $([0,1]^d)^n$. And let $M_{\rm e}$ denotes a median of $\mathfrak{R}_{m,n}$. Set
\begin{equation}\label{set:A}\begin{array}{l}\bbA:=\\
\Big\{ \mathfrak{X}_m\in \big([0,1]^d\big)^m, \mathfrak{Y}_n\in \big([0,1]^d\big)^n; \mathfrak{R}_{m,n}(\mathfrak{X}_m,\mathfrak{Y}_n)\leq M_{e}\Big\}.\ena\end{equation}

Following the notations in \cite{Yu}, $H(\bx,\bx')=\#\{i, \bx_i\neq\bx'_i)$ and $\phi_{\bbA}(\bx')+\phi_{\bbA}(\by')=\min\{H(\bx,\bx')+H(\by,\by'): \bx,\by\in\bbA\}$ and $\phi_{\bbA}(\bx')\;\phi_{\bbA}(\by')=\min\{H(\bx,\bx')\;H(\by,\by'): \bx,\by\in\bbA\}$ . Then
\begin{equation}\label{lem13:0.0}\begin{array}{l} \diy \mu^{m+n}\bigg(\Big\{\bx'\in([0,1]^d)^m, \by'\in ([0,1]^n): \phi_{\bbA}(\bx')\;\phi_{\bbA}(\by') \geq t\Big\}\bigg)\\[10pt]
\qquad\qquad \leq 4\exp\Big(\diy\frac{-t}{8(m+n)}\Big).\ena \end{equation}
\end{lemma}

Now, we continue by providing the proof of Theorem \ref{Concentration around the median}. 
Recall (\ref{set:A}) and denote
\beqq\begin{array}{cl} \bbF_{\bx}:=\left\{\bx_i, i=1,\dots,m, \bx_i=\bx'_i\right\},\\[8pt]
 \bbF_{\by}:=\left\{\by_j, j=1,\dots,n, \by_j=\by'_j\right\},\\[5pt]
\hbox{and}\\[5pt]
\bbG_{\bx}:=\left\{\bx_i, i=1,\dots,m, \bx_i\neq\bx'_i\right\},\\[8pt]
 \bbG_{\by}:=\left\{\by_j, j=1,\dots,n, \by_j\neq\by'_j\right\}.
\ena\eeqq
And, for given integer $h$,  define events $\bbB$, $\bbB'$ by
\beqq\begin{array}{cl} \bbB:=\left\{ \Big|\mathfrak{R}_{m,n}(\mathfrak{X}'_m,\mathfrak{Y}'_n)-\mathfrak{R}(\bbF_\bx,\bbF_\by)\Big|\leq c_{\ep,h}\big(\# \bbG_\bx\;\#\bbG_\by\big)^{1-1/d}\right\},\\[10pt]
\bbB':=\left\{ \Big|\mathfrak{R}(\bbF_\bx,\bbF_\by)-\mathfrak{R}_{m,n}(\mathfrak{X}_m,\mathfrak{Y}_n)\Big|\leq c_{\ep,h}\big(\# \bbG_\bx\;\#\bbG_\by\big)^{1-1/d}\right\},\ena \eeqq
where $c_{\ep,h}$ is a constant. By virtue of smoothness property, Lemma \ref{smoothness}, for $\epsilon \geq h^2 \delta^h_{m,n}$ we know $P(\bbB)\geq 2 g(\ep)-1$ and $P(\bbB')\geq 2 g(\ep)-1$. On the other hand we have
\begin{equation*}\begin{array}{l}  \diy\mathfrak{R}_{m,n}(\mathfrak{X}'_m,\mathfrak{Y}'_n)\leq\diy \Big|\mathfrak{R}_{m,n}(\mathfrak{X}'_m,\mathfrak{Y}'_n)-\mathfrak{R}(\bbF_\bx,\bbF_\by)\Big|\\[10pt]
\qquad\qquad+\Big|\mathfrak{R}(\bbF_\bx,\bbF_\by)-\mathfrak{R}_{m,n}(\mathfrak{X}_m,\mathfrak{Y}_n)\Big|+\mathfrak{R}_{m,n}(\mathfrak{X}_m,\mathfrak{Y}_n).\\[10pt]
\qquad= |\varpi'|+|\varpi|+\mathfrak{R}_{m,n}(\mathfrak{X}_m,\mathfrak{Y}_n)\;\;\; \hbox{(say)}.\end{array}\end{equation*}
Moreover $P(\mathfrak{R}_{m,n}(\mathfrak{X}_m,\mathfrak{Y}_n)\leq M_{\rm e})\geq \diy{1}/{2}$. Therefore we can write
\begin{equation}\begin{array}{cl}\diy{1}/{2}\leq P\Big(\mathfrak{R}_{m,n}(\mathfrak{X}'_m,\mathfrak{Y}'_n)\leq M_{\rm e}+|\varpi'|+|\varpi|\Big)\\[10pt]
\leq P\Big(\mathfrak{R}_{m,n}(\mathfrak{X}'_m,\mathfrak{Y}'_n)\leq M_{\rm e}+|\varpi'|+|\varpi|\big|\;\bbB\cap\bbB'\Big) P(\bbB\cap\bbB')\\[10pt]
+P(\bbB^{\rm c}\cup \bbB'^{\rm c}). \ena \end{equation}
So, we obtain
\beqq\begin{array}{l} \diy P\Big(\mathfrak{R}_{m,n}(\mathfrak{X}'_m,\mathfrak{Y}'_n)\leq M_{\rm e}+4\ep\;\big(\# \bbG_\bx\;\#\bbG_\by\big)^{1-1/d}\Big)\\[10pt]
\qquad\qquad\geq \big(1/2-1+P(\bbB\cap\bbB')\big)\big/P(\bbB\cap\bbB')\\[10pt]
\qquad\qquad\qquad= \diy1-\Big(\big(2\;P(\bbB\cap\bbB')\big)^{-1}\Big).\ena\eeqq
Note that $P(\bbB\cap\bbB')=P(\bbB)\;P(\bbB')\geq \big(2\;g(\ep)-1\big)^2$. Now, we easily claim that 
\begin{equation}\label{eq:new1}1-\Big(\big(2\;P(\bbB\cap\bbB')\big)^{-1}\Big) \geq 1-\Big(\big(2\;(2\;g(\ep)-1)^2\big)^{-1}\Big).\end{equation} 
Thus 
\begin{equation*}\begin{array}{l} P\Big(\mathfrak{R}_{m,n}(\mathfrak{X}'_m,\mathfrak{Y}'_n)\leq M_{\rm e}+4\ep\;\big(\# \bbG_\bx\;\#\bbG_\by\big)^{1-1/d}\Big)\\[10pt]
\qquad\geq 1-\Big(\big(2\;(2\;g(\ep)-1)^2\big)^{-1}\Big) .\ena\end{equation*}
On the other word, calling $\phi_{\bbA}(\bx')$ and $\phi_{\bbA}(\by')$ in Lemma \ref{Eq:isoperimetry}, we get
\beq\label{eq752:1.1} \begin{array}{l} P\Big(\mathfrak{R}_{m,n}(\mathfrak{X}'_m,\mathfrak{Y}'_n)\leq M_{\rm e}+4\ep\;\big(\phi_{\bbA}(\bx')\;\phi_{\bbA}(\by')\big)^{1-1/d}\Big)\\[10pt]
\qquad\geq 1-\Big(\big(2\;(2\;g(\ep)-1)^2\big)^{-1}\Big).\ena\eeq
Furthermore, denote event 
$$\bbC:=\big\{\mathfrak{R}_{m,n}(\mathfrak{X}'_m,\mathfrak{Y}'_n)\leq M_{\rm e}+4\ep\;\big(\phi_{\bbA}(\bx')\;\phi_{\bbA}(\by')\big)^{1-1/d}\big\}.$$
Then we have
\begin{equation}\label{eq75:1.1}\begin{array}{l}
P\big(\mathfrak{R}_{m,n}(\mathfrak{X}_m,\mathfrak{Y}_n)\geq M_e+t\big)=\mu^{m+n}\big(\mathfrak{R}_{m,n}(\mathfrak{X}'_m,\mathfrak{Y}'_n)\geq M_e+t\big)\\
\\
\quad=\mu^{m+n}(\big(\mathfrak{R}_{m,n}(\mathfrak{X}'_m,\mathfrak{Y}'_n)\geq M_e+t\big)\big|\bbC)P(\bbC)\\
\\
\qquad\quad+\diy\mu^{m+n}(\big(\mathfrak{R}_{m,n}(\mathfrak{X}'_m,\mathfrak{Y}'_n)\geq M_e+t\big)\big|\bbC^{\rm c})P(\bbC^{\rm c})\\
\\
\quad\leq \diy\mu^{m+n}\Big(\big(\phi_{\bbA}(\bx')\;\phi_{\bbA}(\by')\big)^{1-1/d}\geq \diy\frac{t}{4\ep}\Big)P(\bbC)\\
\\
\qquad\quad+\diy\mu^{m+n}(\big(\mathfrak{R}_{m,n}(\mathfrak{X}'_m,\mathfrak{Y}'_n)\geq M_e+t\big)\big|\bbC^{\rm c})P(\bbC^{\rm c})\\
\\
\hbox{Using}\;\; P(\bbC)=1-P(\bbC^{\rm c})\\
\\
\qquad=\diy\mu^{m+n}\Big(\big(\phi_{\bbA}(\bx')\;\phi_{\bbA}(\by')\big)^{1-1/d}\geq \diy\frac{t}{4\ep}\Big)\\[10pt]
\quad\qquad+P(\bbC^{\rm c})\bigg\{\diy\mu^{m+n}(\big(\mathfrak{R}_{m,n}(\mathfrak{X}'_m,\mathfrak{Y}'_n)\geq M_e+t\big)\big|\bbC^{\rm c})\\[10pt]
\qquad\qquad\qquad-\mu^{m+n}\Big(\big(\phi_{\bbA}(\bx')\;\phi_{\bbA}(\by')\big)^{1-1/d}\geq \diy\frac{t}{4\ep}\Big)\bigg\}.\\
\ena\end{equation}
Define set $\bbK_t=\Big\{ \big(\phi_{\bbA}(\bx')\;\phi_{\bbA}(\by')\big)^{1-1/d}\geq \diy\frac{t}{4\ep}\Big\}$, so 
\beqq \begin{array}{l}\diy\mu^{m+n}\big(\mathfrak{R}_{m,n}(\mathfrak{X}'_m,\mathfrak{Y}'_n)\geq M_e+t \big|\bbC^{\rm c}\big)\\
\\
\quad=\diy\mu^{m+n}\big(\mathfrak{R}_{m,n}(\mathfrak{X}'_m,\mathfrak{Y}'_n)\geq M_e+t \big|\bbC^{\rm c}, \bbK_t\big) \mu^{m+n}(\bbK_t)\\
\\
\qquad+\diy\mu^{m+n}(\big(\mathfrak{R}_{m,n}(\mathfrak{X}'_m,\mathfrak{Y}'_n)\geq M_e+t\big)\big|\bbC^{\rm c}, \bbK^{\rm c}_t)\mu^{m+n}(\bbK^{\rm c}_t).\ena \eeqq
Since 
$$\diy\mu^{m+n}\big(\mathfrak{R}_{m,n}(\mathfrak{X}'_m,\mathfrak{Y}'_n)\geq M_e+t \big|\bbC^{\rm c}, \bbK_t\big)=1,$$
and 
\beqq\begin{array}{l}\diy\mu^{m+n}\big(\mathfrak{R}_{m,n}(\mathfrak{X}'_m,\mathfrak{Y}'_n)\geq M_e+t\big|\bbC^{\rm c}, \bbK^{\rm c}_t\big)\\[10pt]
\qquad\qquad=\diy\mu^{m+n}\big(\mathfrak{R}_{m,n}(\mathfrak{X}'_m,\mathfrak{Y}'_n)\geq M_e+t\big).\ena\eeqq
Consequently, from (\ref{eq75:1.1}) one can write 
\def\bbA{\mathbb{A}} 
\beq\begin{array}{l} \diy P\big(\mathfrak{R}_{m,n}(\mathfrak{X}_m,\mathfrak{Y}_n)\geq M_e+t\big)\\[10pt]
\qquad\qquad\leq \diy\mu^{m+n}\Big(\big(\phi_{\bbA}(\bx')\;\phi_{\bbA}(\by')\big)^{1-1/d}\geq \frac{t}{4\ep}\Big) \\[10pt]
\quad+P(\bbC^{\rm c}) \Big\{\diy\mu^{m+n}\big(\mathfrak{R}_{m,n}(\mathfrak{X}'_m,\mathfrak{Y}'_n)\geq M_e+t\big)\mu^{m+n}(\bbK_t^{\rm c})\Big\}\\[10pt]
\qquad\qquad\leq \diy\mu^{m+n}\Big(\big(\phi_{\bbA}(\bx')\;\phi_{\bbA}(\by')\big)^{1-1/d}\geq \frac{t}{4\ep}\Big) \\[10pt]
\quad+ \Big(\big(2\;(2\;g(\epsilon)-1)^2\big)^{-1}\Big) P\big(\mathfrak{R}_{m,n}(\mathfrak{X}_m,\mathfrak{Y}_n)\geq M_e+t\big) . \ena \eeq
Last inequality implies by owing to (\ref{eq752:1.1}) and $\mu^{m+n}(\bbK^{\rm c}_t)\leq 1$. For $g(\ep)\geq \diy{1}/{2}+{1}/{\big(2\sqrt{2}\big)}$, we have
$$1- \Big(\big(2\;(2\;g(\epsilon)-1)^2\big)^{-1}\Big)\geq 0,$$
or equivalently this holds true when $\ep\geq\diy (2h\sqrt{2}\;\delta^h_{m,n})\big/(\sqrt{2}-1)$. Further for $h\geq 7$ we have 
\begin{equation}h^2\delta^h_{m,n}\geq \diy (2h\sqrt{2}\;\delta^h_{m,n})\big/(\sqrt{2}-1),\end{equation}
therefore $P\big(\mathfrak{R}_{m,n}(\mathfrak{X}_m,\mathfrak{Y}_n)\geq M_e+t\big)$ is less than and equal to 
\begin{equation}
\bigg(1- \Big(\big(2\;(2\;g(\epsilon)-1)^2\big)^{-1}\Big) \bigg)^{-1}
 \diy\mu^{m+n}\Big(\big(\phi_{\bbA}(\bx')\;\phi_{\bbA}(\by')\big)^{1-1/d}\geq \frac{t}{4\ep}\Big).\end{equation}
By virtue of Lemma \ref{Eq:isoperimetry}, finally we obtain
\begin{equation} \begin{array}{l} \diy P\big(\mathfrak{R}_{m,n}(\mathfrak{X}_m,\mathfrak{Y}_n)\geq M_e+t\big)\\[8pt]
\leq 4\;\bigg(1- \Big(\big(2\;(2\;g(\epsilon)-1)^2\big)^{-1}\Big) \bigg)^{-1}\exp\Big(\diy\frac{-t^{d/(d-1)}}{8(4\epsilon)^{d/d-1}(m+n)}\Big).\ena\end{equation}
Similarly we can derive the same bound on $ P\big(\mathfrak{R}_{m,n}(\mathfrak{X}_m,\mathfrak{Y}_n)\leq M_e-t\big)$, so we obtain 
\begin{equation} \label{eq:appendix.theorem5}\begin{array}{l}P\Big(\big|\mathfrak{R}_{m,n}-M_e\big|\geq t \Big)\\[8pt]
\qquad\quad\leq C'_{m,n}(\epsilon,h)
\exp\Big(\diy\frac{-t^{d/(d-1)}}{8(4\ep)^{d/(d-1)}(m+n)}\Big). \ena\end{equation}
where
\begin{equation}\label{Apen:eq:concen:ep} \begin{array}{l}
C'_{m,n}(\ep,h)\\
\qquad =8\bigg(1-2^{-1}\;\Big(1-\diy\frac{2h\;\;O\big(h^{d-1}(m+n)^{1/d}\big)}{\ep}\Big)^{-2}\bigg)^{-1}.
\ena\end{equation}
We will analyze (\ref{eq:appendix.theorem5}) together with Theorem \ref{concentration}. 
The Next lemma will be employed in the Theorem \ref{concentration}'s proof. 
\vspace{2pt}
\def\ep{\epsilon}
\def\ep{\epsilon}
\begin{lemma}\label{lem.Deviation of the Mean and Median}
{\rm (Deviation of the Mean and Median)}  Consider $M_e$ as a median of $\mathfrak{R}_{m,n}$. Then for $\epsilon\geq h^2\delta^h_{mn,}$ and given $h\geq 7$, we have
\begin{equation}\Big|\bbE\big[\mathfrak{R}_{m,n}(\mathfrak{X}_m,\mathfrak{Y}_n)\big]-M_e\Big|\leq C_{m,n}(\epsilon,h)\;(m+n)^{(d-1)/d},\end{equation}
where $C_{m,n}(\ep,h)$ is a constant depending on $\ep$, $h$, $m$, and $n$ by
\beq C_{m,n}(\ep,h)=\diy C\;\bigg(1- \Big(\big(2\;(2\;g(\epsilon)-1)^2\big)^{-1}\Big) \bigg)^{-1},\eeq
\end{lemma}
where $C$ is a constant and 
$$\delta^h_{m,n}=O\big(h^{d-1}(m+n)^{1/d}\big),\;\; \hbox{and}\;\; g(\epsilon)=1-\diy\frac{h\;\delta^h_{m,n}}{\ep}.$$

We conclude this part by pursuing our primary intension which has been the Theorem \ref{concentration}'s proof.  Observe from Theorem \ref{Concentration around the median}, (\ref{Concentration:Mediam}), that
\beqq \begin{array}{l} P\Big(\Big|\mathfrak{R}_{m,n}-\bbE[\mathfrak{R}_{m,n}]\Big|\geq t+C_{m,n}(\ep,l)\big(m+n)^{(d-1)/d} \Big)\\[8pt]
\qquad \leq P\bigg(\big|\mathfrak{R}_{m,n}-M_e\big|+\big|\bbE[\mathfrak{R}_{m,n}]-M_e\big|\\[8pt]
\qquad\qquad\qquad\geq t+C_{m,n}(\ep,l)\big(m+n)^{(d-1)/d}\bigg)\\[8pt]
\qquad\leq P\Big(\big|\mathfrak{R}_{m,n}-M_e\big|\geq t\Big)\\[8pt]
\leq \diy 8\;\bigg(1- \Big(\big(2\;(2\;g(\epsilon)-1)^2\big)^{-1}\Big) \bigg)^{-1}
\;\exp\Big(\diy\frac{-t^{d/(d-1)}}{8(4\ep)^{d/d-1}(m+n)}\Big). \ena \eeqq
Note that the last bound is derived by (\ref{Concentration:Mediam}). The rest of the proof is as the following: When $t\geq 2C_{m,n}(\ep,h)\big(m+n)^{(d-1)/d}$ we use 
$$\Big(t-C_{m,n}(\ep,h)\big(m+n)^{(d-1)/d}\Big)^{d/(d-1)}\geq \Big(t/2\Big)^{d/(d-1)}.$$
 Therefore it turns out that
\begin{equation} \begin{array}{l}P\Big(\Big|\mathfrak{R}_{m,n}-\bbE[\mathfrak{R}_{m,n}]\Big|\geq t \Big)\\[8pt]
\leq 8\;\bigg(1- \Big(\big(2\;(2\;g(\epsilon)-1)^2\big)^{-1}\Big) \bigg)^{-1}
\exp\Big(\diy\frac{-t^{d/(d-1)}}{8(8\ep)^{d/(d-1)}(m+n)}\Big). \ena\end{equation}
On the other word, there exist constants $C'_{m,n}(\ep,h)$ depending on $m,\;n$, $\ep$, and $h$ such that
\begin{equation}\label{Apen:Ineq:concentration}\begin{array}{l}  P\Big(\Big|\mathfrak{R}_{m,n}-\bbE[\mathfrak{R}_{m,n}]\Big|\geq t \Big)\\
\qquad\leq C'_{m,n}(\ep,h) \exp\left(\diy\frac{-(t/(2\ep))^{d/(d-1)}}{(m+n)\;\tilde{C}}\right),  \ena\end{equation}
where $\tilde{C}=8 (4)^{d/(d-1)}$. 

To verify the behavior of bound (\ref{Apen:Ineq:concentration}) in terms of $\epsilon$, observe (\ref{eq:appendix.theorem5}) first; It is not hard to see that this function is decreasing in $\epsilon$. However, the function 
$$\diy\exp\left(\diy\frac{-(t/(2\ep))^{d/(d-1)}}{(m+n)\tilde{C}}\right)$$
increases in $\epsilon$. Therefore, one can not immediately infer that the bound in (\ref{Ineq:concentration}) is monotonic with respect to $\epsilon$.
For fixed $N=n+m$, $d$, and $h$ the first and second derivatives of the bound (\ref{Ineq:concentration}) with respect to $\ep$ are quite complicated functions. So deriving an explicit optimal solution for the minimization problem with the objective function (\ref{Ineq:concentration}) is not feasible. However, in sequel we discuss that under condition when $t$ is not much larger than $N=m+n$ this bound becomes convex with respect to $\epsilon$. Set
\begin{equation}\begin{array}{l} 
\diy K(\ep)=\diy C'_{m,n}(\ep,h) \; \exp\Big(\diy\frac{-B(t)}{\epsilon^{d/(d-1)}}\Big),\end{array}\end{equation}
where $C'_{m,n}$ is given in (\ref{eq:concen:ep}) and $$B(t)=\diy\frac{t^{d/(d-1)}}{8\;(8)^{d/(d-1)}(N)}.$$
By taking the derivative with respect to $\ep$, we have
\begin{equation}\begin{array}{l}
\diy\frac{d K(\ep)}{d\ep}=K(\ep)\left(\diy\frac{d}{d\ep}\big(\log C'_{m,n}\big) +\diy\frac{B(t)\; d/(d-1)}{\ep^{(2d-1)/(d-1)}}\right).
\end{array}\end{equation}
where 
\begin{equation}\begin{array}{l}
\diy\frac{d}{d\ep}\big(\log C'_{m,n}\big)=\diy\frac{-4\; a_h\; \ep}{(\ep-2a_h)(8 a_h^2-8\ep a_h+\ep^2)},
\end{array}\end{equation}
where $a_h=h\delta^h_{m,n}$. The second derivative $K(\ep)$ with respect to $\ep$ after simplification is given as
\begin{equation}\label{second.derivative}\begin{array}{l}
\diy\frac{d^2}{d\ep^2} K(\ep)
=\left(\diy\frac{-4\; a_h\; \ep}{(\ep-2a_h)(8 a_h^2-8\ep a_h+\ep^2)}+\diy\frac{B(t)\; \bar{d}}{\ep^{\bar{d}+1}}\right)^2\\
\\
+K(\ep)\left(\diy\frac{8a_h\; (8a_h^3+\ep^2(\ep-5a_h))}{(8a_h^2-8a_h\ep+\ep^2)^2(\ep-2a_h)^2}-\diy\frac{B(t)\bar{d}(\bar{d}+1)}{\ep^{\bar{d}+2}}\right),
\end{array}\end{equation}
where $\bar{d}=d/(d-1)$. The first term in (\ref{second.derivative}) and $K(\ep)$ are non-negative, so $K(\ep)$ is convex if the second term in the second line of (\ref{second.derivative}) is non-negative. We know that $\ep\geq h^2\delta^h_{m,n}=h\;a_h$, when $h=7$ we can parameterize $\epsilon$ by setting it equal to $\gamma a_h$ where $\gamma\geq 7$. After simplification, $K(\ep)$ is convex if 
\begin{equation}\begin{array}{l}
a_h^{\bar{d}-1}\Big(\gamma^{\bar{d}-1}+3\gamma^{\bar{d}-2}\Big)
+B(t)\bar{d}(\bar{d}+1)\\
\times \bigg\{a_h^{-1}\Big(-32\gamma^{-6}+64\gamma^{-5}-48\gamma^{-4}+8\gamma^{-3}-\diy\frac{7}{2}\gamma^{-2}+2\gamma^{-1}-\diy\frac{1}{8}\Big)\\
+a_h^{-2}\Big(32\gamma^{-6}-64\gamma^{-5}+40\gamma^{-4}+8\gamma^{-3}+\diy\frac{1}{2}\gamma^{-2}\Big)\bigg\}\geq 0.
\end{array}\end{equation}
This is implied if
\begin{equation}\label{convex.inequl}\begin{array}{l}
0\leq B(t)\bar{d}(\bar{d}+1)\;a_h^{-1}\\
\times \Big(-32\gamma^{-6}+64\gamma^{-5}-48\gamma^{-4}+8\gamma^{-3}-\diy\frac{7}{2}\gamma^{-2}+2\gamma^{-1}-\diy\frac{1}{8}\Big),
\end{array}\end{equation}
such that $\gamma\geq 7$. One can easily check that as $\gamma\rightarrow \infty$, then  $(\ref{convex.inequl})$ tends to $ -\diy\frac{1}{8}B(t)\bar{d}(\bar{d}+1)\;a_h^{-1}$. This term can be negligible unless we have $t$ that is much larger than $N=m+n$ with the threshold depending on $d$. Here by setting $B(t)/a_h=1$ a rough threshold $t=O\big(7^{d-1}(m+n)^{1-1/d^{2}}\big)$ depending on $d$, $m+n$ is proposed. Therefore minimizing (\ref{eq:appendix.theorem5}) and (\ref{Apen:Ineq:concentration}) with respect to $\ep$ when optimal $h=7$ is a convex optimization problem. Denote $\ep^*$ the solution of the convex optimization problem (\ref{convexoptimization}). By plugging optimal $h$ ($h=7$) and $\epsilon$ ($\ep=\ep^*)$ in (\ref{eq:appendix.theorem5}) and (\ref{Apen:Ineq:concentration}) we derive (\ref{Concentration:Mediam}) and (\ref{Ineq:concentration}), respectively. 

In this Appendix we also analyze the bound numerically. By simulation, we observed that lower $h$ i.e. $h=7$ is the optimal value experimentally. Indeed, this can be verified by the Theorem \ref{Concentration:Mediam}'s proof. We address the reader to Lemma \ref{varianceD} in Appendix \ref{AppendixD} and Supplementary Material where as $h$ increases, the lower bound for the probability increases, too. In other words, for fixed $N=m+n$ and $d$ the lowest $h$ implies the maximum bound in (\ref{variancD}). For this, we set $h=7$ in our experiments. We vary the dimension $d$ and sample size $N=m+n$ in relatively large and small ranges. In Table 2 we solve (\ref{convexoptimization}) for various values of $d$ and $N=m+n$.
We also compute the lower bound for $\ep$ i.e. $7^{d+1}N^{1/d}$ per experiment. In Table 2, we observe that as we have higher dimension the optimal value  $\ep^*$ equals the $\ep$ lower bound $h^{d+1}N^{1/d}$, but this is not true for smaller dimensions with even relatively large sample size.
\begin {table}[H]\label{Tabl1}
\begin{center}
\scalebox{0.88}{
\begin{tabular}{ |c|c|c|c|c||c|  }
 \hline
 \multicolumn{6}{|c|}{Concentration bound (11)} \\
 \hline
$d$ & $N=m+n$  & $\epsilon^*$ & $t_0$ & $h^{d+1}N^{1/d}$ & Optimal (11) \\
 \hline
 2 & $10^3$ & $1.1424 \times 10^4$ & $2\times 10^7$ & $1.0847 \times 10^4$& 0.3439\\ 
 4 & $10^4$ & $1.7746\times 10^5$ & $3\times 10^{10}$ & 168070 & 0.0895\\
 5 & 550& $4.7236\times 10^5$ & $10^{10}$ & $4.1559\times 10^5$ &  0.9929\\
 6 & $10^4$ & $3.8727 \times 10^6$ & $2\times 10^{12}$ & $3.8225\times 10^6$ & 0.1637\\
8 &1200 & $9.7899\times 10^{7}$ & $12\times 10^{12}$ & $9.7899\times 10^{7}$ & 0.7176\\
10 & 3500  & $4.4718\times 10^{9}$  &$2\times 10^{15}$  & $4.4718\times 10^{9}$ & 0.4795\\
15 & $10^8$ &$1.1348\times 10^{14}$& $10^{24}$  & $1.1348\times 10^{14}$ & 0.9042 \\
 \hline
\end{tabular}
}
\caption*{{ Table 2: \small $d$, $N$, $\ep^*$ are dimension, total sample size $m+n$, and optimal $\ep$ for the bound in (\ref{Ineq:concentration}). The column $h^{d+1}N^{1/d}$ represents approximately the lower bound for $\ep$ which is our constraint in the minimization problem and our assumption in Theorems \ref{Concentration around the median}, \ref{concentration}. Here we set $h=7$. }}
\end{center}
\end{table}
 To validate our proposed bound in (\ref{Ineq:concentration}), we again set $h=7$ and for $d=4,5,7$ we ran experiments with sample sizes $N=m+n=9000,1100,140$ respectively. Then we solved the minimization problem to derive optimal bound for $t$ in the range $10^{10}[1,3]$. Note that we chose this range to have non-trivial bound for all three curves, otherwise the bounds partly become one. Fig \ref{figA} shows that when $t$ increases in the given range, the optimal curves approach zero. 
\begin{figure}[h]
\centering
  \includegraphics[width=0.85\linewidth]{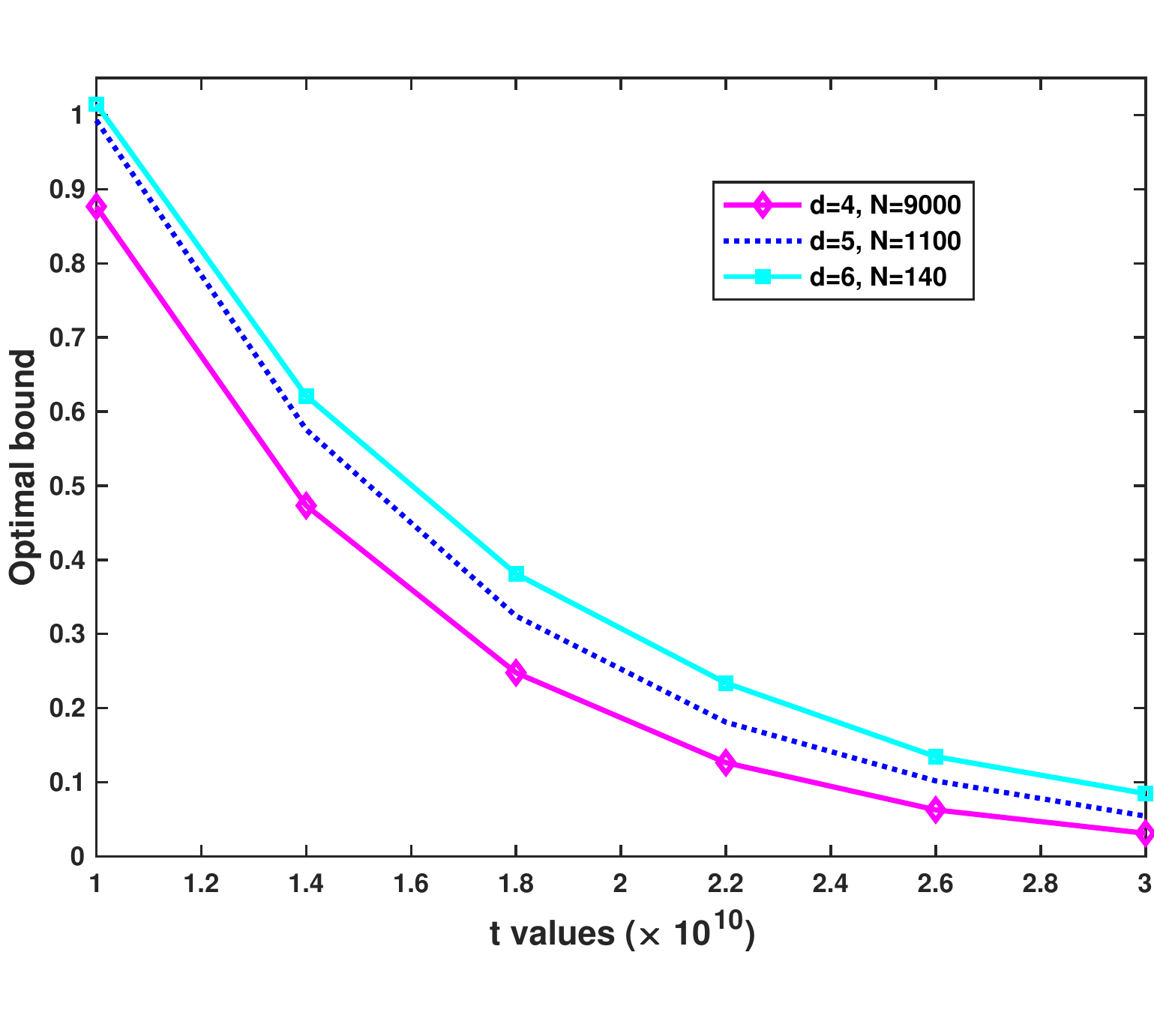}
  \caption{{\small Optimal bound for (\ref{Ineq:concentration}), when $h=7$ versus $t\in 10^{10}[1,3]$. The bound decreases as $t$ grows.}}
  \label{figA}
\end{figure}

To prove the Theorem \ref{prob.variance} in the concentration of $\mathfrak{R}_{m,n}$, Theorem \ref{concentration}, let 
$$\delta= C'_{m,n}(\ep^*) \exp\Big(\diy\frac{-(t/(2\ep^*))^{d/(d-1)}}{(m+n)\;\tilde{C}}\Big),$$
this implies 
\beq t=O\Big(\epsilon^*\;(m+n)^{(d-1)/d}\big(\log\big(C'_{m,n}(\ep^*)\big/\delta)\big)^{(d-1)/d}\Big).\eeq
Then the proofs are completed.




\ifCLASSOPTIONcaptionsoff
  \newpage
\fi



\label{sec:refs}
\bibliographystyle{IEEEbib}
\bibliography{refs}





\newpage
\onecolumn
\begin{center}
{\bf \Large {Supplementary Materials}}
\end{center}

\bigskip
\def\bz{\mathbf{z}}
{\it Lemma \ref{lem:error1}:} 
 Let $g(\bx)$ be a density function with support $[0,1]^d$ and belong to the strong H\"{o}lder class $\Sigma_d^{\rm S}(\eta,L)$, $0<\eta\leq 1$, expressed in Definition \ref{def:strong.Holder}. Also, assume that $P(\bx)$ is a $\eta$-H\"{o}lder smooth function, such that its absolute value is bounded from above by some constants $c$. Define the quantized density function with parameter $l$ and constants $\phi_i$ as
\beq\label{hat:g} \widehat{g}(\bx)=\sum\limits_{i=1}^M \phi_i\mathbf{1}\{\bx\in Q_i\},\;\;\; \hbox{where}\; \phi_i=l^d\;\int\limits_{Q_i} g(\bx)\;\rd\bx, \eeq
and $M=l^d$ and $Q_i=\{\bx,\bx_i:\|\bx-\bx_i\|<l^{-d}\}$. Then 
\beq \diy \int \Big\|\big(g(\bx)-\widehat{g}(\bx)\big) P(\bx)\Big\|\;\rd \bx\leq  O(l^{-d\eta}).\eeq
\begin{IEEEproof}
By the mean value theorem, there exist points $\epsilon_i\in Q_i$ such that
\beqq \phi_i=l^d \;\int\limits_{Q_i} g(\bx)\;\rd \bx=g(\epsilon_i).\eeqq
Using the fact that $g\in \Sigma_d^{\rm S}(\eta,L)$ and $P(\bx)$ is a bounded function, we have
\beqq\begin{array}{ccl} \diy\int\big\| g(\bx)-\widehat{g}(\bx)\big) P(\bx)\big\|\;\rd \bx &=&\diy \sum_{i=1}^{M}\int_{Q_i} \big\|(g(\bx)-\Phi_i)P(\bx)\big\|\rd\bx\\
&=&\diy  \sum_{i=1}^{M}\int_{Q_i} \big\|(g(\bx)-g(\epsilon_i))P(\bx)\big\|\rd\bx\\
&\leq&\diy  c\;L  \sum_{i=1}^{M}\int_{Q_i} g(\bx)\big\|\bx-\epsilon_i\big\|^{\eta}\;\rd\bx.
\ena\eeqq
Here $L$ is the H\"{o}lder constant. As $\bx, \epsilon_i\in Q_i$, a sub-cube with edge length $l^{-1}$, then $\big\|\bx-\epsilon_i\big\|^{\eta}=O(l^{-d\eta})$ and $\diy\sum\limits_{i=1}^M\int\limits_{Q_i} g(\bx)\;\rd\bx=1$. This concludes the proof.
\end{IEEEproof}
\bigskip
{\it Lemma \ref{lem2:3}:}
Let  $\Delta(\bx,\mathcal{S})$ denote the degree of vertex $\bx\in\mathcal{S}$ in the $MST$ over set $\mathcal{S}\subset \bbR^d$ with the $n$ number of vertices. For given function $P(\bx,\bx)$, one yields
\begin{equation}\label{eq2:(10)} \diy\int P(\bx,\bx)  g(\bx) \bbE[\Delta(\bx,\mathcal{S})]  \;\rd \bx=2\;\int P(\bx,\bx) g(\bx)\;\rd\bx+ \varsigma_\eta(l,n),\end{equation}
where for constant $\eta>0$, 
\beq\label{def:O(l)} \varsigma_{\eta}(l,n)=\diy\Big(O\big(l/n\big)-\diy 2\;l^{d}/n\Big)\diy\int g(\bx)P(\bx,\bx)\;\rd\bx+O(l^{-d\eta}).\eeq
\begin{IEEEproof}
Recall notations in Lemma \ref{lem:error1} and 
\beqq \Big|\int g(\bx) P(\bx)\;\rd\bx -\int \widehat{g}(\bx) P(\bx)\;\rd \bx\Big|\leq \diy \int \big|\big(g(\bx)-\widehat{g}(\bx)\big) P(\bx)\big|\;\rd \bx.\eeqq
Therefore by substituting $\widehat{g}$, defined in (\ref{hat:g}), into $g$ with considering its error, we have
 \begin{equation}\label{eq:lem3.2}\begin{array}{l} \diy \int P(\bx,\bx) g(\bx) \bbE[\Delta(\bx,\mathcal{S})]\;\rd \bx \\
\qquad\qquad=\diy \int P(\bx,\bx)\bbE[\Delta(\bx,\mathcal{S})]\sum\limits_{i=1}^M \phi_i \mathbf{1}\{\bx\in Q_i\}\;\rd \bx+O(l^{-d\eta})\\
\qquad\qquad= \diy \sum\limits_{i=1}^M  \phi_i \diy\int_{Q_i} P(\bx,\bx)\bbE[\Delta(\bx, \mathcal{S})]\;\rd\bx+O(l^{-d\eta}).
\end{array}\end{equation}
Here $Q_i$ represents as before in Lemma \ref{lem:error1}, 
so the RHS of (\ref{eq:lem3.2}) becomes 
\beq  \begin{array}{l} \diy \sum\limits_{i=1}^M  \phi_i \diy\int_{Q_i} P(\bx,\bx)\bbE[\Delta(\bx, \mathcal{S}\cap Q_i)]\;\rd\bx + \diy \sum\limits_{i=1}^M  \phi_i \diy\int_{Q_i} P(\bx,\bx)O(l^{1-d}\big/n)+O(l^{-d\eta})\\
\quad =\diy \sum\limits_{i=1}^M \phi_i  P(\bx_i,\bx_i)\diy\frac{1}{M}\diy\int _{Q_i} M\;\bbE[\Delta(\bx, \mathcal{S}\cap Q_i)]\;\rd\bx + \diy \sum\limits_{i=1}^M  \phi_i \diy\int_{Q_i} P(\bx,\bx)O(l^{1-d}\big/n)+2\;O(l^{-d\eta}).
\end{array}\eeq
Now note that $\diy \int_{Q_i}M\; \bbE[\Delta(\bx, \mathcal{S}\cap Q_i)]\;\rd\bx$ is the expectation of $\bbE[\Delta(\bx, S\cap Q_i)]$ over the nodes in $Q_i$, which is equal to $2-\diy\frac{2}{k_i}$, where $k_i=\diy\frac{n}{M}$. Consequently, we have
\beq\begin{array}{l}  \diy \int P(\bx,\bx) g(\bx) \bbE[\Delta(\bx,\mathcal{S})]\;\rd \bx=\diy \left(2-\diy\frac{2\;M}{n}\right)\sum\limits_{i=1}^M \phi_i \; P(\bx_i,\bx_i)\frac{1}{M}+\diy O\left(\frac{l^{1-d}}{n}\right)\sum\limits_{i=1}^M \phi_i \; P(\bx_i,\bx_i)+3\;O(l^{-d\eta})\\
\\
\qquad\quad= \diy 2\int g(\bx) P(\bx,\bx)\;\rd\bx +5\; O(l^{-d\eta)})+\diy M\; \left(O\left(\frac{l^{1-d}}{n}\right)-\left(\frac{2}{n}\right)\right)\int g(\bx) P(\bx,\bx)\;\rd\bx.
\end{array}\eeq
This gives the assertion (\ref{eq2:(10)}).
\end{IEEEproof}
\bigskip
{\it Lemma \ref{lem:2.4.0}:}
Assume that for given $k$, $g_k(\bx)$ is a bounded function belong to $\Sigma^s_d(\eta,L)$. Let $P:\bbR^d\times\bbR^d\mapsto[0,1]$ be a symmetric, smooth, jointly measurable function, such that, given $k$,  for almost every $\bx\in\bbR^d$, $P(\bx,.)$ is measurable with $\bx$ a Lebesgue point of the function $g_k(.)P(\bx,.)$. Assume that the first derivative $P$ is bounded.  For each $k$, let $\BZ_1^k,\BZ_2^k,\dots,\BZ_k^k$ be independent $d$-dimensional variable with common density function $g_k$. Set $\mathfrak{Z}_k=\{\BZ_1^k,\BZ_2^k\dots,\BZ_k^k\}$ and $\mathfrak{Z}_k^{\bx}=\{\bx,\BZ_2^k,\BZ_3^k\dots,\BZ_k^k\}$. Then
\begin{equation}\label{eq:13.1.0} \begin{array}{l}\diy \bbE\bigg[ \sum\limits_{j=2}^k P(\bx,\BZ_j^k)\mathbf{1}\big\{(\bx,\BZ_j^k)\in MST(\mathfrak{Z}_k^{\bx})\big\}\bigg]\\
\qquad=\diy P(\bx,\bx)\;\bbE\big[\Delta(\bx,\mathfrak{Z}_k^{\bx})\big]+\Big\{O\big(k^{-\eta/d}\big)+O\big(k^{-1/d}\big)\Big\}.\ena\end{equation}
\begin{IEEEproof}
Let $\bbB(\bx,r)=\{\by:\|\by-\bx\|_d\leq r\}$. For any positive K, we can obtain:
\begin{equation}\label{lemma:B.5:eq1} 
\begin{array}{l}
\diy\bbE\sum\limits_{j=2}^k\Big|P(\bx, \BZ_j^k)-P(\bx,\bx)\Big|\mathbf{1}\big\{\BZ_j^k\in \bbB\big(\bx,K k^{-1/d}\big)\big\}\\[15pt]
\qquad =\diy (k-1)\int\limits_{\bbB\big(\bx;K k^{-1/d}\big)}\Big|\big(P(\bx,\by)g_k(\by)-P(\bx,\bx)g_k(\bx)\big)+\diy P(\bx,\bx)\big(g_k(\bx)-g_k(\by)\big)\Big|\;\rd\by\\[20pt]
\leq (k-1) \bigg[ \diy\int\limits_{\bbB\big(\bx;K k^{-1/d}\big)}\Big|\big(P(\bx,\by)g_k(\by)-P(\bx,\bx)g_k(\bx)\big)\Big|\rd\by+O\big(k^{-\eta/d}\big)\BV\big(\bbB\big(\bx,K k^{-1/d}\big)\bigg],  \end{array}
\end{equation}
where $\BV$ is the volume of space $\bbB$ which equals to $O(k^{-1})$. Note that the above inequality appears cause $g_k(\bx)\in \diy \Sigma^s_d(\eta,L)$ and $ P(\bx,\bx)\in [0,1]$. The first order Taylor series expansion of $P(\bx,\by)$ around $\bx$ is
\beqq \begin{array}{ccl} P(\bx,\by)&=&P(\bx,\bx)+P^{(1)}(\bx,\bx)\|\by-\bx\|+o\big(\|\by-\bx\|^2\big)\\[10pt]
&=& P(\bx,\bx)+O\big(k^{-1/d}\big)+o\big(k^{-2/d}\big).\ena\eeqq
Then, by recalling the strong H\"{o}lder class, we have
\beqq\begin{array}{ccl} \Big|P(\bx,\by)g_k(\by)-P(\bx,\bx)g_k(\bx)\Big|&=&\diy\Big|\big(P(\bx,\bx)+O(k^{-1/d})\big)\big(g_k(\bx)+O(k^{-\eta/d})\big)-P(\bx,\bx)g_k(\bx)\Big|\\[10pt]
&=& O(k^{-\eta/d})+O(k^{-1/d}). \ena\eeqq 
Hence, the RHS of (\ref{lemma:B.5:eq1}) becomes
\beqq\begin{array}{l} \diy(k-1)\Big[\big(O(k^{-\eta/d})+O(k^{-1/d})\big)\BV\big(\bbB\big(\bx,K k^{-1/d}\big)\big)+\diy O\big(k^{-\eta/d}\big)\BV\big(\bbB\big(\bx,K k^{-1/d}\big)\big)\bigg]\\[10pt]
\qquad\qquad\qquad=\diy (k-1)\Big[O\big(k^{-1-\eta/d}\big)+O\big(k^{-1-1/d}\big)\Big].\ena\eeqq
The expression in (\ref{eq:13.1.0}) can be obtained by choice of $K$.
\end{IEEEproof}
\bigskip
{\it Lemma \ref{lem:2.4}:}
Consider the notations and assumptions in Lemma \ref{lem:2.4.0}.  Then 
\begin{equation}\label{lema:eq:2.4}\begin{array}{l}\diy  \Big|\diy k^{-1} \mathop{\sum\sum}_{\ 1\leq i<j\leq k} P(\BZ_i^k,\BZ_j^k)\mathbf{1}\{(\BZ_i^k,\BZ_j^k)\in MST(\mathfrak{Z}_k)-\diy\int_{\bbR^d}P(\bx,\bx)g_k(\bx)\;\rd\bx\Big|\\[10pt]
\qquad\qquad \quad \leq\diy \varsigma_{\eta}(l,k)+O(k^{-\eta/d})+O(k^{-1/d}).\ena\end{equation}
Here $MST(\mathcal{S})$ denotes the MST graph over nice and finite set $\mathcal{S}\subset \bbR^d$ and $\eta$ is the smoothness H\"{o}lder parameter. Note that $\varsigma_{\eta}(l,k)$ is given as before  in (\ref{def:O(l)}). 
\begin{IEEEproof}
Following notations  in \cite{HP}, let $\Delta(\bx,\mathcal{S})$ denote the degree of vertex $\bx$ in the $MST(\mathcal{S})$ graph. Moreover, let $\bx$ be a Lebesgue point of $g_k$ with $g_k(\bx)>0$. Also let $\mathfrak{Z}_k^{\bx}$ be the point process $\{\bx, \BZ_2^k,\BZ_3^k,\dots, \BZ_k^k\}$. Now by virtue of (\ref{lemma:B.5:eq1}) in Lemma \ref{lem:2.4.0},  we can write
\beq\begin{array}{l}\label{eq:13.1} \bbE\bigg[ \sum\limits_{j=2}^k P(\bx,\BZ_j^k)\mathbf{1}\{(\bx,\BZ_j^k)\in MST(\mathfrak{Z}_k^{\bx})\}\bigg]
=\diy P(\bx,\bx)\;\bbE\big[\Delta(\bx,\mathfrak{Z}_k^{\bx})\big]+\Big\{O\big(k^{-\eta/d}\big)+O\big(k^{-1/d}\big)\Big\}.\ena\eeq
On the other hand it can be seen that
\begin{equation}\begin{array}{l} \diy k^{-1}\bbE\Big[\mathop{\sum\sum}_{1\leq i<j\leq k}P(\BZ_i^k,\BZ_j^k)\mathbf{1}\{(\BZ_i^k,\BZ_j^k)\in MST(\mathfrak{Z}_k)\}\Big]\\[15pt]
\qquad\qquad=\diy\frac{1}{2} \bbE\Big[\sum\limits_{j=2}^k P(\BZ_1^k,\BZ_j^k) \mathbf{1}\{(\BZ_i^k,\BZ_j^k)\in MST(\mathfrak{Z}_k)\}\Big]\\[15pt]
\qquad\qquad=\diy \frac{1}{2} \int g_k(\bx)\; \rd \bx\;\bbE\Big[\sum\limits_{j=2}^k P(\bx,\BZ_j^k)\mathbf{1}\{(\bx,\BZ_j^k)\in MST(\mathfrak{Z}_k)\}\Big].
\end{array}\end{equation}
Recalling (\ref{eq:13.1}), 
\begin{equation}\label{eq:2.17}\begin{array}{cl} =\diy \frac{1}{2} \int g_k(\bx) P(\bx,\bx)\bbE\big[\Delta(\bx,\mathfrak{Z}_k^\bx)\big]\;\rd\bx+O\big(k^{-\eta/d}\big)+O\big(k^{-1/d}\big).\ena\end{equation}
By virtue of Lemma \ref{lem2:3},  (\ref{eq2:(10)}) can be substituted into expression (\ref{eq:2.17}) to obtain (\ref{lema:eq:2.4}). 
\end{IEEEproof}
\bigskip
{\it Theorem \ref{thm:bias}:}
 Assume $\mathfrak{R}_{m,n}:=\mathfrak{R}(\mathfrak{X}_m,\mathfrak{Y}_n)$ denotes the FR test statistic as before. Then the rate for the  bias of the $\mathfrak{R}_{m,n}$ estimator for $0<\eta\leq 1$, $d\geq 2$ is of the form: 
 \begin{equation}\label{bound:Bias} \begin{array}{l}\diy \Big|\frac{\bbE\big[\mathfrak{R}_{m,n}\big]}{m+n}-2pq\int \frac{f_0(\bx) f_1(\bx)}{p f_0(\bx)+q f_1(\bx)}\;\rd\bx \Big| \leq O\big(l^d(m+n)^{-\eta/d}\big)+O(l^{-d\eta}).
\ena \end{equation}
Here $\eta$ is the Holder smoothness parameter. A more explicit form for the bound on the RHS is given in (\ref{final:up.bound232}) below: 
\begin{equation}
\label{final:up.bound232}\begin{array}{ccl}
\Big|\diy\frac{\bbE\big[\mathfrak{R}'_{m,n}(\mathfrak{X}_m,\mathfrak{Y}_n)\big]}{m+n}- \diy\int \frac{2pq f_0(\bx)f_1(\bx)}{pf_0(\bx)+q f_1(\bx)}\;\rd\bx\Big|\leq \diy  O\big(l^d(m+n)^{-\eta/d}\big)\\
\\
\qquad + O\big(l^d(m+n)^{-1/2}\big)+2\;c_1\;l^{d-1} (m+n)^{(1/d)-1}+c_d\;\;2^d\;(m+n)^{-1}\\
\\
-\diy 2\; l^d (m+n)^{-1} \diy\int \frac{2 p q f_0(\bx) f_1(\bx)}{p\;f_0(\bx)+q\;f_1(\bx)}\;\rd\bx+c_2\; (m+n)^{-1} l^{d}\\
\\
+\diy O(l)(m+n)^{-1} \sum\limits_{i=1}^M l^d (a_i)^{-1}\int \frac{2 f_0(\bx) f_1(\bx)}{p\;f_0(\bx)+q\;f_1(\bx)}\;\rd\bx+O(l^{-d\eta})\\
\\
+ \diy O(l) \sum\limits_{i=1}^M l^{d/2} \frac{\sqrt{b_i}}{a_i^2} \int \frac{2 f_0(\bx) f_1(\bx) \big(f_0(\bx)\sqrt{m}+f_1(\bx)\sqrt{n}\big)}{ \big(m f_0(\bx)+n f_1(\bx)\big)^2}\;\rd\bx\\
\\
\qquad +\diy \sum\limits_{i=1}^M 2\; l^{-d/2} \frac{\sqrt{b_i}}{a_i^{2}}\int \diy \frac{f_0(\bx) f_1(\bx)\Big(\alpha_i \beta_i\big(m a_i  f^2_0(\bx)+n b_i  f^2_1(\bx)\big) \Big)^{1/2}}{\big(m f_0(\bx)+n f_1(\bx)\big)^2 (m+n)}\;\rd \bx.
\end{array}
\end{equation}
\begin{IEEEproof}
Assume $M_m$ and $N_n$ be Poisson variables with mean $m$ and $n$, respectively, one independent of another and of $\{\BX_i\}$ and $\{\BY_j\}$. Let also $\mathfrak{X}'_m$ and $\mathfrak{Y}'_n$ be the Poisson processes $\{\BX_1,\dots,\BX_{M_n}\}$ and $\{\BY_1,\dots,\BY_{N_n}\}$. Set $\mathfrak{R}'_{m,n}:=\mathfrak{R}_{m,n}(\mathfrak{X}'_m,\mathfrak{Y}'_n)$. Applying Lemma 1, and (12) cf. \cite{HP}, we can write
\beq \label{eq:5.19}\Big|\mathfrak{R}'_{m,n}-\mathfrak{R}_{m,n}\Big|\leq K_d \big(|M_m-m|+|N_n-n|). \eeq
Here $K_d$ denotes the largest possible degree of any vertex of the MST graph in $\bbR^d$. Moreover by the matter of Poisson variable fact and using stirling approximation, \cite{WW}, we have
\begin{equation}\bbE\big[\big|M_m-m\big|\big]=\diy e^{-m}\diy\frac{m^{m+1}}{m!}\leq  e^{-m}\diy\frac{m^{m+1}}{\sqrt{2\pi}m^{m+1/2}e^{-m}}=O\big(\;m^{1/2}\big). \end{equation}
Similarly $\bbE\big[\big|N_n-n\big|\big]=O(n^{1/2})$. Therefore by (\ref{eq:5.19}), one yields
\beq \begin{array}{l}\bbE[\mathfrak{R}_{m,n}]=\bbE\big[\mathfrak{R}_{m,n}-\mathfrak{R}'_{m,n}\big]+\bbE\big[\mathfrak{R}'_{m,n}\big]
=O\big((m+n)^{1/2}\big)+\bbE\big[\mathfrak{R}'_{m,n}\big].\ena\eeq
Therefore 
\beq\label{eq2:2.23} \diy\frac{ \bbE[\mathfrak{R}_{m,n}]}{m+n}=\diy\frac{\bbE\big[\mathfrak{R}'_{m,n}\big]}{m+n}+O\big((m+n)^{-1/2}\big).\eeq
Hence it will suffice to obtain the rate of convergence of $\bbE\big[\mathfrak{R}'_{m,n}\big]\big/(m+n)$ in the RHS of (\ref{eq2:2.23}). For this, let $m_i$, $n_i$ denote the number of Poisson process samples $\mathfrak{X}'_m$ and $\mathfrak{Y}'_n$ with the FR statistic $\mathfrak{R}'_{m,n}$, falling into partitions $Q'_i$ with FR statistic $\mathfrak{R}'_{m_i,n_i}$. Then by virtue of Lemma \ref{lema:1.1}, we can write 
\beqq\label{Eq:3.2} \bbE\Big[\mathfrak{R}'_{m,n}\Big]\leq \sum\limits_{i=1}^M \bbE\Big[\mathfrak{R}'_{m_i,n_i}\Big]+2\;c_1\;l^{d-1}(m+n)^{1/d}. \eeqq
Note that the Binomial RVs $m_i$, $n_i$ are independent with marginal distributions $m_i\sim B(m,a_i l^{-d})$, $n_i\sim B(n,b_il^{-d})$, where $a_i$, $b_i$ are non-negative constants satisfying, $\forall i,\; a_i\leq b_i$ and  $\sum\limits_{i=1}^{l^d} a_i l^{-d}=\sum\limits_{i=1}^{l^d} b_i l^{-d}=1$. 
Therefore
\begin{equation}\label{Eq:2:2.19}\bbE\Big[\mathfrak{R}'_{m,n}\Big]\leq \sum\limits_{i=1}^M\bbE\Big[\bbE\Big[\mathfrak{R}'_{m_i,n_i}|m_i,n_i\Big]\Big]+2\;c_1\;l^{d-1}(m+n)^{1/d}. \end{equation}

Let us first compute the internal expectation given $m_i$, $n_i$. For this reason, given $m_i$, $n_i$,  let $Z_1^{m_i,n_i}, Z_2^{m_i,n_i}, \dots$ be independent variables with common densities $g_{m_i,n_i}(\bx)=\Big(m_i f_0(\bx)+n_i f_1(\bx)\Big)\big/(m_i+n_i)$, $\bx\in\bbR^d$. Moreover let $L_{m_i,n_i}$ be an independent Poisson variable with mean $m_i+n_i$  Denote $\mathfrak{F}'_{m_i,n_i}=\{Z_1^{m_i,n_i},\dots, Z_{L_{m_i.n_i}}^{m_i,n_i}\}$ a non-homogeneous Poisson of rate $m_if_0+n_if_1$. Let $\mathfrak{F}_{m_i,n_i}$ be the non-Poisson point process $\{Z_1^{m_i,n_i},\dots Z_{m_i+n_i}^{m_i,n_i}\}$. Assign a mark from the set $\{1,2\}$ to each points of $\mathfrak{F}'_{m_i,n_i}$. Let $\widetilde{\mathfrak{X}}'_{m_i}$ be the sets of points marked 1 with each probability $m_i f_0(\bx)\big/ \big(m_i f_0(\bx)+n_i f_i(\bx)\big)$ and let $\widetilde{\mathfrak{Y}}'_{n_i}$ be the set points with mark 2. Note that owing to the marking theorem  \cite{Ki}, $\widetilde{\mathfrak{X}}'_{m_i}$ and $\widetilde{\mathfrak{Y}}'_{n_i}$ are independent Poisson processes with the same distribution as $\mathfrak{X}'_{m_i}$ and $\mathfrak{Y}'_{n_i}$ , respectively. Considering $\widetilde{R}'_{m_i.n_i}$ as FR statistic over nodes in $\widetilde{\mathfrak{X}}'_{m_i}\cup \widetilde{\mathfrak{Y}}'_{n_i}$ we have 
\beqq \bbE\big[\mathfrak{R}'_{m_i,n_i}|m_i,n_i\big]=\bbE\big[\widetilde{\mathfrak{R}}'_{m_i,n_i}|m_i,n_i\big].\eeqq

Again using Lemma 1 and analogous arguments in \cite{HP} along with the fact that $\bbE\big[|M_m+N_n-m-n|\big]=O((m+n)^{1/2})$, we have
\beqq\begin{array}{ccl} \diy\bbE\big[\widetilde{\mathfrak{R}}'_{m_i,n_i}|m_i,n_i\big]=\diy\bbE\Big[\bbE\big[\widetilde{\mathfrak{R}}'_{m_i,n_i}|\mathfrak{F}'_{m_i,n_i}\big]\Big]\\
\\
=\diy\bbE\Big[\mathop{\sum\sum}_{s<j<m_i+n_i}P_{m_i,n_i}(Z_s^{m_i,n_i},Z_j^{m_i,n_i})\mathbf{1}
\big\{(Z_s^{m_i,n_i},Z_j^{m_i,n_i})\in\mathfrak{F}_{m_i,n_i}\big\}\Big]
+O((m_i+n_i)^{1/2})).\\ 
\end{array}\eeqq
Here, 
\beqq \begin{array}{l} \diy P_{m_i,n_i}(\bx,\by):=P_{r}\{\hbox{mark}\; x\neq \hbox{mark}\;y, (\bx,\by)\in \mathfrak{F}'_{m_i,n_i}\}\\
\\
\qquad=\diy\frac{m_if_0(\bx) n_if_1(\by)+n_if_1(\bx)m_if_0(\by)}{\big(m_i f_0(\bx)+n_i f_1(\bx)\big)\big(m_if_0(\by)+n_i f_1(\by)\big)}.\ena\eeqq
By owing  to Lemma \ref{lem:2.4}, we obtain 
\beq\label{B.23}\begin{array}{ccl}\diy \sum\limits_{i=1}^M\bbE_{m_i,n_i}\bbE\Big[\mathop{\sum\sum}_{s<j<m_i+n_i}P_{m_i,n_i}(Z_s^{m_i,n_i},Z_j^{m_i,n_i})
 \mathbf{1}
\big\{(Z_s^{m_i,n_i},Z_j^{m_i,n_i})\in\mathfrak{F}_{m_i,n_i}\big\}\Big]+\diy \sum\limits_{i=1}^M\bbE_{m_i,n_i}\big[O\big((m_i+n_i)\big)^{1/2}\big]\\[15pt]
=\diy\sum\limits_{i=1}^M \bbE_{m_i,n_i}\Big[(m_i+n_i)\int g_{m_i,n_i}(\bx,\bx) P_{m_i,n_i}(\bx,\bx)\;\rd\bx
+\big(\varsigma_\eta(l,m_i,n_i)+O\big((m_i+n_i)^{-\eta/d}\big)\\[15pt]
+O\big((m_i+n_i)^{-1/d}\big)\big)(m_i+n_i)\Big]
+\diy \sum\limits_{i=1}^M\bbE_{m_i,n_i}\big[O\big((m_i+n_i)^{1/2}\big)\big],\end{array}\eeq
where 
\begin{equation*} \begin{array}{l} \varsigma_\eta(l,m_i,n_i)
=\left(O\big(l/(m_i+n_i)\big)-\diy 2\;l^{d}/(m_i+n_i)\right)\diy\int g_{m_i,n_i}(\bx)P_{m_i,n_i}(\bx,\bx)\;\rd\bx+O(l^{-d\eta}).\ena\end{equation*}
The expression in (\ref{B.23}) equals to 
\begin{equation} \label{RHS1}\begin{array}{l}\diy\sum\limits_{i=1}^M \int\bbE_{m_i,n_i}\Big[\frac{2m_i n_i f_0(\bx) f_1(\bx)}{m_i f_0(\bx)+n_i f_1(\bx)}\Big]\;\rd\bx+\diy\sum\limits_{i=1}^M\bbE_{m_i,n_i}\big[ (m_i+n_i)\; \varsigma_\eta(l,m_i,n_i)\big] 
\\
\\
\qquad\qquad+O\big(l^d(m+n)^{1-\eta/d}\big)+O\big(l^d(m+n)^{1/2}\big).\ena\end{equation}
Because of Jensen inequality for concave function:
\beqq\begin{array}{l} \diy\sum\limits_{i=1}^M\bbE_{m_i,n_i}\big[O\big((m_i+n_i)^{1/2}\big)\big]=\sum\limits_{i=1}^M O\big(\bbE[m_i]+\bbE[n_i]\big)^{1/2}\\[10pt]
\qquad\qquad=\diy \sum\limits_{i=1}^M O(m a_i l^{-d}+n b_i l^{-d})^{1/2}=O\big(l^d(m+n)^{1/2}\big).\ena\eeqq
And similarly since $\eta < d$, we have
\beq\label{eq:44:0} \diy\sum\limits_{i=1}^M\bbE_{m_i,n_i}\big[O\big((m_i+n_i)^{1-\eta/d}\big)\big]=O\big(l^d(m+n)^{1-\eta/d}\big),\eeq
and for $d\geq 2$, one yields
\beq\label{eq:45:0} \begin{array}{l} \diy\sum\limits_{i=1}^M\bbE_{m_i,n_i}\big[O\big((m_i+n_i)^{1-1/d}\big)\big]=O\big(l^d(m+n)^{1-1/d}\big)=O\big(l^d(m+n)^{1/2}\big).\ena\eeq
Next, we state the following lemma (Lemma 1 from \cite{HCM} and \cite{HCB}) which will be used in the sequel:
\begin{lemma}\label{lem1:2}
Let $k(x)$ be a continuously differential function of $x\in\bbR$ which is convex and monotone decreasing over $x\geq 0$. Set $k'(x)=\diy\frac{\rd k(x)}{\rd x}$.  Then for any $x_0>0$ we have
\beq\label{eq2:2.1}
k(x_0)+\diy\frac{k(x_0)}{x_0}|x-x_0|\geq k(x)\geq k(x_0)-k'(x_0)|x-x_0|.\eeq
\end{lemma}

Next, continuing the proof of Theorem \ref{bound:Bias}, we attend to find an upper bound for 
\beq  \bbE_{m_i,n_i}\Big[\diy\frac{m_i n_i }{m_i f_0(\bx)+n_i f_1(\bx)}\Big]. \eeq
In order to pursue this aim, In Lemma \ref{lem1:2} consider $k(x)=\diy \frac{1}{x}$ and $x_0=\bbE_{m_i,n_i}\big[m_i f_0(\bx)+n_i f_1(\bx)\big]$, therefore as the function $k(x)$ is decreasing and convex, one can write
\beq\label{eq2:2.28} \begin{array}{ccl} 
\diy\frac{1}{m_i f_0(\bx)+n_i f_1(\bx)} 
\leq\diy\frac{1}{\bbE_{m_i,n_i}\big[m_i f_0(\bx)+n_i f_1(\bx)\big]}
+\diy\frac{\Big|m_i f_0(\bx)+n_i f_1(\bx)-\bbE_{m_i,n_i}\big[m_i f_0(\bx)+n_i f_1(\bx)\big]\Big|}{\bbE^2_{m_i,n_i}\big[m_i f_0(\bx)+n_i f_1(\bx)\big]}.
 \end{array}\eeq
Using the H\"{o}lder inequality implies the following inequality:
\beq\label{eq2:22} \begin{array}{ccl}\diy\bbE_{m_i,n_i}\Big[\frac{m_in_i}{m_i f_0(\bx)+n_i f_1(\bx)}\Big]\leq  \diy\frac{\bbE_{m_i,n_i}[m_i n_i]}{\bbE_{m_i,n_i}\big[m_i f_0(\bx)+n_i f_1(\bx)\big]}\\[20pt]
+\diy\frac{\Big(\bbE_{m_i,n_i}\big[m^2_i n^2_i\big]\Big)^{1/2}}{\bbE^2_{m_i,n_i}\big[m_i f_0(\bx)+n_i f_1(\bx)\big]}
\times\diy\bigg( \bbE_{m_i,n_i}\Big[m_if_0(\bx)+n_i f_1(\bx)-\bbE_{m_i,n_i}\big[m_i f_0(\bx)+n_i f_1(\bx)\big]\Big]^2\bigg)^{1/2}.
\end{array}\eeq
\def\bbV{\mathbb{V}}
As random variables $m_i$, $n_i$ are independent, and because of $\bbV[m_i]\leq m a_i l^{-d}$, $\bbV[n_i]\leq n b_i l^{-d}$,  we can claim that the RHS of (\ref{eq2:22}) becomes less than and equal to 
\beq\begin{array}{l} \diy\frac{ m n a_i b_i l^{-2d}}{m a_i l^{-d} f_0(\bx)+ n b_i l^{-d} f_1(\bx)}+\diy\frac{\Big(\alpha_i \beta_i\big(m a_i l^{-d} f^2_0(\bx)+n b_i l^{-d} f^2_1(\bx)\big) \Big)^{1/2}}{\big(ma_i f_0(\bx)+n b_i f_1(\bx)\big)^2},\ena\eeq
where
\beqq\begin{array}{ccl} \alpha_i=m a_i l^{d}\;(1-a_i l^{-d})+m^2 a_i^2, \\
\\
\beta_i=n b_i l^d\;(1-b_i l^{-d})+n^2 b_i^2 .\ena \eeqq
Going back to (\ref{Eq:2:2.19}), we have
\begin{equation}\begin{array}{l}
\bbE\Big[{\mathfrak{R}'}_{m,n}(\mathfrak{X}_m,\mathfrak{Y}_n)\Big]\leq
 \diy \sum\limits_{i=1}^M \;a_i b_i l^{-d}\int\diy\frac{2\;m n f_0(\bx) f_1(\bx)}{m a_i f_0(\bx)+n b_i f_1(\bx)}\;\rd\bx\\[15pt]
+ \diy\sum\limits_{i=1}^M2\int \diy \frac{f_0(\bx) f_1(\bx)\Big(\alpha_i \beta_i\big(m a_i l^{-d} f^2_0(\bx)+n b_i l^{-d} f^2_1(\bx)\big) \Big)^{1/2}}{\big(ma_i f_0(\bx)+n b_i f_1(\bx)\big)^2}\;\rd \bx\\[15pt]
\quad+\diy\sum\limits_{i=1}^M\bbE_{m_i,n_i}\big[(m_i+n_i)\; \varsigma_\eta(l,m_i,n_i) \big]+ O\big(l^d(m+n)^{1-\eta/d}\big)\\[15pt]
\qquad+\diy O\big(l^d(m+n)^{1/2}\big) +2c_1\;l^{d-1}(m+n)^{1/d}.
\end{array}\end{equation}
Finally, owing to $a_i\leq b_i$ and $\diy\sum\limits_{i=1}^M b_i l^{-d}=1$, when $\diy\frac{m}{m+n}\rightarrow p$, we have
\begin{equation}\label{Up.bound} \begin{array}{l}
\diy\frac{\bbE\Big[{\mathfrak{R}'}_{m,n}(\mathfrak{X}_m,\mathfrak{Y}_n)\Big]}{m+n}\leq \diy\int \frac{2\;p q f_0(\bx)f_1(\bx)}{p f_0(\bx)+q f_1(\bx)} \;\rd \bx \\
\\
\quad+ \diy\sum\limits_{i=1}^M2\int \diy \frac{f_0(\bx) f_1(\bx)\Big(\alpha_i \beta_i\big(m a_i l^{-d} f^2_0(\bx)+n b_i l^{-d} f^2_1(\bx)\big) \Big)^{1/2}}{\big(ma_i f_0(\bx)+n b_i f_1(\bx)\big)^2 (m+n)}\;\rd \bx\\
\\
\quad+\diy \frac{1}{m+n} \diy\sum\limits_{i=1}^M\bbE_{m_i,n_i}\big[ (m_i+n_i)\;\varsigma_\eta(l,m_i,n_i)\big]+ O\big(l^d(m+n)^{-\eta/d}\big)\\
\\
\qquad+O\big(l^d(m+n)^{-1/2}\big)+2c_1\; l^{d-1}\;(m+n)^{(1/d)-1}.
\end{array}\end{equation}

Passing to the Definition \ref{def:dual},  ${{\rm MST}}^*$, and Lemma \ref{lem:2.5}, similar discussion as above, consider the Poisson processes samples and  the FR statistic under the union of samples, denoted by ${\mathfrak{R}'}^*_{m,n}$, and superadditivity of dual $\mathfrak{R}^*_{m,n}$,  we have 
\beq \begin{array}{l}\bbE\Big[{\mathfrak{R}'}^*_{m,n}(\mathfrak{X}_m,\mathfrak{Y}_n)\Big]\geq \diy\sum\limits_{i=1}^M \bbE\Big[{\mathfrak{R}'}^*_{m_i,n_i}\big((\mathfrak{X}_m,\mathfrak{Y}_n)\cap Q_i\big)\Big]-\;c_2\;l^d\\[15pt]
\quad=\diy \sum\limits_{i=1}^M \bbE_{m_i,n_i}\Big[\bbE\big[{\mathfrak{R}'}^*_{m_i,n_i}\big((\mathfrak{X}_m,\mathfrak{Y}_n)\cap Q_i\big)|m_i,n_i\big]\Big]- c_2\;l^d\\[15pt]
\quad \geq\diy  \sum\limits_{i=1}^M \bbE_{m_i,n_i}\Big[\bbE\big[{\mathfrak{R}'}_{m_i,n_i}\big((\mathfrak{X}_m,\mathfrak{Y}_n)\cap Q_i\big)|m_i,n_i\big]\Big]- c_2\;l^d, \end{array}\eeq
the last line is derived from Lemma \ref{lem:2.5}, (ii), inequality (\ref{dual:eq:1.1}). Owing to the Lemma \ref{lem:2.4}, (\ref{eq:44:0}) and (\ref{eq:45:0}), one obtains
\begin{equation}\label{Eq:2.19} \begin{array}{l}\bbE\Big[{\mathfrak{R}'}^*_{m,n}(\mathfrak{X}_m,\mathfrak{Y}_n)\Big]\geq \diy \sum\limits_{i=1}^M \int \bbE_{m_i,n_i}\Big[\diy\frac{2m_i n_i f_0(\bx) f_1(\bx)}{m_i f_0(\bx)+n_i f_1(\bx)}\Big]\;\rd\bx\\[15pt]
 -\diy\sum\limits_{i=1}^M\bbE_{m_i,n_i}\big[ (m_i+n_i)\;\varsigma_\eta(l,m_i,n_i)\big]-O\big(l^d(m+n)^{1-\eta/d}\big)
- O\big(l^d(m+n)^{1/2}\big) -c_2\;l^d.\end{array}\end{equation}

Furthermore,  by using the Jenson's inequality we get
\beqq\begin{array}{ccl} \diy\bbE_{m_i,n_i}\Big[\frac{m_i n_i}{m_i f_0(\bx)+n_i f_1(\bx)}\Big]\geq \diy \frac{\bbE[m_i]\bbE[n_i]}{\bbE[m_i] f_0(\bx)+\bbE[n_i] f_1(\bx)}
=\diy\frac{l^{-d}\big(m a_i n b_i\big)}{m a_i f_0(\bx)+n b_i f_1(\bx)}.\ena\eeqq
Therefore since $a_i\leq b_i$, we can write
\begin{equation}\label{eq2:2.37} \begin{array}{ccl} \diy\bbE_{m_i,n_i}\Big[\frac{m_i n_i}{m_i f_0(\bx)+n_i f_1(\bx)}\Big]\geq \diy \frac{l^{-d} m n\; a_{i}b_{i}}{b_i\big(m f_0(\bx)+n f_1(\bx)\big)}
= \diy \frac{l^{-d} m n\; a_{i}}{\big(m f_0(\bx)+n f_1(\bx)\big)}. \ena\end{equation}
Consequently, the RHS of  (\ref{Eq:2.19}) becomes greater than or equal to  
\begin{equation} \begin{array}{cl}\diy\sum\limits_{i=1}^M a_i\;l^{-d}\int \diy\frac{2mn f_0(\bx)f_1(\bx)}{m f_0(\bx)+n f_1(\bx)}\;\rd\bx\\[15pt]
\quad-\diy\sum\limits_{i=1}^M\bbE_{m_i,n_i}\big[(m_i+n_i)\; \varsigma_\eta(l, m_i,n_i) \big]-O\big(l^d(m+n)^{1-\eta/d}\big)
 - O\big(l^d(m+n)^{1/2}\big) -c_2\;l^d.\end{array}\end{equation}
Finally, since $\diy\sum\limits_{i=1}^M a_i l^{-d}=1$ and  $\diy\frac{m}{m+n}\rightarrow p$, we have
\begin{equation} \begin{array}{l} \diy\frac{\bbE\Big[{\mathfrak{R}'}^*_{m,n}(\mathfrak{X}_m,\mathfrak{Y}_n)\Big]}{m+n}\geq \diy\int \diy\frac{2 p q f_0(\bx)f_1(\bx)}{p f_0(\bx)+q f_1(\bx)}\;\rd\bx - (m+n)^{-1}\diy\sum\limits_{i=1}^M\bbE_{m_i,n_i}\big[(m_i+n_i)\;\varsigma(l,m_i,n_i)\big]\\[20pt]
\hspace{90pt} -O\big(l^d(m+n)^{-\eta/d}\big)
- O\big(l^d(m+n)^{-1/2}\big) -c_2\;l^d (m+n)^{-1}.\ena \end{equation}
By definition of the dual $\mathfrak{R}^*_{m,n}$ and (i) in Lemma \ref{lem:2.5}, 
\beq \diy\frac{\bbE\Big[{\mathfrak{R}'_{m,n}}(\mathfrak{X}_m,\mathfrak{Y}_n)\Big]}{m+n}+\frac{c_d\;2^d}{m+n}\geq \frac{\bbE\Big[{\mathfrak{R}'}^*_{m,n}(\mathfrak{X}_m,\mathfrak{Y}_n)\Big]}{m+n},\eeq
we can imply
\beq\label{Low2:bound} \begin{array}{l} \diy\frac{\bbE\Big[{\mathfrak{R}'}_{m,n}(\mathfrak{X}_m,\mathfrak{Y}_n)\Big]}{m+n}\geq \diy\int \diy\frac{2 p q f_0(\bx)f_1(\bx)}{p f_0(\bx)+q f_1(\bx)}\;\rd\bx  - (m+n)^{-1}\diy\sum\limits_{i=1}^M\bbE_{m_i,n_i}\big[(m_i+n_i)\; \varsigma_\eta(l,m_i,n_i)\big]\\[20pt]
\qquad-O\big(l^d(m+n)^{-\eta/d}\big) - O\big(l^d(m+n)^{-1/2}\big)
-c_2\;l^d (m+n)^{-1}-c_d\;2^d\;(m+n)^{-1}.\ena \eeq
\vspace{5pt}

The combination of two lower and upper bounds (\ref{Low2:bound}) and (\ref{Up.bound}), yields the following result
\begin{equation}\label{final:up.bound}\begin{array}{cl}
\Big|\diy\frac{\bbE\big[\mathfrak{R}'_{m,n}(\mathfrak{X}_m,\mathfrak{Y}_n)\big]}{m+n}- \diy\int \frac{2pq f_0(\bx)f_1(\bx)}{pf_0(\bx)+q f_1(\bx)}\;\rd\bx\Big|\leq \diy O\big(l^d(m+n)^{-\eta/d}\big)
 + O\big(l^d(m+n)^{-1/2}\big)+2\;c_1\;l^{d-1}\; (m+n)^{(1/d)-1}\\[10pt]
+c_d\;2^d\;(m+n)^{-1}+\diy c_2\; (m+n)^{-1}\; l^{d}
+\diy \frac{1}{m+n} \diy\sum\limits_{i=1}^M\bbE_{m_i,n_i}\big[(m_i+n_i)\; \varsigma_\eta(l, m_i,n_i)\big]\\[11pt]
+\diy \sum\limits_{i=1}^M2\int \diy \frac{f_0(\bx) f_1(\bx)\Big(\alpha_i \beta_i\big(m a_i l^{-d} f^2_0(\bx)+n b_i l^{-d} f^2_1(\bx)\big) \Big)^{1/2}}{\big(ma_i f_0(\bx)+n b_i f_1(\bx)\big)^2 (m+n)}\;\rd \bx.
\end{array}\end{equation}

Recall $\varsigma_\eta(l,m_i,n_i)$, then we obtain
\beq \begin{array}{l} \diy\sum\limits_{i=1}^M \bbE_{m_i,n_i}\Big[(m_i+n_i)\; \varsigma_\eta(l, m_i,n_i)\Big]
=\diy\sum\limits_{i=1}^M O(l) \int \bbE\Big[\diy\frac{2m_i n_i f_0(\bx) f_1(\bx)}{(m_i+n_i)(m_i f_0(\bx)+n_i f_1(\bx))}\Big]\;\td\bx\\[15pt]
\quad \quad- \diy 2\;l^{d}\sum\limits_{i=1}^M \int \bbE\Big[\frac{2 m_i n_i f_0(\bx) f_1(\bx)}{(m_i+n_i) (m_i f_0(\bx) + n_i f_1(\bx))}\Big]\;\rd\bx
+O(l^{-\eta})\sum\limits_{i=1}^M \bbE_{m_i,n_i}[m_i+n_i]. \ena \eeq
In addition, we have
\beq\begin{array}{l}\diy \bbE_{m_i,n_i}\Big[\frac{2 m_i n_i f_0(\bx) f_1(\bx)}{(m_i+n_i) (m_i f_0(\bx) + n_i f_1(\bx))}\Big]
\geq \diy\frac{1}{m+n} \bbE_{m_i,n_i}\Big[\frac{2 m_i n_i f_0(\bx) f_1(\bx)}{ (m_i f_0(\bx) + n_i f_1(\bx))}\Big].\ena\eeq
This implies
\beq\begin{array}{l} \diy\sum\limits_{i=1}^M \int \bbE\Big[\frac{2 m_i n_i f_0(\bx) f_1(\bx)}{(m_i+n_i) (m_i f_0(\bx) + n_i f_1(\bx))}\Big]\;\rd\bx
\geq \diy\int \frac{2pq f_0(\bx) f_1(\bx)}{p f_0(\bx)+q f_1(\bx)}\;\rd \bx. \ena\eeq
Note that the above inequality is derived from (\ref{eq2:2.37}) and $\diy\frac{m}{m+n}\rightarrow p$. Further, 
\begin{equation} \begin{array}{l}\diy \frac{1}{m+n}\sum\limits_{i=1}^M O(l) \int  \bbE_{m_i,n_i}\Big[\frac{2 m_i n_i f_0(\bx) f_1(\bx)}{(m_i+n_i) (m_i f_0(\bx) + n_i f_1(\bx))}\Big]\;\rd\bx\\[15pt]
\qquad\quad \leq 
\diy\sum\limits_{i=1}^M O(l) \int  \bbE_{m_i,n_i}\Big[\frac{2 m_i n_i f_0(\bx) f_1(\bx)}{(m_i+n_i)^2 (m_i f_0(\bx) + n_i f_1(\bx))}\Big]\;\rd\bx\\[15pt]
\qquad\qquad\leq \diy\sum\limits_{i=1}^M O(l) \int  \bbE_{m_i,n_i}\Big[\frac{2 f_0(\bx) f_1(\bx)}{ (m_i f_0(\bx) + n_i f_1(\bx))}\Big]\;\rd\bx.\ena\end{equation}
The last line holds because of $m_i n_i\leq (m_i+n_i)^2$. Going back to (\ref{eq2:2.28}), we can give an upper bound for the RHS of above inequality as
\beqq \begin{array}{cl}\diy\bbE_{m_i,n_i}\Big[\big(m_i f_0(\bx)+n_i f_1(\bx)\big)^{-1}\Big]\\
\\
\quad\leq \diy\big(m a_i l^{-d} f_0(\bx)+n b_i l^{-d} f_1(\bx)\big)^{-1}
+\Big(\bbE_{m_i,n_i}\Big|m_i f_0(\bx)+n_i f_1(\bx)-
 \Big(\bbE[m_i] f_0(\bx)+\bbE[n_i] f_1(\bx)\Big|\Big)\Big/\big(m a_i l^{-d} f_0(\bx)+n b_i l^{-d}f_1(\bx)\big)^2.\ena\eeqq
Note that we have assumed $a_i\leq b_i$ and by using H\"{o}lder inequality we write
\begin{equation}\begin{array}{l} \diy\bbE_{m_i,n_i}\Big[\big(m_i f_0(\bx)+n_i f_1(\bx)\big)^{-1}\Big]
\leq l^d (a_i)^{-1} \big(m f_0(\bx)+n f_1(\bx)\big)^{-1}
 +\Big(f_0(\bx) \sqrt{\bbV(m_i)}+f_1(\bx)\sqrt{\bbV(n_i)}\Big)\Big/\big(a_i^2 l^{-d}(m f_0(\bx)+nf_1(\bx))^2\big)\\
 \\
\quad\qquad \leq  l^d (a_i)^{-1} \big(m f_0(\bx)+n f_1(\bx)\big)^{-1}
+ l^{-d/2} \sqrt{b_i}\Big(f_0(\bx)\sqrt{m}+f_1(\bx)\sqrt{n}\Big)\Big/\big(a_i^2 l^{-d}(m f_0(\bx)+n f_1(\bx))^2\big). \ena\end{equation}
As result, we have
\begin{equation}\label{setcounter3} \begin{array}{l}
 \diy\sum\limits_{i=1}^M O(l) \int  \bbE_{m_i,n_i}\Big[\frac{2 f_0(\bx) f_1(\bx)}{ (m_i f_0(\bx) + n_i f_1(\bx))}\Big]\;\rd\bx \\[15pt]
\quad\qquad\leq \diy \sum\limits_{i=1}^M O(l) \int  l^d (a_i)^{-1} \diy\frac{2 f_0(\bx) f_1(\bx)}{m f_0(\bx)+n f_1(\bx)}\;\rd\bx\\[15pt]
\quad\qquad\qquad+ \diy \sum\limits_{i=1}^M O(l) \int l^{-d/2}\sqrt{b_i}\; \frac{2 f_0(\bx) f_1(\bx) \big(f_0(\bx)\sqrt{m}+f_1(\bx)\sqrt{n}\big)}{a_i^2 l^{-d} \big(m f_0(\bx)+n f_1(\bx)\big)^2}\;\rd\bx . \end{array}\end{equation}
As consequence, owing to (\ref{final:up.bound}), for $0<\eta\leq 1$, $d\geq2$ which implies  $\eta\leq d-1$, we can derive (\ref{final:up.bound232}). So, the proof can be concluded by giving the summarized bound in (\ref{bound:Bias}).
\end{IEEEproof}
\bigskip
{\it Lemma \ref{varianceD}:}
For $h=1,2,\dots$, let $\delta^h_{m,n}$ be the function $c\;h^{d-1}(m+n)^{1/d}$. Then for $\epsilon>0$, we have 
\begin{equation} \label{variancD}P\Big(\mathfrak{R}_{m,n}(\mathfrak{X}_m,\mathfrak{Y}_n)\leq \sum\limits_{i=1}^{h^d}\mathfrak{R}_{m_i,n_i}(\mathfrak{X}_{m_i},\mathfrak{Y}_{n_i})+2\ep\Big)\geq \diy\frac{\ep-\delta^h_{m,n}}{\ep}.\end{equation}
Note that in case $\ep\leq \delta^h_{m,n}$ the above claimed inequality is trivial. 
\begin{IEEEproof}
Consider the cardinality of the set of all edges of ${\rm MST}\big(\bigcup\limits_{i=1}^{h^d} Q_i\big)$ which intersect two different subcubes $Q_i$ and $Q_j$, $|D|$. Using the Markov inequality we can write
\beqq P\Big(|D|\geq \ep\Big)\leq \diy \frac{\bbE(|D|)}{\ep}, \eeqq
where $\epsilon>0$. Since $\bbE|D|\leq c\;h^{d-1}(m+n)^{1/d}:=\delta^h_{m,n}$, therefore for $\ep> \delta^h_{m,n}$ and $h=1,2,\dots$:
\beqq P\Big(|D|\geq \ep\Big)\leq \diy \frac{\delta^h_{m,n}}{\ep}. \eeqq
And if $Q_i$, $i=1,\dots h^d$ is a partition of $[0,1]^d$ into congruent subcubes of edge length $1/h$, then
\begin{equation} \label{lem9:eq2.1} \begin{array}{l}P\bigg(\diy\sum\limits_{i=1}^{h^d} \mathfrak{R}_{m_i,n_i}(\mathfrak{X}_m,\mathfrak{Y}_n\cap Q_i)+2|D|\geq\sum\limits_{i=1}^{h^d} \mathfrak{R}_{m_i,n_i}(\mathfrak{X}_m,\mathfrak{Y}_n\cap Q_i)+2\ep\bigg)
\leq \diy \frac{\delta^h_{m,n}}{\ep}. \ena\end{equation}
This implies
\begin{equation}\label{lem9:eq2.2} \begin{array}{l}P\bigg(\diy\sum\limits_{i=1}^{h^d} \mathfrak{R}_{m_i,n_i}(\mathfrak{X}_m,\mathfrak{Y}_n\cap Q_i)+2|D|\leq\diy\sum\limits_{i=1}^{h^d} \mathfrak{R}_{m_i,n_i}(\mathfrak{X}_m,\mathfrak{Y}_n\cap Q_i)+2\ep\bigg)\geq 1-\diy\frac{\delta^h_{m,n}}{\ep}. \ena\end{equation}
By subadditivity (\ref{subadd:R}), we can write
\beqq \diy \mathfrak{R}_{m,n}(\mathfrak{X}_m,\mathfrak{Y}_n)\leq\diy\sum\limits_{i=1}^{h^d} \mathfrak{R}_{m_i,n_i}(\mathfrak{X}_m,\mathfrak{Y}_n\cap Q_i)+2|D|,\eeqq
and this along with (\ref{lem9:eq2.2}) establishes (\ref{variancD}).
\end{IEEEproof}
\bigskip
\def\bbB{\mathbb{B}}
{\it Lemma \ref{growth bound}:}
{\rm (Growth bounds for $\mathfrak{R}_{m,n}$)} Let $\mathfrak{R}_{m,n}$ be the FR statistic. Then for given non-negative $\ep$,  such that $\ep\geq h^2\; \delta^h_{m,n}$, with at least probability $g(\ep):=1-\diy\frac{h\;\delta^h_{m,n}}{\ep}$,  $h=2,3,\dots$, we have
\beq\label{Growth bound} \mathfrak{R}_{m,n}(\mathfrak{X}_m,\mathfrak{Y}_n)\leq {c}''_{\ep,h}\;\big(\#\mathfrak{X}_m \;\#\mathfrak{Y}_n\big)^{1-1/d}.\eeq
Here  $ {c}''_{\ep,h}=O\left(\diy\frac{\ep}{h^{d-1}-1}\right)$ depending only on $\ep$, $h$. Note that for $\ep< h^2\; \delta^h_{m,n}$, the claim is trivial. 
\begin{IEEEproof}
Without loss of generality consider the unit cube $[0,1]^d$. For given $h$, if $Q_i$, $i=1,\dots h^d$ is a partition of $[0,1]^d$ into congruent subcubes of edge length $1/h$ then by Lemma \ref{varianceD}, we have
\begin{equation} \label{lem10:eq1} P\Big(\mathfrak{R}_{m,n}(\mathfrak{X}_m,\mathfrak{Y}_n)\leq \sum\limits_{i=1}^{h^d}\mathfrak{R}_{m_i,n_i}(\mathfrak{X}_{m_i},\mathfrak{Y}_{n_i})+2\ep\Big)\geq \diy\frac{\ep-\delta^h_{m,n}}{\ep}.\end{equation}
We apply the induction methodology on $\# \mathfrak{X}_m$ and $\#\mathfrak{Y}_n$. Set $c:=\sup\limits_{\bx,\by\in[0,1]^d} \mathfrak{R}_{m,n}(\{\bx,\by\})$ which is finite according to assumption. Moreover, set $c_2:=\diy\frac{2\ep}{h^{d-1}-1}$ and $c_1:=c+d\;h^{d-1}c_2$. Therefore, it is sufficient to show that for all $(\mathfrak{X}_m,\mathfrak{Y}_n)\in [0,1]^d$ with at least probability $g(\ep)$
\beq \mathfrak{R}_{m,n}(\mathfrak{X}_m,\mathfrak{Y}_n)\leq c_1 \big(\# \mathfrak{X}_m\;\#\mathfrak{Y}_n\big)^{(d-1)/d}. \eeq
Alternatively as for the induction hypothesis we assume the stronger bound 
\beq \mathfrak{R}_{m,n}(\mathfrak{X}_m,\mathfrak{Y}_n)\leq c_1 \big(\# \mathfrak{X}_m\;\#\mathfrak{Y}_n\big)^{(d-1)/d}-c_2, \eeq
holds whenever $\# \mathfrak{X}_m<m $ and $\# \mathfrak{Y}_n<n$ with at least probability $g(\ep)$. Note that $d\geq 2$, $\ep>0$ and $c_1$, $c_2$ both depend on $\epsilon$, $h$. Hence
\beqq c_1-c_2= c+c_2\big(d\;h^{d-1}-1\big)\geq c+c_2\big(h^{d-1}-1\big)= c+2\ep\geq c.\eeqq
which implies $P(\mathfrak{R}_{m,n}\leq c_1-c_2)\geq P(\mathfrak{R}_{m,n}\leq c)$. Also we know that $P(\mathfrak{R}_{m,n}\leq c)=1\geq g(\ep)$, therefore, the induction hypothesis holds particularly $\#\mathfrak{X}_m=1$ and $\#\mathfrak{Y}_n=1$. Now consider the partition $Q_i$ of $[0,1]^d$, therefore for all $1\leq i\leq h^d$ we have $m_i:= \#(\mathfrak{X}_m\cap Q_i)<m$ and $n_i:= \#(\mathfrak{Y}_n\cap Q_i)<n$ and thus by induction hypothesis one yields with at least probability $g(\ep)$
\beq\label{lem10:eq2:59} \mathfrak{R}_{m_i,n_i}(\mathfrak{X}_m,\mathfrak{Y}_n\cap Q_i)\leq c_1 \;(m_i\;n_i)^{1-1/d}-c_2.\eeq
Set $\bbB$ the event $\big\{ {\rm all}\; i\; :\;\mathfrak{R}_{m_i,n_i}\leq c_1\;(m_i\;n_i)^{1-1/d}-c_2\big\}$ and $\bbB_i$ stands with the event $\big\{ \mathfrak{R}_{m_i,n_i}\leq c_1\;(m_i\;n_i)^{1-1/d}-c_2\big\}$. From (\ref{lem10:eq1}) and  since $Q_i$'s are partitions which implies 
\beqq\begin{array}{ccl} P(\bbB)=\big(P(\bbB_i)\big)^{h^d}\leq P(\bbB_i),\;\;
P(\bbB^{\rm c})=P(\bigcup\limits_{i=1}^{l^d}\bbB_i^{\rm c})\leq \diy\sum\limits_{i=1}^{h^d} P(\bbB_i^{\rm c})\leq h^d\big(1-g(\ep)\big),\;\;
\hbox{and}\;\;\;  P(\bbB)=\prod\limits_{i=1}^{h^d}P(\bbB_i)\geq \big(g(\ep)\big)^{h^d}.\ena\eeqq
So, we obtain
\beqq \begin{array}{ccl}  \diy\frac{\ep-\delta^h_{m,n}}{\ep}\leq P\Big(\mathfrak{R}_{m,n}\leq \sum\limits_{i=1}^{h^d}\mathfrak{R}_{m_i,n_i}(\mathfrak{X}_{m_i},\mathfrak{Y}_{n_i})+2\ep\big|\bbB\Big)P(\bbB)
+ P\Big(\mathfrak{R}_{m,n}\leq \sum\limits_{i=1}^{h^d}\mathfrak{R}_{m_i,n_i}(\mathfrak{X}_{m_i},\mathfrak{Y}_{n_i})+2\ep\big|\bbB^{\rm c}\Big)P(\bbB^{\rm c})\\[15pt]
\leq  P\Big(\mathfrak{R}_{m,n}\leq \diy\sum\limits_{i=1}^{l^d}\mathfrak{R}_{m_i,n_i}(\mathfrak{X}_{m_i},\mathfrak{Y}_{n_i})+2\ep\big|\bbB\Big)P(\bbB)+P(\bbB^{\rm c}).
\end{array}\eeqq
Equivalently 
\beqq  \begin{array}{l}P\Big(\mathfrak{R}_{m,n}\leq \diy\sum\limits_{i=1}^{h^d}\mathfrak{R}_{m_i,n_i}(\mathfrak{X}_{m_i},\mathfrak{Y}_{n_i})+2\ep\big|\bbB\Big)
\geq \big(1-\diy\frac{\delta^h_{m,n}}{\ep}-1+P(\bbB)\big)\big/ P(\bbB)=1-\diy\frac{\delta^h_{m,n}}{\ep\; P(\bbB)}. \ena\eeqq
In fact in this stage we want to show that 
\beqq \diy 1-\diy\frac{\delta^h_{m,n}}{\ep\;P(\bbB)}\geq g(\ep)\;\;\;\hbox{or}\;\;\; P(\bbB)\geq \diy\frac{\delta^h_{m,n}}{\ep\;(1-g(\ep))}.\eeqq
Since $P(\bbB)\geq \big(g(\ep)\big)^{h^d}$ therefore it is sufficient to derive that $\big(g(\ep)\big)^{h^d}\geq  \diy\frac{\delta^h_{m,n}}{\ep\;(1-g(\ep))}$. Indeed for given $g(\ep)=\Big(\diy\frac{\ep-h\;\delta^h_{m,n}}{\ep}\Big)$ we have $g(\ep)\leq \diy\frac{\ep-\delta^h_{m,n}}{\ep} $ hence $\diy\frac{\delta^h_{m,n}}{\ep\;(1-g(\ep))}=\diy\frac{1}{h}\leq 1$. Further, we know $\diy\frac{1}{h}\leq 1-\diy\frac{1}{h^{(1/h^{d})}}$ and since $\ep\geq h^2\;\delta^h_{m,n}$ this implies $\diy\frac{h\;\delta^h_{m,n}}{\ep}\leq \diy\frac{1}{h}$ and consequently 
$$ \diy\frac{h\;\delta^h_{m,n}}{\ep}\leq 1-\diy\frac{1}{h^{h^{-d}}}.$$
Or
$$ g(\ep)^{h^d}=\diy\Big(\frac{\ep-h\;\delta^h_{m,n}}{\ep}\Big)^{h^d}\geq \frac{1}{h}=\diy\frac{\delta^h_{m,n}}{\ep\;(1-g(\ep))}.$$
This implies the fact that for $\ep\geq h^2\delta^h_{m,n}$
\beqq \begin{array}{ccl} P\Big(\mathfrak{R}_{m,n}\leq \sum\limits_{i=1}^{h^d}\big(c_1(m_i n_i)^{1-1/d}-c_2\big)+2\ep\Big)\geq g(\ep),\;\;\;
\hbox{where}\;\;\; g(\ep)=\diy\frac{\ep-h\;\delta^h_{m,n}}{\ep}.\ena\eeqq
Now let $\gamma:=\#\{i: m_i, n_i>0\}$ and using H\"{o}lder inequality gives 
\begin{equation}\label{eqB:lem9.1} P\Big(\diy \mathfrak{R}_{m,n}(\mathfrak{X}_m,\mathfrak{Y}_n)\leq
\diy c_1 \gamma^{1/d}(m\;n)^{1-1/d}-\gamma c_2+c_2 \;(h^{d-1}-1)\Big)\geq \diy g(\ep). \end{equation}
Next, we just need to show that $ c_1 \gamma^{1/d}(m\;n)^{1-1/d}-\gamma c_2+c_2 \;(h^{d-1}-1)$ in (\ref{eqB:lem9.1}) is less than or equal to $c_1 (m\;n)^{1-1/d}-c_2$, which is equivalent to show
\beqq c_2\big(h^{d-1}-\gamma\big)\leq c_1 (m\;n)^{1-1/d}(1-\gamma^{1/d}). \eeqq
We know that $m,n\geq 1$ and $c_1\geq d\;h^{d-1} c_2$, so it is sufficient to get
\beq\label{EqB:gamma1} c_2\big(h^{d-1}-\gamma\big)\leq  d\;h^{d-1} c_2 (1-\gamma^{1/d}),\eeq
choose $t$ as $\gamma=t\;h^d$, then $0<t\leq 1$, so (\ref{EqB:gamma1}) becomes
\beq (h^{-1}-t)\geq d\;h^{-1}\big(1-h\; t^{1/d}).\eeq
Note that the function $ d\;h^{-1}\big(1-h\; t^{1/d})+t -h^{-1}$ has a minimum at $t=1$ which implies  (\ref{EqB:gamma1}) and subsequently (\ref{Growth bound}). Hence the proof is completed.
\end{IEEEproof}
\bigskip
{\it Lemma \ref{smoothness}:}
{\rm (Smoothness for $\mathfrak{R}_{m,n}$)} Given observations of 
$$\mathfrak{X}_m:=(\mathfrak{X}_{m'},\mathfrak{X}_{m''})=\{\BX_1,\dots,\BX_{m'},\BX_{m'+1},\dots,\BX_m\},$$
 such that $m'+m''=m$ and  $\mathfrak{Y}_n:=(\mathfrak{Y}_{n'},\mathfrak{Y}_{n''})=\{\BY_1,\dots,\BY_{n'},\BY_{n'+1},\dots,\BY_n\}$, where $n'+n''=n$, denote $\mathfrak{R}_{m,n}(\mathfrak{X}_m,\mathfrak{Y}_n)$ as before, the number of edges of ${\rm MST}(\mathfrak{X}_m,\mathfrak{Y}_n)$ which connect a point of $\mathfrak{X}_m$ to a point of $\mathfrak{Y}_n$. 
Then for  integer $h\geq 2$,  for all $(\mathfrak{X}_n, \mathfrak{Y}_m)\in[0,1]^d$, $\ep\geq h^2\;\delta^h_{m,n}$ where $\delta^h_{m,n}=O\big(h^{d-1}(m+n)^{1/d}\big)$, we have
\begin{equation} \begin{array}{l}\label{Smoothness}P\bigg(\Big|\mathfrak{R}_{m,n}(\mathfrak{X}_m,\mathfrak{Y}_n)-\mathfrak{R}_{m',n'}(\mathfrak{X}_{m'},\mathfrak{Y}_{n'})\Big|
\leq \tilde{c}_{\ep,h} \;\big(\#\mathfrak{X}_{m''}\;\#\mathfrak{Y}_{n''}\Big)^{1-1/d}\bigg)\geq \diy 1-\diy\frac{2h\;\delta^{h}_{m,n}}{\ep}. \ena\end{equation}
 where $\tilde{c}_{\ep,h}=O\left(\diy\frac{\ep}{h^{d-1}-1}\right)$. For the case $\ep< h^2\;\delta^h_{m,n}$ this holds trivially.  
\begin{IEEEproof}
We begin with removing the edges which contain a vertex in $\mathfrak{X}_{m''}$ and $\mathfrak{Y}_{n''}$ in minimal spanning tree on $(\mathfrak{X}_{m},\mathfrak{Y}_n)$. Now since each vertex has bounded degree, say $c_d$, we can generate a subgraph in which has at most $c_d (\#\mathfrak{X}_{m''}+\#\mathfrak{Y}_{n''})$ components. Next choose one vertex from each component and form the minimal spanning tree on these vertices, assuming all of them can be considered in FR test statistic, we can write
\beq\begin{array}{l} \mathfrak{R}_{m,n}(\mathfrak{X}_m,\mathfrak{Y}_n) 
\leq\diy \mathfrak{R}_{m',n'}(\mathfrak{X}_{m'},\mathfrak{Y}_{n'})+c''_{\ep,h}\big(c^2_d\;\#\mathfrak{X}_{m''}\;\#\mathfrak{Y}_{n''}\big)^{1-1/d},\\
\\
\hbox{or equivalently}\\
\qquad\qquad\qquad \leq \diy \mathfrak{R}_{m',n'}(\mathfrak{X}_{m'},\mathfrak{Y}_{n'})+c^h_{\ep1}\big(\;\#\mathfrak{X}_{m''}\;\#\mathfrak{Y}_{n''}\big)^{1-1/d},\ena\eeq
with probability at least $g(\ep)$, where $g(\ep)$ is as in Lemma \ref{growth bound}. Note that this expression is obtained from  Lemma \ref{growth bound}. In this stage, it remains to show that with at least probability $g(\ep)$ 
\beq\label{Eq1:smoothness} \mathfrak{R}_{m,n}(\mathfrak{X}_m,\mathfrak{Y}_n)\geq \mathfrak{R}_{m',n'}(\mathfrak{X}_{m'},\mathfrak{Y}_{n'}) - \tilde{c}_{\ep,h}\;\big(\# \mathfrak{X}_{m''}\;\#\mathfrak{Y}_{n''}\big)^{1-1/d}.\eeq
Which again by using the method before, with at least probability $g(\ep)$, one derives
\beqq \begin{array}{l} \mathfrak{R}_{m',n'}(\mathfrak{X}_{m'},\mathfrak{Y}_{n'})
\leq \diy \mathfrak{R}_{m,n}(\mathfrak{X}_m,\mathfrak{Y}_n)+\hat{c}_{\ep,h}\;\big(c_d^2\;(\# \mathfrak{X}_{m''}\;\# \mathfrak{Y}_{n''})\big)^{1-1/d},\\
\\
\hbox{or equivalently} \\
\qquad\qquad\quad\leq  \diy \mathfrak{R}_{m,n}(\mathfrak{X}_m,\mathfrak{Y}_n)+c^h_{\ep 2}\;\big(\# \mathfrak{X}_{m''}\;\# \mathfrak{Y}_{n''}\big)^{1-1/d}.\ena\eeqq
Letting $\tilde{c}_{\ep,h}=\max\{c^h_{\ep1},c^h_{\ep 2}\}$  implies (\ref{Eq1:smoothness}). So
\begin{equation} \label{Smoothness}\begin{array}{l} P\bigg(\Big|\mathfrak{R}_{m,n}(\mathfrak{X}_m,\mathfrak{Y}_n)-\mathfrak{R}_{m',n'}(\mathfrak{X}_{m'},\mathfrak{Y}_{n'})\big|\geq \tilde{c}_{\ep,h} \;\big(\#\mathfrak{X}_{m''}\;\#\mathfrak{Y}_{n''}\Big)^{1-1/d}\bigg)\leq 2-2\;g(\ep), \ena\end{equation}
Hence, the smoothness is given with at least probability $2\;g(\ep)-1$ as in the statement of Lemma \ref{smoothness}.
\end{IEEEproof}
\bigskip
{\it Lemma \ref{Eq:isoperimetry}:}
{\rm (Semi-Isoperimetry)} Let $\mu$ be a measure on $[0,1]^d$; $\mu^n$ denotes the product measure on space $([0,1]^d)^n$. And let $M_{\rm e}$ denotes a median of $\mathfrak{R}_{m,n}$. Set
\begin{equation}\begin{array}{l}\bbA:=
\Big\{ \mathfrak{X}_m\in \big([0,1]^d\big)^m, \mathfrak{Y}_n\in \big([0,1]^d\big)^n; \mathfrak{R}_{m,n}(\mathfrak{X}_m,\mathfrak{Y}_n)\leq M_{e}\Big\}.\ena\end{equation}
Then
\begin{equation}\label{lem13:0.0}\begin{array}{l} \diy \mu^{m+n}\bigg(\Big\{\bx'\in([0,1]^d)^m, \by'\in ([0,1]^n): \phi_{\bbA}(\bx')\;\phi_{\bbA}(\by') \geq t\Big\}\bigg)\leq 4\exp\Big(\diy\frac{-t}{8(m+n)}\Big).\ena \end{equation}
\begin{IEEEproof}
Let $\phi_{\bbA}(\bz')=\min\{H(\bz,\bz'), \bz\in \bbA\}$. Using Proposition 6.5 in \cite{Yu}, isoperimetric inequality, we have 
\begin{equation}\label{Lem13:1.1} \mu^{m+n}\bigg(\Big\{\bz'\in([0,1]^d)^{m+n} : \phi_{\bbA}(\bz')\geq t\Big\}\bigg)\leq 4\exp\Big(\diy\frac{-t^2}{8(m+n)}\Big).\end{equation}
Furthermore, we know that 
$$\Big(\phi_{\bbA}(\bx')+\phi_{\bbA}(\by')\Big)^2\geq \phi_{\bbA}(\bx')\;\phi_{\bbA}(\by'),$$
hence
\begin{equation}\label{lem13:1.2}\begin{array}{l}\diy\mu^{m+n}\bigg(\Big\{(\bx'\in([0,1]^d)^m, \by'\in ([0,1]^n): \phi_{\bbA}(\bx')\phi_{\bbA}(\by')\geq t\Big\}\bigg)\\[15pt]
\qquad\leq \diy \mu^{m+n}\bigg(\Big\{(\bx'\in([0,1]^d)^m, \by'\in ([0,1]^n):\big(\phi_{\bbA}(\bx')+\phi_{\bbA}(\by')\big)^2\geq t\Big\}\bigg)\\[15pt]
\qquad\quad=\diy \mu^{m+n}\bigg(\Big\{(\bx'\in([0,1]^d)^m, \by'\in ([0,1]^n):\phi_{\bbA}(\bx')+\phi_{\bbA}(\by')\geq \sqrt{t}\Big\}\bigg).\ena\end{equation}
The last equality in (\ref{lem13:1.2}) achieves because of $\phi_{\bbA}(\bx'),\phi_{\bbA}(\by')\geq 0$ and note that $\phi_{\bbA}(\bz')\geq \phi_{\bbA}(\bx')+\phi_{\bbA}(\by')$. Therefore
\beqq \begin{array}{l} \diy \mu^{m+n}\bigg(\Big\{(\bx'\in([0,1]^d)^m, \by'\in ([0,1]^n): \phi_{\bbA}(\bx')+\phi_{\bbA}(\by')\geq \sqrt{t}\Big\}\bigg)\\[15pt]
\qquad\quad\leq \diy \mu^{m+n}\bigg(\Big\{(\bz'\in([0,1]^d)^{m+n} : \phi_{\bbA}(\bz') \geq \sqrt{t}\Big\}\bigg).\ena\eeqq
By recalling  (\ref{Lem13:1.1}), we derive the bound (\ref{lem13:0.0}).
\end{IEEEproof}
\bigskip
\def\ep{\epsilon}
{\it Lemma \ref{lem.Deviation of the Mean and Median}:}
{\rm (Deviation of the Mean and Median)}  Consider $M_e$ as a median of $\mathfrak{R}_{m,n}$. Then for given $g(\epsilon)=1-\diy\frac{h\;\delta^h_{m,n}}{\ep}$, and $\delta^h_{m,n}=O\big(h^{d-1}(m+n)^{1/d}\big)$ such that for $h\geq 7$, $\epsilon\geq h^2 \delta^h_{m,n}$ we have
\begin{equation}\Big|\bbE\big[\mathfrak{R}_{m,n}(\mathfrak{X}_m,\mathfrak{Y}_n)\big]-M_e\Big|\leq C_{m,n}(\epsilon,h)\;(m+n)^{(d-1)/d},\end{equation}
where $C_{m,n}(\ep,h)$ stands with a form depends on $\ep$, $h$, $m$, $n$ as
\beq C_{m,n}(\ep,h)=\diy C\;\bigg(1- \Big(\big(2\;(2\;g(\epsilon)-1)^2\big)^{-1}\Big) \bigg)^{-1},\eeq
where $C$ is a constant. 
\begin{IEEEproof} Following the analogous arguments in \cite{PPS} and \cite{Yu}, we have
\begin{equation}\label{last:eq} \begin{array}{l}\Big|\bbE\big[\mathfrak{R}_{m,n}(\mathfrak{X}_m,\mathfrak{Y}_n)\big]-M_e\Big|\leq \bbE\Big|\mathfrak{R}_{m,n}(\mathfrak{X}_m,\mathfrak{Y}_n)-M_e\Big|
=\diy\int_0^\infty P\Big(\Big|\mathfrak{R}_{m,n}(\mathfrak{X}_m,\mathfrak{Y}_n)-M_e\Big|\geq t\Big)\;\rd t \\[15pt]
\qquad\leq \diy 8\;\bigg(1- \Big(1\Big/\big(2\;(2\;g(\epsilon)-1)^2\big)\Big) \bigg)^{-1}
\diy\int_0^\infty  \exp\Big(\diy\frac{-t^{d/(d-1)}}{8(4\ep)^{d/d-1}(m+n)}\Big)\;\rd t\\[15pt]
\qquad\qquad=\diy C\;\bigg(1- \Big(\big(2\;(2\;g(\epsilon)-1)^2\big)^{-1}\Big) \bigg)^{-1}\;(m+n)^{(d-1)/d},
\ena\end{equation}
where $g(\ep)=1-\Big(h\; O\big(h^{d-1}(m+n)^{1/d}\big)\Big)\big/\ep$. The inequality in (\ref{last:eq}) is implied from Theorem \ref{Concentration around the median}. Hence, the proof is completed. 
\end{IEEEproof}
\end{document}